\documentclass[letterpaper,twocolumn,10pt]{article}

\usepackage{chancery}
\newcommand{\chancery}[1]{{\fontfamily{pzc}\selectfont #1}}

\usepackage{usenix-2020-09}
\usepackage{amsmath,amssymb}
\usepackage[utf8]{inputenc}
\usepackage{epsfig,xspace,url,xcolor}
\usepackage{amsmath}
\usepackage{mathtools}
\usepackage{amsthm}
\usepackage{tikz}
\usetikzlibrary{arrows.meta,calc,positioning, patterns}
\usepackage{caption}

\usepackage{graphicx}
\usepackage[inline]{enumitem}
\usepackage{tikz}
\usepackage{thmtools}
\usepackage{thm-restate}
\usepackage{scalerel}
\usepackage[ruled, lined, linesnumbered, commentsnumbered, noend]{algorithm2e}
\usepackage[noend]{algpseudocode}
\usepackage{graphicx}
\usepackage{blindtext}
\usepackage{url}
\usepackage{subcaption}
\usepackage{mathtools}
\usepackage{xcolor}
\usepackage{color, colortbl}
\usepackage{tablefootnote}
\usetikzlibrary{positioning, shapes.geometric, shadows}
\usepackage{graphbox}
\usepackage{url}

\usepackage[normalem]{ulem}
\usepackage{wrapfig}
\usepackage{wrapstuff}
\usepackage{breakurl}

\SetKwInOut{Input}{Input}
\SetKwInOut{Output}{Output}
\SetKwFunction{PullForward}{PullForward}
\SetKwFunction{ResetAndEstablish}{ResetAndEstablish}
\SetKwFunction{CanReach}{CanReach}
\SetKwFunction{Establish}{Establish}
\SetKwFunction{Release}{Release}
\SetKwProg{Fn}{Function}{:}{end}

\usepackage[subtle]{savetrees}
\usepackage{blindtext}

\newcommand{\name}{{\chancery{Tetris}}\xspace}

\usepackage{titlesec}
\titlespacing*{\section}{0ex}{2ex plus .2ex minus .2ex}{1ex plus .2ex minus .2ex}
\titlespacing*{\subsection}{0ex}{1ex plus .2ex minus .2ex}{1ex plus .2ex minus .2ex}

\usepackage{enumitem}

\definecolor{darkgreen}{rgb}{0.01, 0.75, 0.24}
\definecolor{darkcyan}{rgb}{0.0, 0.55, 0.55}
\definecolor{mycolor}{rgb}{0.01, 0.75, 0.24}

\definecolor{mygray}{gray}{0.9}
\definecolor{myred}{rgb}{0.75, 0, 0}

\definecolor{myblue}{HTML}{3ba9d1}
\definecolor{myyellow}{HTML}{f2e094}
\definecolor{mydarkred}{HTML}{c27676}
\definecolor{mygreen}{HTML}{7dc276}

\definecolor{alg1}{rgb}{0.855, 1, 0.824}
\definecolor{alg2}{rgb}{1, 0.996, 0.824}
\definecolor{alg3}{rgb}{0.824, 0.914, 1}

\mathchardef\hyphen="2D

\newcommand{\myitem}[1]{\vspace{1mm}\noindent\textbf{#1}}

\newcommand{\remove}[1]{}

\usepackage{xcolor}
\definecolor{light-gray}{gray}{0.9}
\usepackage{tcolorbox}

\newcommand{\cref}[1]{Chapter~\ref{#1}}

\newcommand{\Demand}{\ensuremath{\mathcal{D}}\xspace}
\newcommand{\Trans}{\ensuremath{\hat{\Demand}}}
\makeatletter
\newcommand{\oset}[2]{%
  \mathop{#2}\limits^{
  \vbox to-.5\ex@{\kern-2\ex@
   \hbox{$\scriptstyle#1$}\vss}}}
\makeatother

\DeclareSymbolFont{extraup}{U}{zavm}{m}{n}
\DeclareMathSymbol{\varheart}{\mathalpha}{extraup}{86}
\DeclareMathSymbol{\vardiamond}{\mathalpha}{extraup}{87}

\newcommand{\SNB}{SNB\xspace}
\newcommand{\RNB}{RNB\xspace}
\newcommand{\B}{Blocking\xspace}

\newcommand*\largeblackcircled[1]{\tikz[baseline=(char.base)]{
            \node[scale=1.4,shape=circle,draw, fill,inner sep=3.5pt,color=black!80] (char) {};
            \node[scale=1.1] (char) {\textcolor{myyellow}{\footnotesize{\textbf{#1}}}}}}

\newcommand{\takeaway}[1]{{\textcolor{black}{$\blacksquare$ \textbf{\textit{Takeaway:}}} \textit{#1}}}

\begin{document}

\title{Tetris: Circuit Scheduling for Rearrangeably Non-Blocking Photonic Interconnects}

\author{
    {\rm Eliezer Amponsah\footnotemark[1]}\\
    Purdue University
    \and
    {\rm Deeksha P Rao\footnotemark[1]}\\
    Purdue University
    \and
    {\rm Vamsi Addanki}\\
    Purdue University
}

\sloppy
\maketitle

\begingroup
\renewcommand{\thefootnote}{\fnsymbol{footnote}}
\footnotetext[1]{Co-first authors with equal contribution.}
\endgroup

\thispagestyle{plain}
\pagestyle{plain}

\begin{abstract}
Reconfigurable photonic interconnects are emerging as a promising communication architecture for next-generation distributed computing. Yet, most circuit schedulers are designed around an idealized view of the interconnect as either blocking or strictly non-blocking. Practical scalable designs are often rearrangeably non-blocking (RNB), with connections routed through networks of internal $2\times2$ switches. This changes the scheduling problem fundamentally: establishing a new connection can force existing connections to be rerouted, trigger state changes across multiple internal switches, and impose reconfiguration delay on otherwise unrelated traffic.

We present \name, a circuit scheduling algorithm for RNB photonic interconnects. \name builds on two observations. First, All-to-All demands are often not doubly stochastic, leaving a small number of endpoints as communication bottlenecks. Second, reconfiguration delay can be large enough to change which connection should be scheduled next. \name prioritizes bottleneck endpoints using their remaining communication and reconfiguration work, while selecting and routing matchings to preserve ongoing connections whenever possible. Matchings ensure progress, but connections are scheduled independently, allowing completed connections to be replaced without matching-wide barriers and incurring delay only at switches whose states change. Our simulation and hardware-emulation results show that \name reduces All-to-All demand completion time by up to $6.6$x over Birkhoff--von Neumann-based scheduling and by $30$\% over Sunflow. More broadly, RNB interconnects raise new questions in multi-tenant scheduling and routing for partial reconfiguration, which we discuss at the end of the paper.
\end{abstract}

\begin{figure*}[t]
\centering
\resizebox{\linewidth}{!}{%
\begin{tikzpicture}[
    x=1.15cm,
    y=1.25cm,
    >=stealth,
    port/.style={circle,fill=black,inner sep=1.4pt},
    sw/.style={
        draw,
        rounded corners=2pt,
        minimum width=9mm,
        minimum height=16mm,
        fill=gray!8,
        line width=0.8pt
    },
    swchanged/.style={
        draw,
        rounded corners=2pt,
        minimum width=9mm,
        minimum height=16mm,
        fill=orange!28,
        line width=1.2pt
    },
    baselink/.style={gray!65,line width=0.9pt},
    flowA/.style={blue!70!black,line width=2.6pt},
    flowB/.style={red!75!black,line width=2.6pt},
    flowC/.style={green!50!black,line width=2.6pt},
    lab/.style={font=\small},
    title/.style={font=\small\bfseries}
]

\begin{scope}[shift={(0,0)}]

    \node[title] at (3,4.0) {Existing circuits};

    \foreach \y/\name in {3/1,2/2,1/3,0/4} {
        \coordinate (LI\name) at (0,\y);
        \coordinate (LO\name) at (6,\y);

        \node[port] at (LI\name) {};
        \node[port] at (LO\name) {};

        \node[lab,left=3pt] at (LI\name) {$i_{\name}$};
        \node[lab,right=3pt] at (LO\name) {$o_{\name}$};
    }

    \node[sw] at (1.4,2.5) {};
    \node[sw] at (1.4,0.5) {};

    \node[sw] at (3.0,2.5) {};
    \node[sw] at (3.0,0.5) {};

    \node[sw] at (4.6,2.5) {};
    \node[sw] at (4.6,0.5) {};

    \draw[baselink] (LI1) -- (1.1,3);
    \draw[baselink] (LI2) -- (1.1,2);
    \draw[baselink] (LI3) -- (1.1,1);
    \draw[baselink] (LI4) -- (1.1,0);

    \draw[baselink] (1.7,3) -- (2.7,3);
    \draw[baselink] (1.7,2) -- (2.7,1);
    \draw[baselink] (1.7,1) -- (2.7,2);
    \draw[baselink] (1.7,0) -- (2.7,0);

    \draw[baselink] (3.3,3) -- (4.3,3);
    \draw[baselink] (3.3,2) -- (4.3,1);
    \draw[baselink] (3.3,1) -- (4.3,2);
    \draw[baselink] (3.3,0) -- (4.3,0);

    \draw[baselink] (4.9,3) -- (LO1);
    \draw[baselink] (4.9,2) -- (LO2);
    \draw[baselink] (4.9,1) -- (LO3);
    \draw[baselink] (4.9,0) -- (LO4);

    \draw[flowA]
        (LI1)
        -- (1.1,3)
        -- (1.7,2)
        -- (2.7,1)
        -- (3.3,1)
        -- (4.3,2)
        -- (4.9,2)
        -- (LO2);

    \node[lab,blue!70!black,above]
        at (0.4,3)
        {$A:i_1\!\rightarrow\!o_2$};

    \draw[flowB]
        (LI3)
        -- (1.1,1)
        -- (1.7,1)
        -- (2.7,2)
        -- (3.3,2)
        -- (4.3,1)
        -- (4.9,1)
        -- (LO3);

    \node[lab,red!75!black,below]
        at (0.4,0.95)
        {$B:i_3\!\rightarrow\!o_3$};

\end{scope}

\draw[->,very thick,darkgray]
    (6.85,1.5) -- (8.35,1.5)
    node[
        midway,
        above=5pt,
        lab,
        align=center,
        fill=white,
        inner sep=2pt
    ]
    {admit $C$\\$i_2\!\rightarrow\!o_4$};

\begin{scope}[shift={(9.15,0)}]

    \node[title] at (3,4.0) {After partial reconfiguration};

    \foreach \y/\name in {3/1,2/2,1/3,0/4} {
        \coordinate (RI\name) at (0,\y);
        \coordinate (RO\name) at (6,\y);

        \node[port] at (RI\name) {};
        \node[port] at (RO\name) {};

        \node[lab,left=3pt] at (RI\name) {$i_{\name}$};
        \node[lab,right=3pt] at (RO\name) {$o_{\name}$};
    }

    \node[sw]        at (1.4,2.5) {};
    \node[swchanged] at (1.4,0.5) {};

    \node[swchanged] at (3.0,2.5) {};
    \node[sw]        at (3.0,0.5) {};

    \node[sw]        at (4.6,2.5) {};
    \node[swchanged] at (4.6,0.5) {};

    \draw[baselink] (RI1) -- (1.1,3);
    \draw[baselink] (RI2) -- (1.1,2);
    \draw[baselink] (RI3) -- (1.1,1);
    \draw[baselink] (RI4) -- (1.1,0);

    \draw[baselink] (1.7,3) -- (2.7,3);
    \draw[baselink] (1.7,2) -- (2.7,1);
    \draw[baselink] (1.7,1) -- (2.7,2);
    \draw[baselink] (1.7,0) -- (2.7,0);

    \draw[baselink] (3.3,3) -- (4.3,3);
    \draw[baselink] (3.3,2) -- (4.3,1);
    \draw[baselink] (3.3,1) -- (4.3,2);
    \draw[baselink] (3.3,0) -- (4.3,0);

    \draw[baselink] (4.9,3) -- (RO1);
    \draw[baselink] (4.9,2) -- (RO2);
    \draw[baselink] (4.9,1) -- (RO3);
    \draw[baselink] (4.9,0) -- (RO4);

    \draw[flowA]
        (RI1)
        -- (1.1,3)
        -- (1.7,2)
        -- (2.7,1)
        -- (3.3,1)
        -- (4.3,2)
        -- (4.9,2)
        -- (RO2);

    \node[lab,blue!70!black,above]
        at (0.4,3)
        {$A$ preserved};

    \draw[flowC]
        (RI2)
        -- (1.1,2)
        -- (1.7,3)
        -- (2.7,3)
        -- (3.3,2)
        -- (4.3,1)
        -- (4.9,0)
        -- (RO4);

    \node[lab,green!50!black,above]
        at (0.3,2)
        {$C$ new};

    \draw[flowB]
        (RI3)
        -- (1.1,1)
        -- (1.7,0)
        -- (2.7,0)
        -- (3.3,0)
        -- (4.3,0)
        -- (4.9,1)
        -- (RO3);

    \node[lab,red!75!black,below]
        at (0.3,1)
        {$B$ rerouted};

\end{scope}

\begin{scope}[shift={(2.2,-1.0)}]

    \draw[flowA] (0,0) -- (0.65,0);
    \node[lab,right] at (0.75,0) {Preserved circuit};

    \draw[flowB] (4.2,0) -- (4.85,0);
    \node[lab,right] at (4.95,0) {Rerouted circuit};

    \draw[flowC] (8.35,0) -- (9.0,0);
    \node[lab,right] at (9.10,0) {New circuit};

    \node[swchanged,minimum width=6mm,minimum height=7mm]
        at (12.55,0) {};
    \node[lab,right] at (12.95,0)
        {Switch state changed};

\end{scope}

\end{tikzpicture}%
}
\caption{
A new circuit can disrupt unrelated ongoing traffic in a rearrangeably
non-blocking (RNB) interconnect.
Initially, $A:i_1\!\rightarrow\!o_2$ and $B:i_3\!\rightarrow\!o_3$ are active.
Admitting the endpoint-disjoint connection $C:i_2\!\rightarrow\!o_4$
requires changing internal $2\times2$ switch states.
$A$ can retain its path, but $B$ must be rerouted despite sharing no endpoint
with $C$, incurring additional reconfiguration delay.
}
\label{fig:rnb-dependency}
\vspace{-4mm}
\end{figure*}
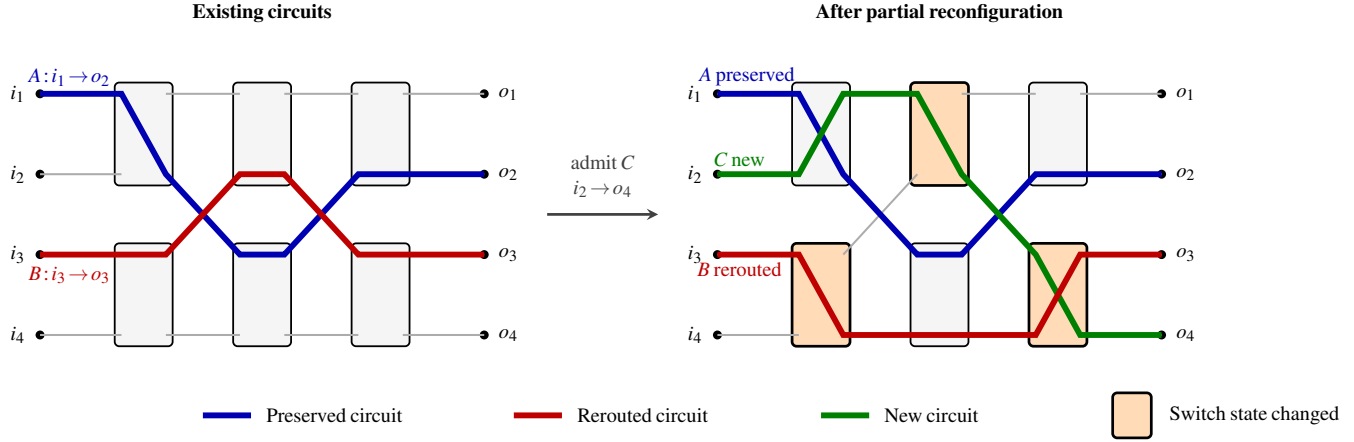

\section{Introduction}

The rapid growth of AI/ML and other communication-intensive workloads is placing unprecedented pressure on datacenter networks. As compute capacity continues to scale, communication is increasingly becoming the bottleneck. Conventional electrical packet-switched networks face growing challenges in bandwidth density, power consumption, and cabling complexity as clusters expand. Recent advances in silicon photonics~\cite{TeraPhy,swithfabric,piloss,4x4,bogaerts2020programmable,10.1145/3718958.3750468} have renewed interest in reconfigurable photonic interconnects for datacenters~\cite{missionapollo,case_server,light_will,Helios,Sirius,Rotornet,10.1145/3452296.3472900,285119,10.1145/3603269.3604836,305352,316650,316652}. By establishing high-bandwidth optical circuits directly between communicating endpoints, photonic interconnects offer an attractive path toward scalable, low-latency, and energy-efficient communication.

This flexibility comes with a fundamental challenge: photonic interconnects must be explicitly reconfigured as communication demands change. Each reconfiguration incurs a non-negligible delay and can temporarily disrupt communication. Application performance thus depends not only on which circuits are established, but also on when they are established, how long they remain active, and which parts of the interconnect must change to realize them. Circuit scheduling is consequently central to extracting the performance benefits of reconfigurable photonic interconnects.

Existing circuit schedulers broadly fall into two categories: demand-oblivious~\cite{Sirius,Rotornet,MARS,Shale,Breaking_VLB} and demand-aware~\cite{vermilion,Helios,ProjecToR,Jupiter}. Demand-oblivious schedulers cycle through predetermined configurations, simplifying control but sacrificing throughput when traffic deviates from that sequence~\cite{MARS,Shale,10.1145/3519935.3520020,Sirius}. Demand-aware schedulers instead adapt the interconnect to the observed communication matrix~\cite{Liu2015SchedulingNetworks,BojjaVenkatakrishnan2016CostlyTheorems,ProjecToR,Helios}, typically by constructing a sequence of matchings and assigning each a duration.

Demand-matrix decomposition is a central problem in demand-aware scheduling. Classical approaches include Birkhoff--von Neumann (BvN) decomposition~\cite{BIRKHOFF1946TresLineal}, maximum-weight matching~\cite{Kuhn1955TheProblem,Jonker1987AProblems,Crouse2016OnAlgorithms}, Eclipse~\cite{BojjaVenkatakrishnan2016CostlyTheorems}, and numerous variants~\cite{vermilion,Liu2015SchedulingNetworks,Bianco2016SchedulingFabrics}. Much of the literature focuses on selecting, ordering, and timing these matchings to reduce completion time while amortizing reconfiguration delay.

A key assumption behind nearly all of this work is that the underlying interconnect is either blocking (\B) or strictly non-blocking (\SNB). In a \B interconnect, each reconfiguration interrupts all communication, forcing schedulers to organize decisions around globally synchronized pauses~\cite{Liu2015SchedulingNetworks,BojjaVenkatakrishnan2016CostlyTheorems}. At the other extreme, \SNB schedulers assume that a new connection between idle endpoints can be established without disturbing existing circuits~\cite{BFF,Sunflow,Interruptible}. This makes partial reconfiguration straightforward: completed connections can be replaced while ongoing ones continue transmitting.

Real scalable photonic interconnects do not fit either abstraction cleanly. Consider a Bene\v{s} interconnect: it can realize any input-output permutation, yet establishing a new circuit may require rerouting existing circuits through its internal $2\times2$ switches. As Figure~\ref{fig:rnb-dependency} illustrates, even a new connection between otherwise idle endpoints can force an unrelated active circuit onto a different path, changing multiple switch states and introducing additional reconfiguration delay. Such interconnects are rearrangeably non-blocking (\RNB): they retain full permutation capability, but only by allowing existing routes to move. This makes circuit scheduling inherently dependent on interconnect routing.

This distinction fundamentally changes the scheduling problem. Communication demand matrices in modern distributed workloads are rarely doubly stochastic and are often highly skewed: a small number of heavily loaded endpoints can determine completion time, while many other connections finish much earlier. These early completions create natural opportunities for partial reconfiguration by replacing completed connections while preserving ongoing ones. In an \RNB interconnect, however, whether such an update is inexpensive depends on how both the existing and new connections are routed. A seemingly local change can disturb long-running traffic elsewhere in the interconnect. Existing schedulers largely ignore these routing dependencies, leaving a gap between the abstractions used by circuit schedulers and the behavior of practical scalable photonic interconnects.

In this paper, we revisit circuit scheduling for rearrangeably non-blocking photonic interconnects with \name, a scheduler that exploits two properties existing approaches largely overlook: communication demand is often skewed, leaving a few endpoints as completion-time bottlenecks, and reconfiguration delay can change which connection should be scheduled next. \name prioritizes endpoints based on their remaining communication and reconfiguration work, selects matchings that guarantee progress on these bottlenecks, and routes them to preserve ongoing connections whenever the interconnect permits. Crucially, matchings provide only the progress guarantee; individual connections are scheduled independently according to the physical paths and switch states they use. This allows \name to partially reconfigure the interconnect without imposing matching-wide barriers or unnecessarily interrupting unrelated traffic. Our evaluation results show that \name reduces All-to-All demand completion time in Mixture-of-Expert workloads by up to $30\%$ compared with Sunflow~\cite{Sunflow} and Best First Fit (BFF)~\cite{BFF}, and by up to $6\times$ compared with BvN, while approaching the performance of an ideal \SNB scheduler on realistic workloads.

More broadly, \RNB interconnects raise systems questions that idealized switching models largely hide, including multi-tenant interference and routing strategies for partial reconfiguration. We briefly discuss these directions at the end of the paper and their implications for future photonic interconnects and circuit schedulers.

In summary, our key contributions are as follows:
\begin{itemize}[label=\small{$\blacksquare$}]
    \item We identify a fundamental mismatch between existing circuit scheduling abstractions and practical rearrangeably non-blocking photonic interconnects, where scheduling decisions are tightly shaped by internal routing and switch-state changes.

    \item We present \name, a greedy circuit scheduler for \RNB interconnects that prioritizes bottleneck endpoints using their remaining communication and reconfiguration work. \name selects matchings that guarantee progress, routes them to preserve ongoing connections whenever possible, and schedules individual connections independently to enable partial reconfiguration without matching-wide barriers.

    \item We evaluate \name using simulation and hardware emulation. \name reduces All-to-All demand completion time by up to $30\%$ compared with Sunflow~\cite{Sunflow} and Best First Fit (BFF)~\cite{BFF}, and by up to $6\times$ compared with BvN, while approaching the performance of an ideal \SNB scheduler on realistic workloads.
\end{itemize}

\section{Motivation}
\label{sec:bg-motive}

We motivate the need for circuit scheduling that explicitly accounts for rearrangeably non-blocking (\RNB) photonic interconnects. We first describe the switch architecture and practical assumptions in this work (\S\ref{subsec:architecture}). We then show how skewed All-to-All demands create substantial idle periods under conventional matching-based schedulers (\S\ref{subsec:challenges-scheduling}), motivating partial reconfiguration as a way to recover this lost capacity. We next explain why partial reconfiguration is fundamentally more challenging in \RNB interconnects (\S\ref{subsec:partial-reconfig}), where establishing a new circuit can disrupt ongoing traffic through internal routing dependencies. These observations lead directly to the design goals of \name (\S\ref{subsec:goals}).

\subsection{Switch Architecture and Assumptions}
\label{subsec:architecture}

The benefit of fine-grained partial reconfiguration depends on two hardware properties: reconfiguration delay and the ability to modify individual connections without disturbing others. Slow-reconfiguring optical switches, such as the MEMS-based switches in Google's Jupiter network~\cite{Jupiter}, operate on millisecond timescales. Their configurations are typically held for long periods, making frequent connection-level updates unattractive relative to communication time. Recent photonic switching technologies can reconfigure orders of magnitude faster, bringing circuit scheduling into a regime where individual reconfiguration decisions can directly affect demand completion time.

We consider chip-to-chip photonic interconnects connecting a modest number of compute devices. Each chip has one outgoing transmit and one incoming receive link, each with bandwidth $b$. The interconnect consists of independently configurable $2\times2$ switches, each operating in a \emph{bar} or \emph{cross} state under thermal or electrical control driven by external circuitry. An end-to-end circuit is established by configuring a sequence of switches along a path from a transmitter to a receiver; reconfiguring a connection therefore requires changing some subset of switches along that path.

How these switches are interconnected determines both interconnect flexibility and hardware cost. Strictly non-blocking (\SNB) architectures provide the strongest flexibility: any idle transmitter-receiver pair can be connected without disturbing existing circuits. At high radix, however, this flexibility requires more switching elements and often longer optical paths. Rearrangeably non-blocking (\RNB) interconnects offer a more scalable alternative. They can realize any input-output permutation, but a new connection cannot always be established while keeping all existing circuits fixed~\cite{Clos1953ASO,10.1145/321439.321449}; some active circuits may need to move to different internal paths, as illustrated in Figure~\ref{fig:rnb-dependency}.

\begin{figure}[!h]
\centering
\includegraphics[width=0.8\linewidth]{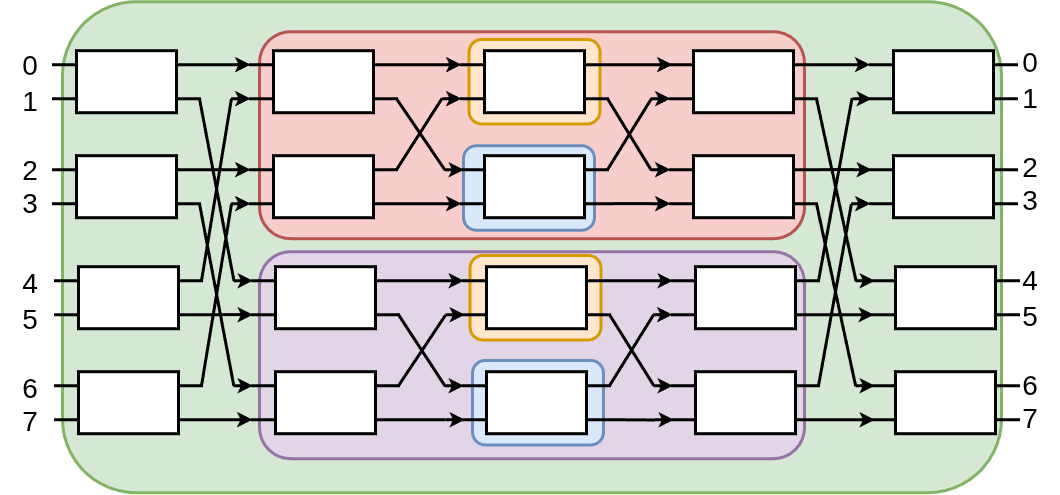}
\caption{Recursive construction of an $8\times8$ Bene\v{s}.}
\label{fig:benes}
\vspace{-3mm}
\end{figure}

The Bene\v{s} interconnect is particularly attractive in this design space. An $n\times n$ Bene\v{s} interconnect realizes arbitrary permutations using only $n\cdot\log_2 n-n/2$ switching elements across $2\cdot\log_2 n-1$ stages, reducing hardware complexity relative to several \SNB alternatives. Its recursive construction, shown in Figure~\ref{fig:benes}, builds larger interconnects from smaller Bene\v{s} networks and naturally scales to higher radix. Table~\ref{tab:topo} summarizes these trade-offs across representative switching topologies. We use Bene\v{s} as our representative \RNB interconnect throughout this paper.

\begin{figure*}
\centering
\begin{subfigure}{0.19\linewidth}
\centering
\includegraphics[trim=0cm 1.5cm 0cm 2cm, clip, width=1\linewidth]{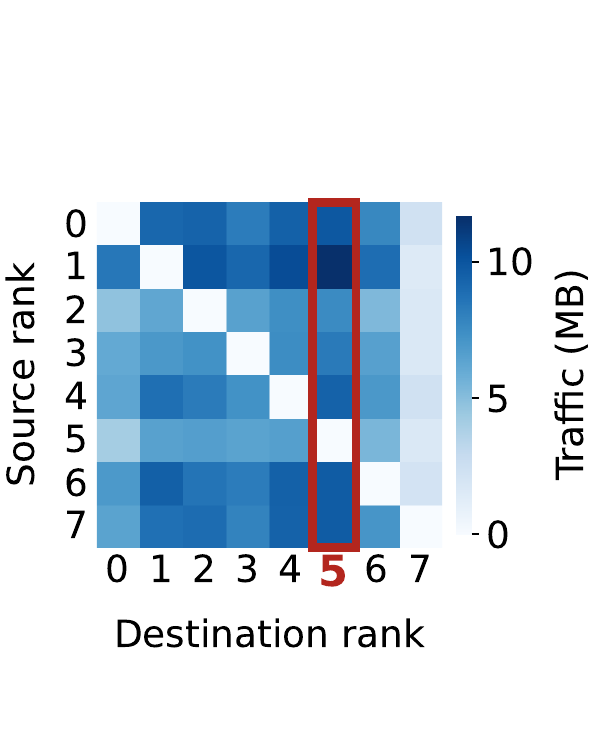}
\caption{Layer $28$}
\end{subfigure}\hfill
\begin{subfigure}{0.19\linewidth}
\centering
\includegraphics[trim=0cm 1.5cm 0cm 2cm, clip, width=1\linewidth]{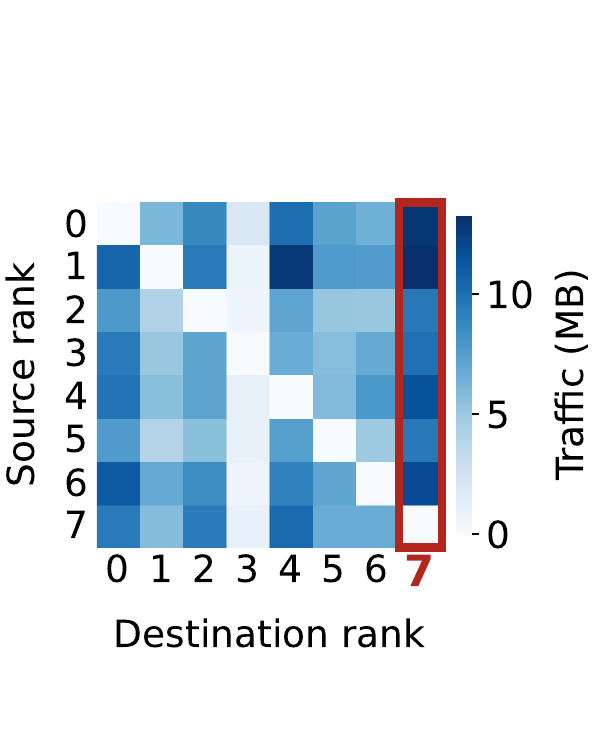}
\caption{Layer $29$}
\end{subfigure}\hfill
\begin{subfigure}{0.19\linewidth}
\centering
\includegraphics[trim=0cm 1.5cm 0cm 2cm, clip, width=1\linewidth]{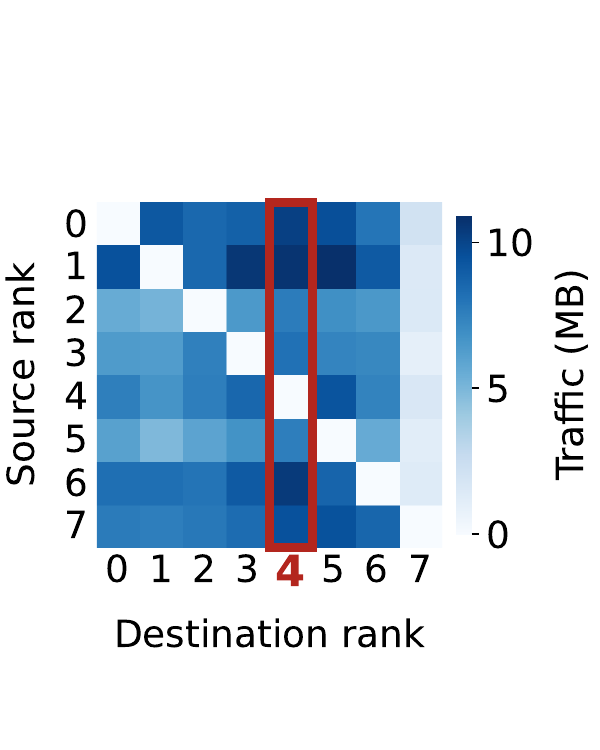}
\caption{Layer $30$}
\label{fig:traffic-layer30}
\end{subfigure}\hfill
\begin{subfigure}{0.19\linewidth}
\centering
\includegraphics[trim=0cm 1.5cm 0cm 2cm, clip, width=1\linewidth]{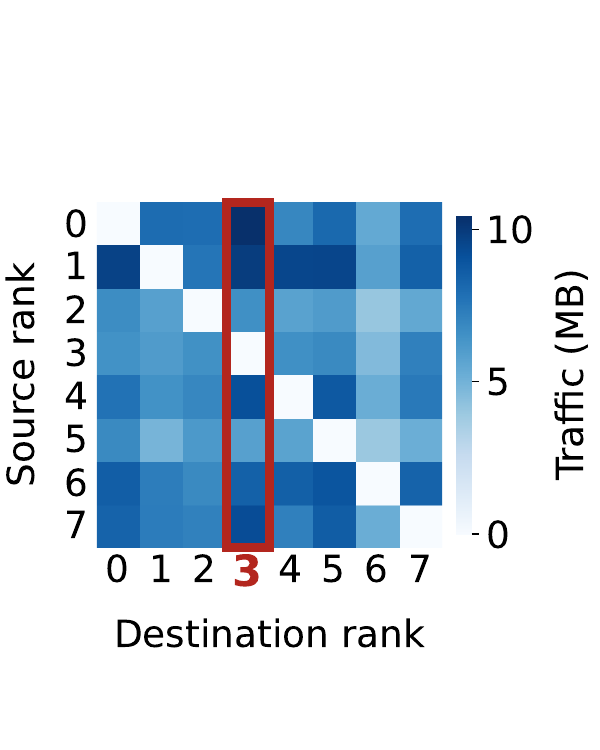}
\caption{Layer $31$}
\end{subfigure}\hfill
\begin{subfigure}{0.19\linewidth}
\centering
\includegraphics[trim=0cm 1.5cm 0cm 2cm, clip, width=1\linewidth]{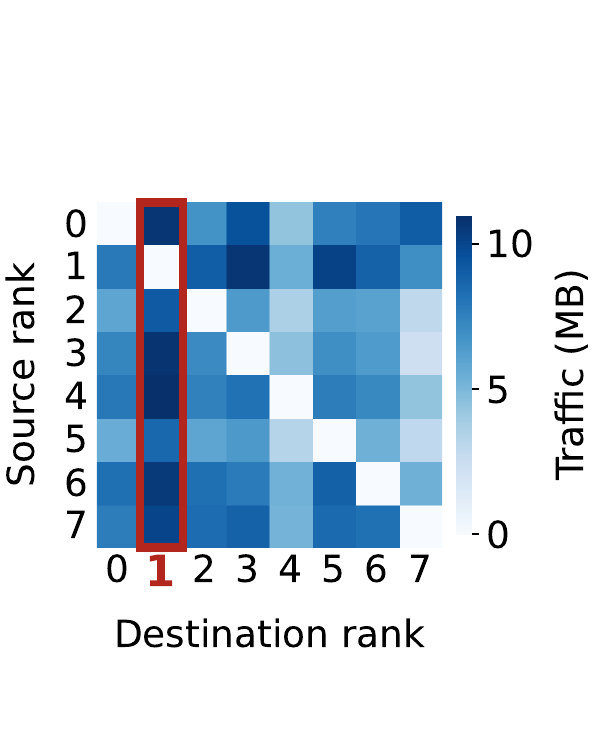}
\caption{Layer $32$}
\end{subfigure}\hfill
\vspace{-2mm}
\caption{All-to-All traffic matrices from consecutive layers of a Mixtral-$8$x$22$B workload. The demand is rarely doubly stochastic and is often highly skewed, with a small number of receivers emerging as communication bottlenecks.}
\label{fig:traffic-matrices}
\vspace{-4mm}
\end{figure*}

The scalability of \RNB interconnects comes with an important scheduling consequence. Endpoint availability alone no longer determines whether a partial reconfiguration is possible. A new connection may require changing internal $2\times2$ switches already used by unrelated traffic, forcing those circuits to be rerouted or interrupted. Circuit scheduling must thus account not only for which transmitters and receivers are available, but also for the internal paths occupied by ongoing connections. We return to this interaction in \S\ref{subsec:partial-reconfig}.

\smallskip
\takeaway{Fast-reconfiguring photonic interconnects make fine-grained circuit scheduling practical, but scalable designs increasingly require circuit schedulers to reason about the interconnect architecture itself.}

\subsection{Challenges of Scheduling All-to-All Traffic}
\label{subsec:challenges-scheduling}

Circuit scheduling has traditionally been approached through demand-matrix decomposition, where the communication demand is decomposed into a sequence of permutations that are executed as matchings on the interconnect. Modern All-to-All workloads challenge several assumptions behind this approach.

\myitem{Traffic matrices are rarely doubly stochastic:}
Figure~\ref{fig:traffic-matrices} shows All-to-All traffic matrices from consecutive layers of Mixtral-$8\times22$B. The demand is highly asymmetric: GPUs send and receive different amounts of traffic, and a small number of receivers often emerge as clear communication bottlenecks. In contrast, a doubly stochastic demand matrix has equal row and column sums. Classical BvN decomposition relies on this structure. Applying BvN to an asymmetric demand matrix requires augmenting or normalizing the demand so that all transmitters and receivers have equal total load~\cite{Sinkhorn1964AMatrices,316654,Liu2015SchedulingNetworks}. The additional entries do not correspond to useful communication and appear as idle capacity when the resulting permutations are executed on the interconnect. Although BvN provides a clean decomposition for doubly stochastic demand, real All-to-All traffic can produce schedules with significant unused capacity.

\myitem{BvN requires excessive reconfigurations:}
BvN decomposition repeatedly extracts a permutation from the remaining demand and subtracts the largest duration supported by that permutation. This process can fragment the original demand across $O(n^2)$ matchings in the worst case. When each matching is separated by a reconfiguration barrier, a schedule with $K$ matchings incurs roughly $K$ reconfiguration events in addition to the transmission time. This overhead becomes increasingly important as reconfiguration delay grows. Figure~\ref{fig:bvn-timeline} illustrates this effect: the demand in Figure~\ref{fig:traffic-layer30} is spread across many short matchings, repeatedly interrupting transmission to reconfigure the interconnect.

\myitem{Reducing reconfigurations creates scheduling bubbles:}
An alternative is to construct fewer, longer-lived matchings. MaxWeight decomposition, for example, prioritizes high-demand connections and significantly reduces the number of reconfigurations compared with BvN. Fewer matchings, however, do not eliminate the asymmetry of the underlying traffic. Within each matching, connection durations can differ significantly, while a blocking scheduler cannot advance to the next matching until the longest connection completes. Connections that finish earlier leave their transmitters and receivers idle for the remainder of the matching. We refer to these intervals as \emph{scheduling bubbles}. Figure~\ref{fig:maxweight-timeline} shows this behavior for the same Mixtral traffic matrix: MaxWeight avoids much of BvN's reconfiguration overhead, but large portions of the interconnect remain idle while each matching waits for its bottleneck connection to finish.

\smallskip
\takeaway{Skewed All-to-All traffic creates a simple trade-off: many short matchings waste time on reconfiguration, while fewer long matchings leave ports idle waiting for bottleneck connections to finish.}

\subsection{Opportunities for Partial Reconfiguration}
\label{subsec:partial-reconfig}

\begin{figure}[!t]
\vspace{0pt}
\centering

\begin{subfigure}[t]{0.48\linewidth}
\vspace{0pt}
\centering
\includegraphics[width=\linewidth]{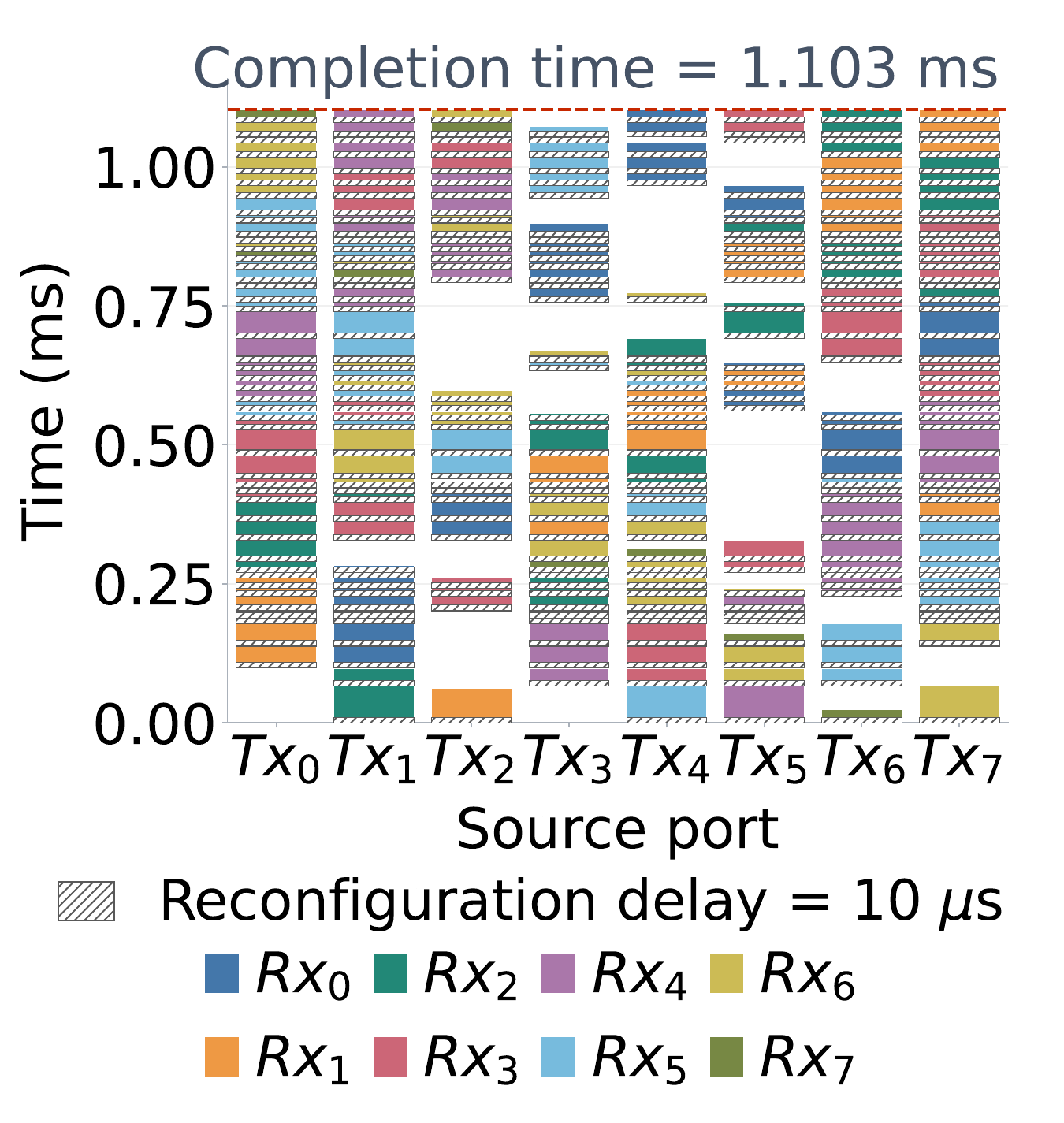}
\caption{BvN}
\label{fig:bvn-timeline}
\end{subfigure}
\hfill
\begin{subfigure}[t]{0.48\linewidth}
\vspace{0pt}
\centering
\includegraphics[width=\linewidth]{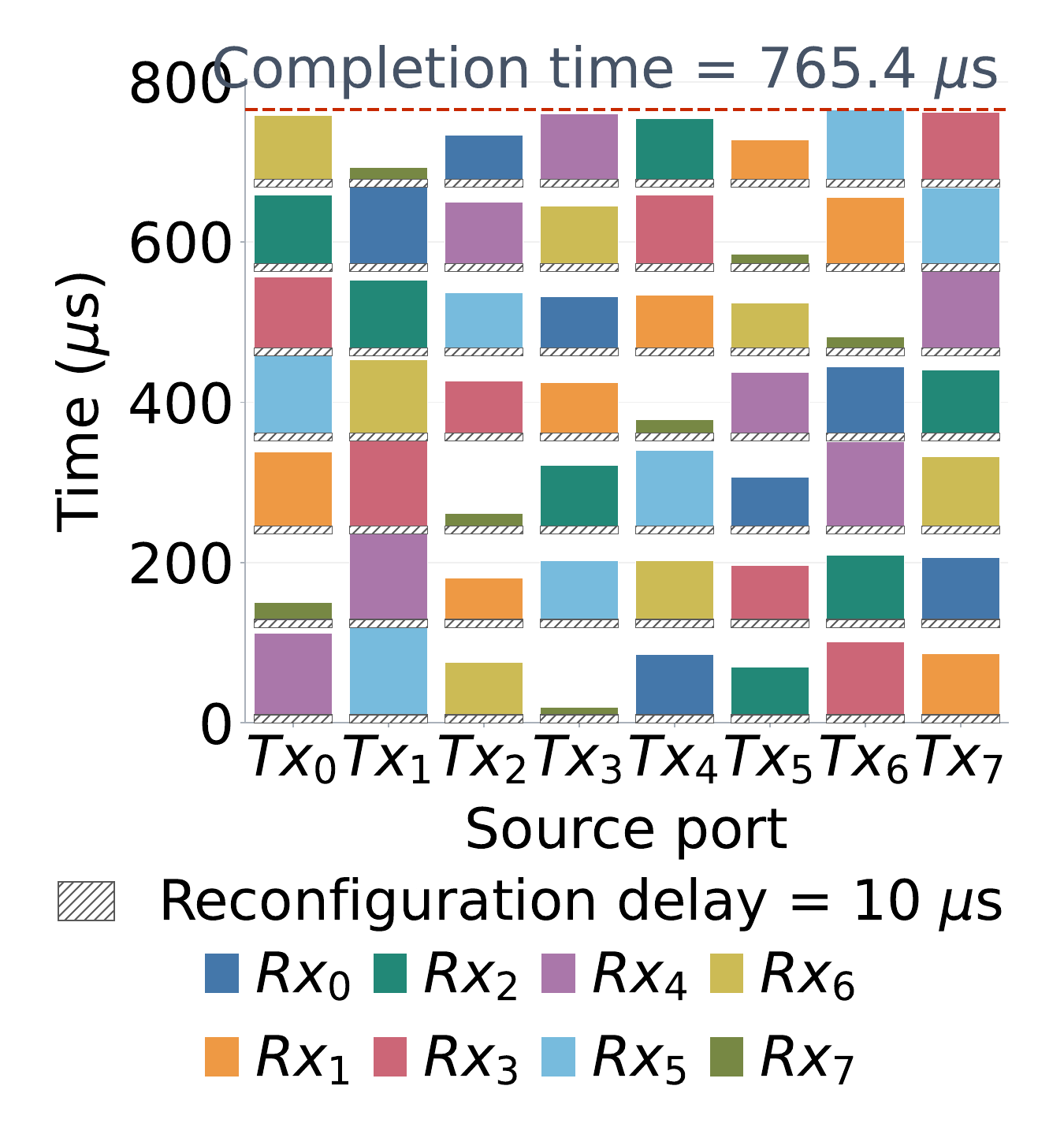}
\caption{MaxWeight}
\label{fig:maxweight-timeline}
\end{subfigure}
\vspace{-2mm}
\caption{BvN and MaxWeight, two widely studied circuit schedulers, both use blocking, per-matching reconfiguration barriers. (a) BvN incurs many reconfigurations and long idle periods when matchings are highly asymmetric, as is common for non-doubly-stochastic traffic (Figure~\ref{fig:traffic-matrices}). (b) MaxWeight reduces reconfigurations, but short-lived connections still finish early, leaving parts of the interconnect idle until the bottleneck connection in each matching completes.}
\label{fig:timeline-bvn-maxweight}
\vspace{-4mm}
\end{figure}

One way to fill the scheduling bubbles in Figure~\ref{fig:timeline-bvn-maxweight} is to avoid waiting for an entire matching to complete. As soon as a connection finishes, its now-idle transmitter and receiver can be reconfigured to serve remaining demand while longer-running connections continue transmitting. This \emph{partial reconfiguration} removes the matching-wide barrier and allows connections from different matchings to overlap in time.

\myitem{Partial reconfiguration can reduce scheduling bubbles:}
Schedulers such as Best First Fit (BFF)~\cite{BFF}, Sunflow~\cite{Sunflow}, and NEIP~\cite{Interruptible} exploit this opportunity by tracking when individual transmitters and receivers become available and immediately assigning them new connections. For the highly skewed traffic discussed in \S\ref{subsec:challenges-scheduling}, this can recover capacity that would otherwise remain idle while waiting for the longest connection in a matching. Partial reconfiguration breaks the matching-wide barrier, allowing completed connections to be replaced while longer-running connections continue transmitting.

\myitem{Bottleneck endpoints dictate completion time:} \label{subsubsec:bottleneck}
Not all connections contribute equally to overall completion time. With one transmit and one receive link of bandwidth $b$ per endpoint, any schedule is lower bounded by the time required to drain the most heavily loaded transmitter or receiver:
$
    T_{\min}
    =
    \frac{1}{b}
    \max\left(
        \max_i \sum_j \Demand_{ij},
        \max_j \sum_i \Demand_{ij}
    \right)
$,
where $\Demand_{ij}$ denotes the demand from source $i$ to destination $j$. We refer to any endpoint attaining this maximum as a \emph{bottleneck endpoint}. Figure~\ref{fig:traffic-matrices} highlights such bottlenecks in red. No schedule can complete before the demand associated with its bottleneck endpoint is drained. This makes bottleneck progress especially important for skewed All-to-All traffic: many connections may finish early, but delaying traffic through a bottleneck endpoint directly increases demand completion time. A partial-reconfiguration scheduler should use newly available capacity while progressing these bottleneck endpoints continuously.

\myitem{Partial reconfiguration can slow down unrelated traffic:}
Existing partial-reconfiguration schedulers largely assume a strictly non-blocking (\SNB) interconnect, where a new connection between idle endpoints can be established without disturbing ongoing circuits. This assumption breaks in an \RNB interconnect. As discussed in \S\ref{subsec:architecture}, establishing a new connection may require changing internal $2\times2$ switches already used by an active circuit, rerouting or interrupting otherwise unrelated traffic, including traffic through a bottleneck endpoint.

A second subtlety is that the global bottleneck endpoint in $\Demand$ need not be the bottleneck within every matching. Consider a naive policy that triggers partial reconfiguration whenever the global bottleneck finishes its current sub-flow. That decision may interrupt a different connection that is the local bottleneck of the current matching. The interrupted connection must then be re-established later, incurring another reconfiguration delay, even though the global bottleneck continues making progress. In an \RNB interconnect, later routing choices can further restrict which residual demands can be partially reconfigured. An effective scheduler must account for both global bottleneck progress and the remaining work of connections affected by a routing change.

\begin{figure}[!t]
\vspace{0pt}
\centering

\begin{subfigure}[t]{0.48\linewidth}
\vspace{0pt}
\centering
\includegraphics[width=\linewidth]{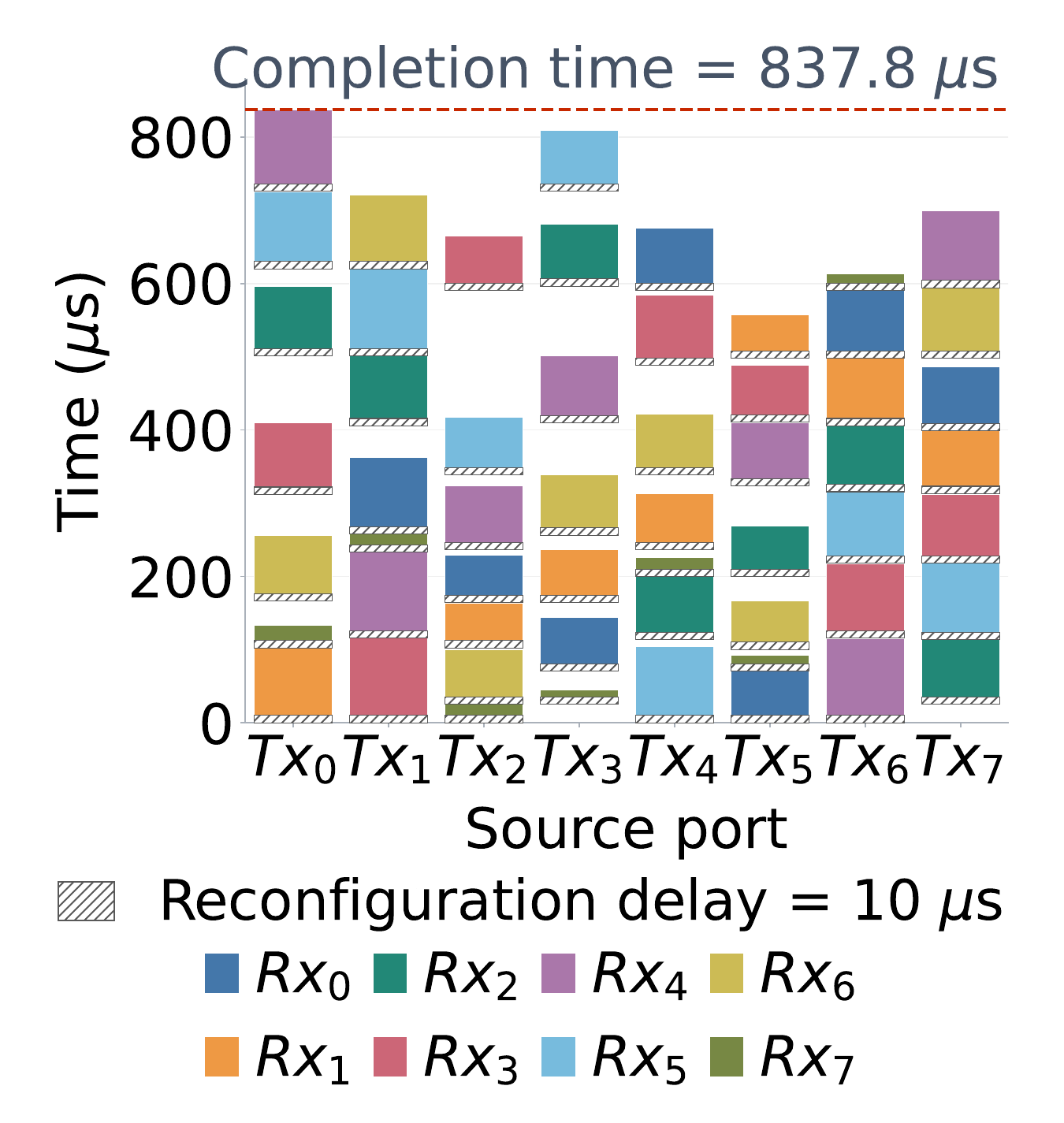}
\caption{Best First Fit (RNB)}
\end{subfigure}
\hfill
\begin{subfigure}[t]{0.48\linewidth}
\vspace{0pt}
\centering
\includegraphics[width=\linewidth]{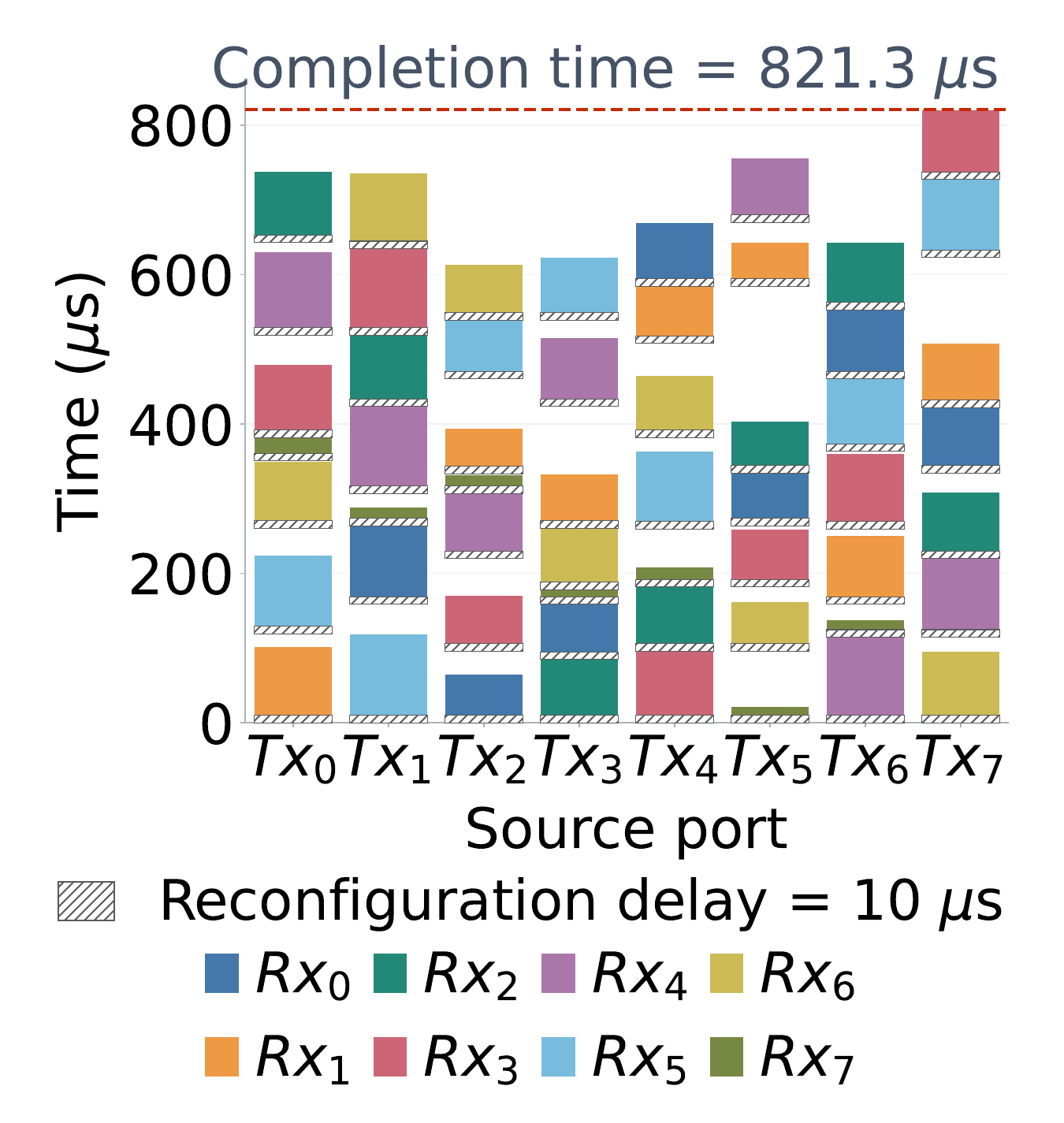}
\caption{Sunflow (RNB)}
\end{subfigure}
\vspace{-2mm}
\caption{Best First Fit~\cite{BFF} and Sunflow~\cite{Sunflow} exploit partial reconfiguration but assume a strictly non-blocking interconnect. When mapped onto an \RNB interconnect, both degrade for different reasons. The purple dashed line marks $Rx_4$, the bottleneck endpoint. BFF fails to prioritize it at approximately $T=500 \mu$s, increasing completion time, while Sunflow makes better bottleneck progress but delays otherwise unrelated connections through its routing choices. Effective partial reconfiguration on \RNB interconnects must account for both endpoint availability and occupied internal paths.}
\label{fig:partial-reconfig-rnb}
\vspace{-4mm}
\end{figure}

Figure~\ref{fig:partial-reconfig-rnb} illustrates both failure modes. When BFF and Sunflow are mapped onto an \RNB interconnect, routing constraints interfere with their original scheduling policies. In this example, both perform worse than MaxWeight with blocking, per-matching reconfiguration barriers (Figure~\ref{fig:timeline-bvn-maxweight}). BFF delays the global bottleneck endpoint, $Rx_4$, while Sunflow assumes an \SNB-style free port and consequently pushes otherwise unrelated flows later under the \RNB routing constraints. The benefit of partial reconfiguration therefore depends on which active connections must be disturbed and how much work remains on them.

\smallskip
\takeaway{Partial reconfiguration can fill the idle gaps left by matching-based schedules, but doing so effectively requires preserving bottleneck progress while accounting for routing dependencies of the underlying interconnect.}

\subsection{Lessons Learned and Design Goals}
\label{subsec:goals}

The observations above suggest that circuit scheduling on \RNB interconnects must jointly reason about three factors: bottleneck progress, reconfiguration cost, and interconnect routing. Optimizing any one of these in isolation can delay demand completion. This leads to three questions that guide the design of \name.

\smallskip
\noindent
\textit{\largeblackcircled{Q1} How should the scheduler prioritize bottleneck progress while accounting for remaining communication work and reconfiguration delay?}

\smallskip
\noindent
\textit{\largeblackcircled{Q2} How should selected connections be routed to minimize disruption to ongoing circuits?}

\smallskip
\noindent
\textit{\largeblackcircled{Q3} How can the scheduler allow connections to be partially reconfigured independently while guaranteeing progress on the remaining demand?}

\section{\LARGE\name}
\label{sec:tetris}

\name takes a communication demand matrix and produces a circuit schedule for an \RNB Bene\v{s} interconnect. As with conventional demand-aware schedulers, the schedule is computed from the demand matrix before communication begins. Unlike a decomposition that simply produces a sequence of matchings, \name also determines how selected connections are routed through the interconnect and when they can be established. At runtime, the interconnect executes this schedule using partial reconfiguration, establishing new circuits as previously scheduled connections complete.

Our design follows directly from the three questions in \S\ref{subsec:goals}. \name repeatedly performs three steps: \emph{(1)} select a matching that guarantees progress on the remaining demand while prioritizing bottleneck endpoints, \emph{(2)} route the selected connections through the Bene\v{s} interconnect while accounting for paths and switch states already committed by the schedule, and \emph{(3)} assign each routed connection an independent start time based on when its path can be established. The key distinction from conventional decomposition is that each matching is only a \emph{planning group}, not an execution barrier. Connections selected in the same matching need not start or finish together, and connections from different matchings may overlap in time.

For ease of reference, Table~\ref{tab:notation} summarizes the main notation used throughout this section. Let $\Demand_{uv}$ denote the number of bytes transmitted from source $u$ to destination $v$, and let $b$ denote the bandwidth of each transmit and receive link. We express each demand in units of transmission time as $\Trans_{uv}=\Demand_{uv}/b$, and maintain a residual transmission-time matrix $R$, initialized as $R=\Trans$. We denote the reconfiguration delay of an internal $2\times2$ switch by $\alpha_r$. The transmit and receive sides of a device are treated as distinct endpoints, even when they belong to the same physical device.

Algorithm~\ref{alg:tetris-construction} summarizes the schedule construction. The schedule $\mathcal{S}$ records each selected connection, its Bene\v{s} path, and the switch reconfigurations required to establish it. In each iteration, \name selects a matching $M$ from the residual demand, jointly routes its connections, and schedules each connection independently. Once a demand $(u,v)$ is added to the schedule, its residual demand is removed from $R$.

\setlength{\textfloatsep}{0.2cm}
\setlength{\floatsep}{0.2cm}
\begin{algorithm}[t]
\small
\DontPrintSemicolon
\SetAlgoLined
\caption{\name schedule construction}
\label{alg:tetris-construction}

\KwIn{Demand matrix $\Demand$, link bandwidth $b$, reconfiguration delay $\alpha_r$, Bene\v{s} interconnect $G$}
\KwOut{Circuit schedule $\mathcal{S}$}

$\Trans\gets\Demand/b$\;
$R\gets\Trans$\;
$\mathcal{S}\gets$ empty schedule on $G$\;

\While{$R$ contains a positive entry}{
    $M\gets\textsc{SelectMatching}(G,R,\alpha_r,\mathcal{S})$\;
    $P\gets\textsc{Route}(G,M,R,\mathcal{S},\emptyset)$\;

    \ForEach{$(u,v)\in M$ in selection order}{
        $\textsc{Schedule}(\mathcal{S},P_{uv},u,v,R_{uv},\alpha_r)$\;
        $R_{uv}\gets0$\;
    }
}

\Return{$\mathcal{S}$}
\end{algorithm}

\begin{figure*}[t]
\centering
\begin{subfigure}{0.24\linewidth}
\includegraphics[width=1\linewidth]{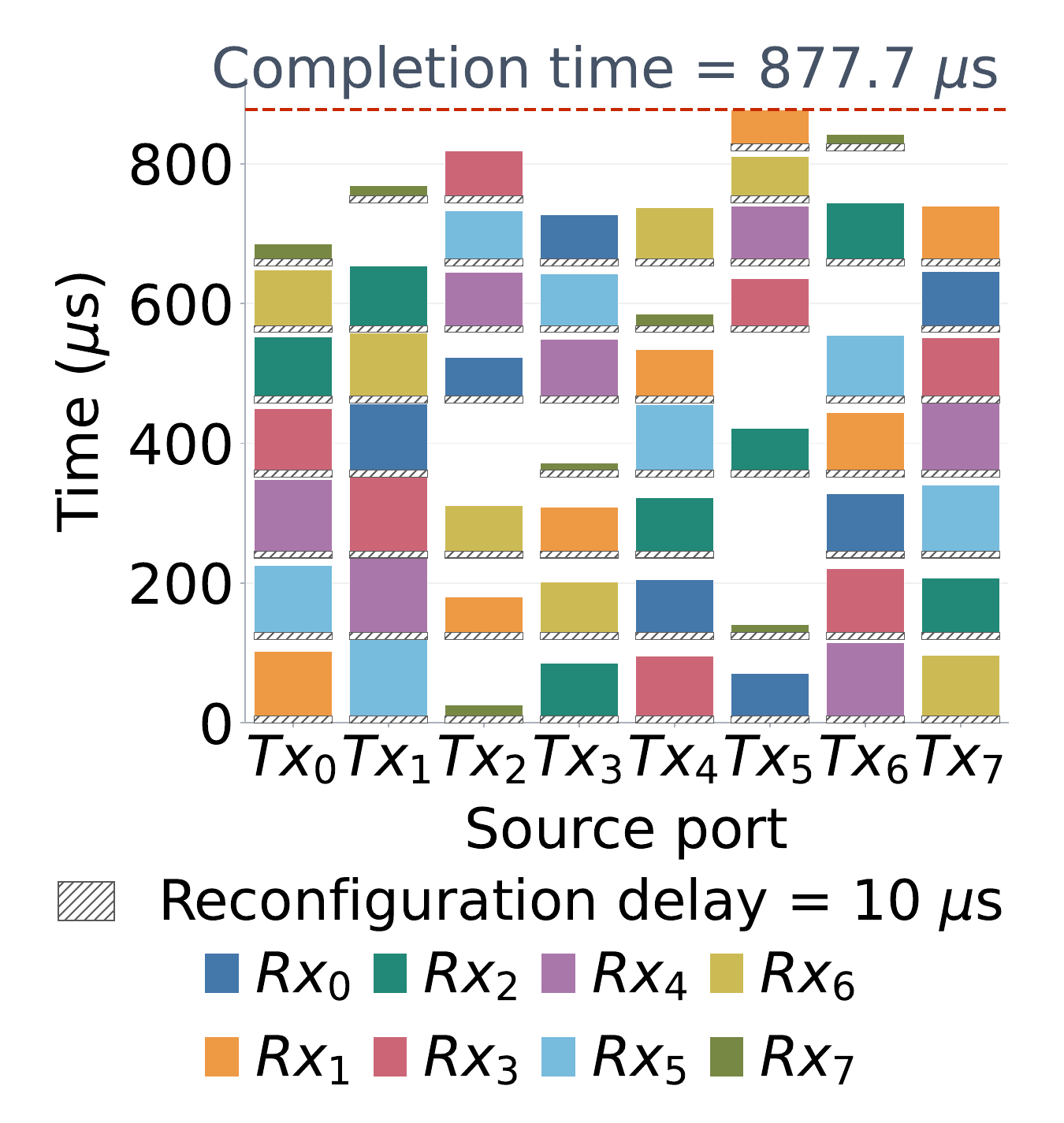}
\caption{\name w/o partial reconfig.}
\label{fig:tetris-without-partial-reconfig}
\end{subfigure}
\begin{subfigure}{0.24\linewidth}
\includegraphics[width=1\linewidth]{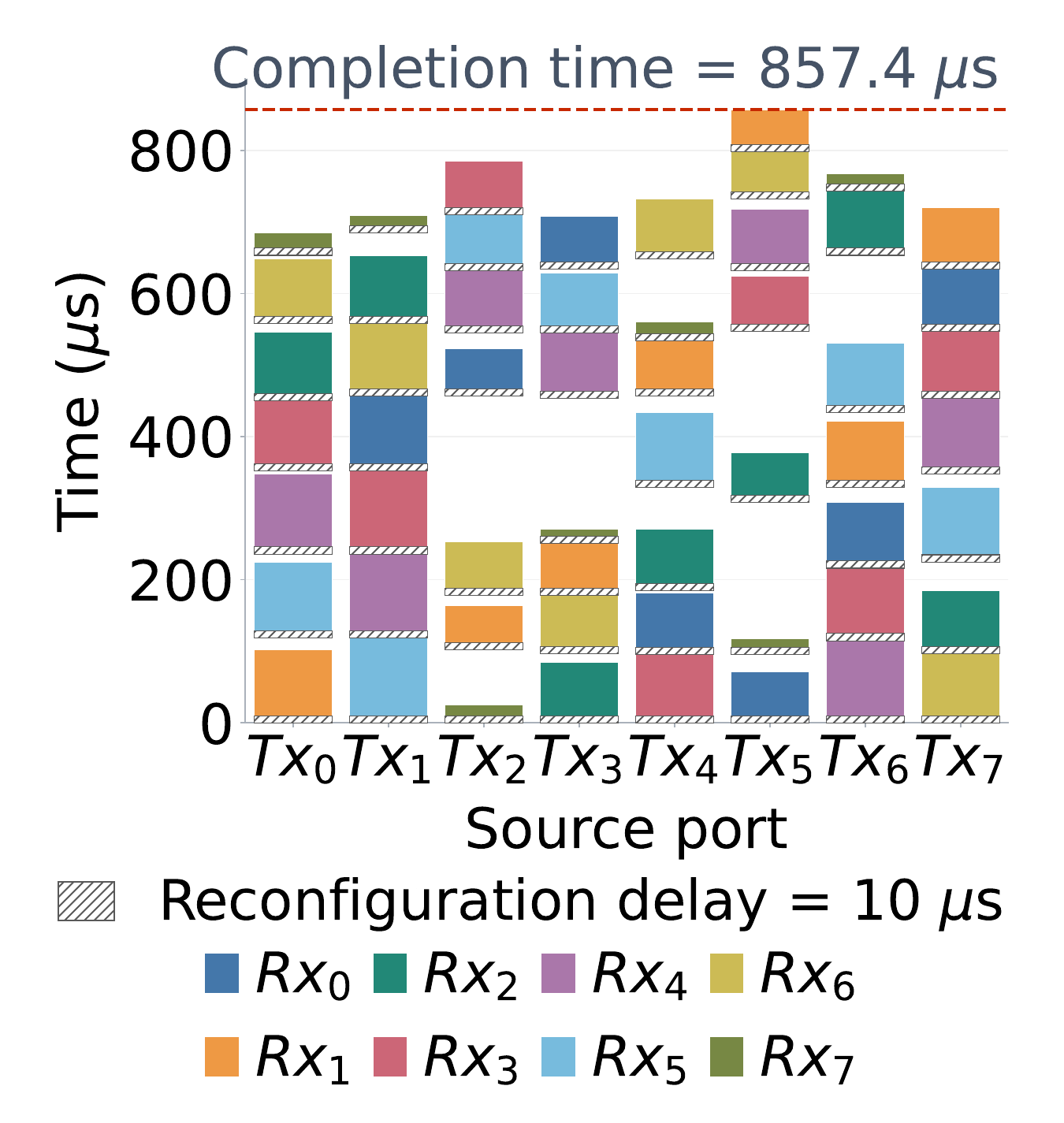}
\caption{\name w/o \RNB routing}
\label{fig:tetris-without-rnb-routingawareness}
\end{subfigure}
\begin{subfigure}{0.24\linewidth}
\includegraphics[width=1\linewidth]{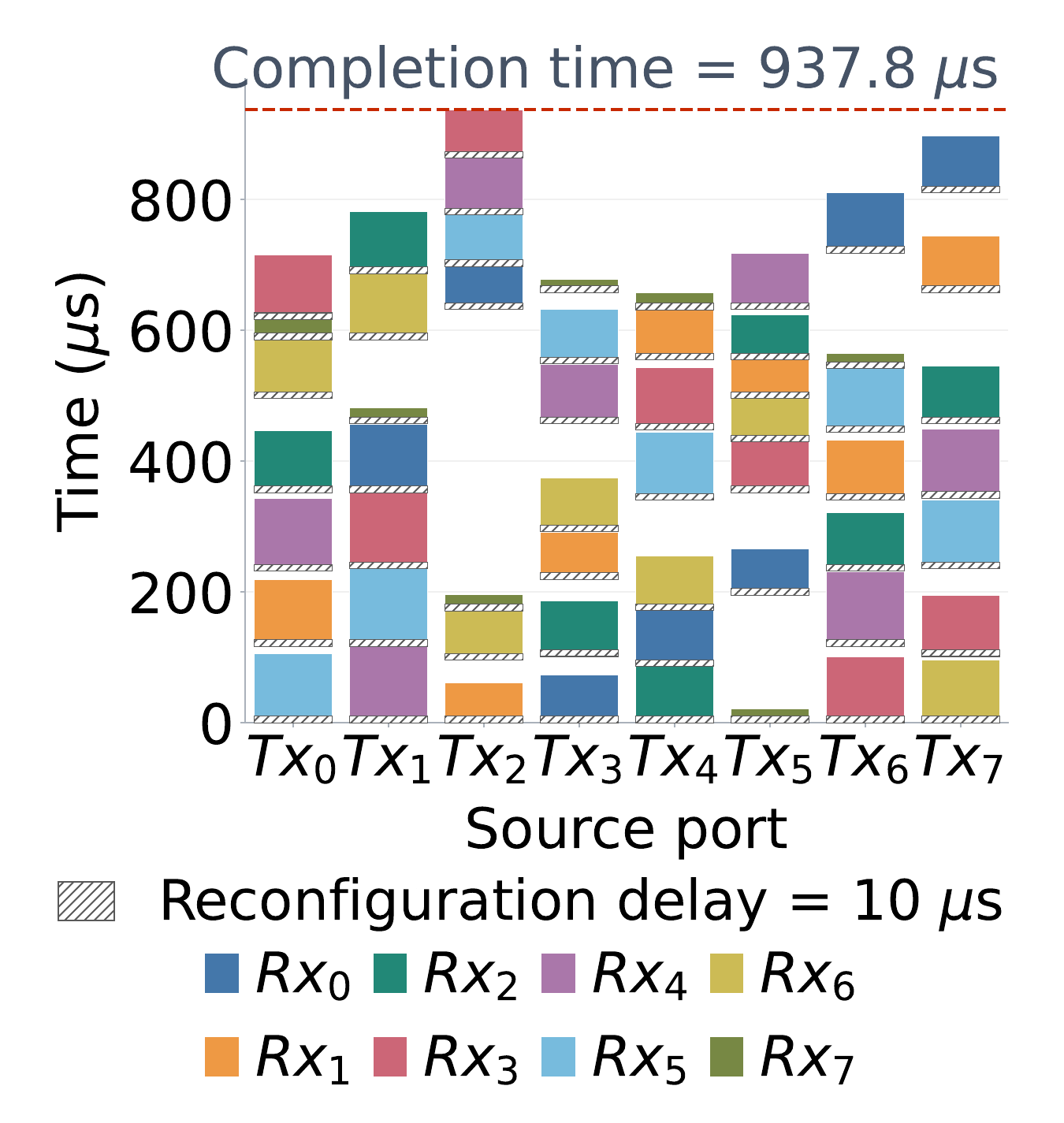}
\caption{\name w/o degree coverage}
\label{fig:tetris-without-degree}
\end{subfigure}
\begin{subfigure}{0.24\linewidth}
\includegraphics[width=1\linewidth]{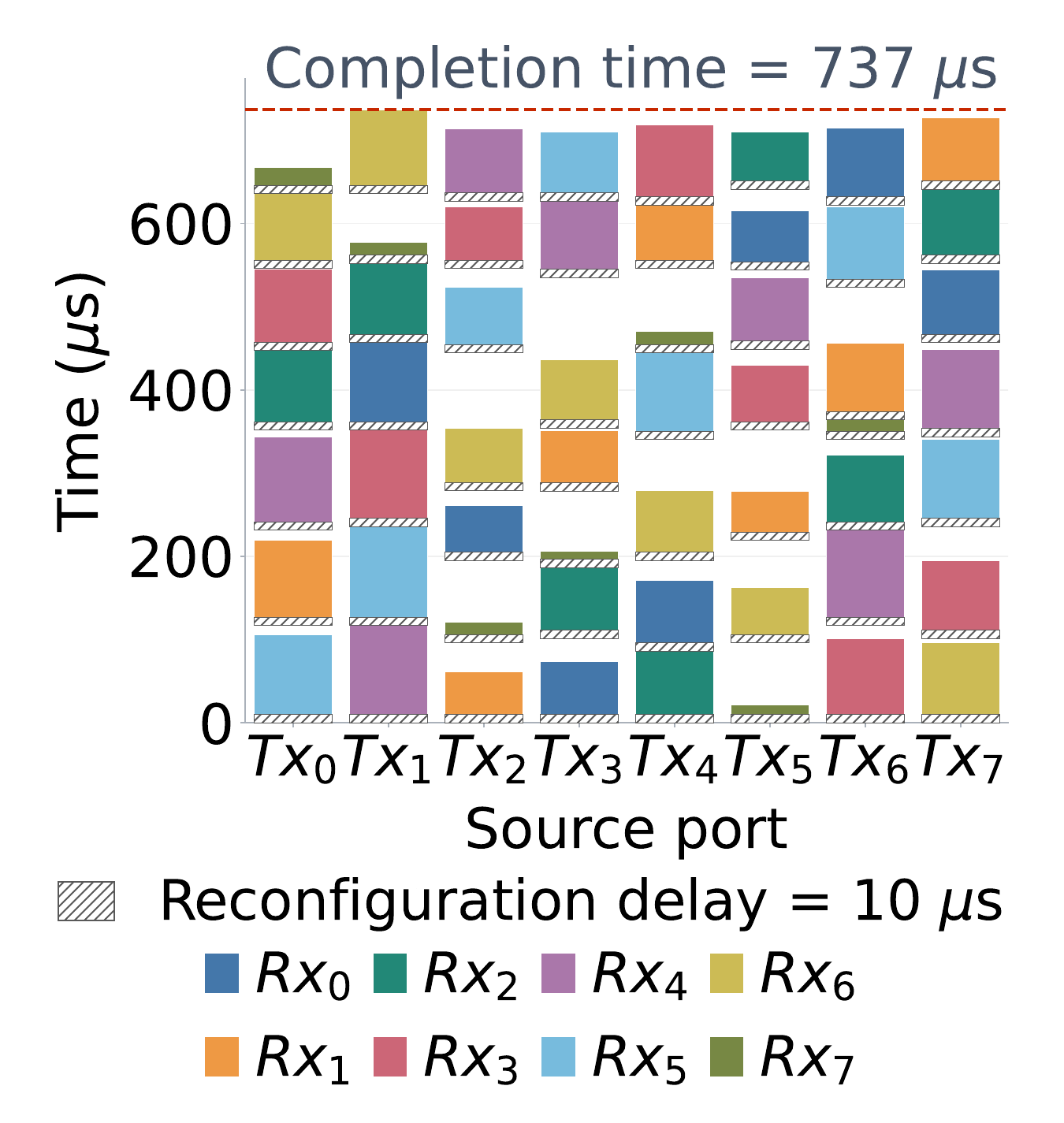}
\caption{\name}
\label{fig:tetris-full}
\end{subfigure}
\vspace{-3mm}
\caption{Impact of \name's three scheduling mechanisms on the same communication demand with $\alpha_r=10\,\mu$s. (a) Matching-wide execution barriers leave ports idle after individual connections finish. (b) Partial reconfiguration without \RNB-aware routing introduces additional delays when new circuits interact with ongoing paths. (c) Without maximum-degree coverage, progress on the residual demand can be delayed. (d) \name combines degree coverage, bottleneck-driven matching selection, \RNB-aware routing, and partial reconfigurations.}
\label{fig:tetris-timeline}
\vspace{-5mm}
\end{figure*}

Figure~\ref{fig:tetris-timeline} gives an overview of how the three design choices affect the resulting schedule. Without partial reconfiguration, short connections finish early but remain separated by matching-wide barriers (Figure~\ref{fig:tetris-without-partial-reconfig}). Allowing partial reconfiguration without accounting for \RNB routing introduces additional delays when newly established paths interfere with ongoing connections (Figure~\ref{fig:tetris-without-rnb-routingawareness}). Removing degree coverage can postpone progress on endpoints with many remaining connections (Figure~\ref{fig:tetris-without-degree}). The complete \name design combines all three mechanisms and achieves the shortest completion time (Figure~\ref{fig:tetris-full}). 

\subsection{Bottleneck-Driven Matching Selection}
\label{subsec:tetris-selection}

The first step addresses \textit{\largeblackcircled{Q1}}: how should \name balance bottleneck progress against communication and reconfiguration cost? Neither demand size nor interconnect readiness alone is sufficient. Prioritizing only large demands can select connections that cannot be established until much later under the current schedule. Conversely, always selecting whichever connection can start earliest can repeatedly postpone heavily loaded endpoints and increase demand completion time. \name separates these concerns: residual degree provides a progress guarantee, while cost estimates based on remaining endpoint work and interconnect readiness determine which feasible connections should make that progress.

\myitem{Estimating the remaining endpoint work:}
For an endpoint~$e$, let $E_R(e)$ denote the set of incident demands with positive residual transmission time. We define
\begin{equation}
    \overbrace{d_e=|E_R(e)|}^{\substack{\textit{residual}\\\textit{degree}}},
    \qquad
    W_e=
    \sum_{(u,v)\in E_R(e)}
    \overbrace{R_{uv}}^{\substack{\textit{transmission}\\\textit{time}}}
    +
    \underbrace{\alpha_r\cdot d_e}_{\substack{\textit{reconfiguration}\\\textit{allowance}}}.
    \label{eq:remaining-work}
\end{equation}
Here, $d_e$ is the number of unscheduled connections incident on endpoint~$e$, while $W_e$ estimates its remaining communication and reconfiguration work. The first term captures the total residual transmission time incident on $e$. The second assigns one reconfiguration allowance, $\alpha_r$, to each remaining connection. This is an endpoint-level estimate: the exact reconfiguration delay depends on the eventual routes and on which internal switch states can be reused.

Let $a_e$ denote the completion time of the latest connection involving endpoint~$e$ already added to the schedule. We use $a_e+W_e$ as a baseline cost estimate for the endpoint's completion time under the current scheduling decisions. It combines the work already committed to $\mathcal{S}$ with the work that remains, while ignoring additional waiting that may be introduced by future matching and routing decisions. The estimate is optimistic, but provides a consistent measure of the completion pressure associated with each endpoint.

\myitem{Guaranteeing progress on the residual demand:}
The positive entries of $R$ form a bipartite graph between transmit and receive endpoints. Since an endpoint can participate in at most one connection in a matching, the maximum residual degree lower bounds the number of additional matchings required to cover all remaining demands. Let $\delta=\max_e d_e$ and $C=\{e\mid d_e=\delta\}$. The set $C$ contains every endpoint with the current maximum residual degree. \name requires each planning matching to cover every endpoint in $C$. Once such a matching is selected, each maximum-degree endpoint loses one incident demand, reducing the maximum residual degree from $\delta$ to $\delta-1$.

Let $\Delta$ denote the maximum endpoint degree before the first matching is selected. Since the maximum residual degree decreases by one in every iteration, \name constructs exactly $\Delta$ planning matchings before all demands are assigned. These matchings provide the progress guarantee only; they do not become synchronized execution rounds. Figure~\ref{fig:tetris-without-degree} illustrates the effect of removing this constraint: connections continue to be scheduled, but progress on the residual demand can become poorly balanced and completion time increases.

\myitem{Estimating the cost of candidate connections:}
Degree coverage identifies which endpoints must make progress, but not which incident connection should be chosen for each endpoint. \name uses the schedule constructed so far to distinguish among these choices.

For every residual demand $(u,v)$, \name performs a read-only path query to estimate when that connection could begin if considered independently. Let $p_{uv}$ denote the resulting candidate path and $s_{uv}=\textsc{Start}(\mathcal{S},p_{uv})$ its estimated start time. The query accounts for paths, switch states, and reconfiguration intervals already committed by $\mathcal{S}$. Candidate connections are evaluated independently at this stage, so their estimated paths may conflict with one another and are not reserved. Joint routing is performed only after the matching has been selected.

For candidate connection $(u,v)$, we define the cost estimate
\begin{equation}
    c_{uv}
    =
    s_{uv}
    -
    \alpha_r
    +
    \max(W_u,W_v).
    \label{eq:connection-cost}
\end{equation}
The estimate combines two effects. The term $s_{uv}$ captures how soon the connection appears realizable given the paths and switch states already committed by the schedule, while $\max(W_u,W_v)$ captures the larger remaining workload of its two endpoints. We subtract $\alpha_r$ to avoid counting the immediate reconfiguration allowance once through path readiness and again through $W_u$ or $W_v$.

The cost $c_{uv}$ is a greedy estimate of the consequence of selecting $(u,v)$, not a prediction of its eventual endpoint completion time. It evaluates the connection against the schedule constructed so far, but does not account for conflicts with other connections selected in the same matching or for decisions made in later iterations. Future routing and matching choices can increase the actual completion time. Exact path compatibility is resolved after the matching is selected.

\myitem{Selecting a feasible low-cost matching:}
The cost estimates above rank individual connections, but selecting connections independently can violate the progress guarantee. \name instead identifies the lowest-cost candidate set that still contains a matching covering every endpoint with the current maximum residual degree. Algorithm~\ref{alg:tetris-selection} summarizes the matching selection procedure. We first define $B=\max_e(a_e+W_e)$, the largest endpoint-level baseline cost under the schedule constructed so far. Starting from this baseline, we find the smallest threshold $H\geq B$ such that
$A= \{(u,v)\mid R_{uv}>0,\;c_{uv}\leq H\}$, 
contains a matching that covers every endpoint in $C$, i.e., every endpoint with residual degree $\delta$. Intuitively, \name admits higher-cost connections only when they are needed to preserve the progress guarantee.

Once the candidate set $A$ is fixed, we consider its connections in decreasing order of $\max(W_u,W_v)$, prioritizing connections incident on endpoints with the most remaining work. We break ties by decreasing residual transmission time $R_{uv}$, favoring longer connections when endpoint workloads are comparable. A connection is added only if the remaining candidates can still cover every unmatched endpoint in $C$. This prevents a locally attractive choice from blocking a required endpoint later in the matching.

We find the threshold $H$ using binary search over the candidate cost estimates. At each step, we test feasibility on the residual bipartite demand graph using alternating paths to determine whether every required endpoint can still be covered. This separates two objectives cleanly: maximum-degree coverage guarantees progress on the remaining demand, while endpoint-work and path-readiness estimates determine which connections provide that progress.

\setlength{\textfloatsep}{0.2cm}
\setlength{\floatsep}{0.2cm}
\begin{algorithm}[t]
\small
\DontPrintSemicolon
\SetAlgoLined
\caption{\textsc{SelectMatching}$(G,R,\alpha_r,\mathcal{S})$}
\label{alg:tetris-selection}

\KwIn{Bene\v{s} interconnect $G$, residual transmission times $R$, reconfiguration delay $\alpha_r$, schedule $\mathcal{S}$}
\KwOut{Next planning matching $M$}

$E\gets\{(u,v)\mid R_{uv}>0\}$\;

\Comment{\textcolor{gray}{Residual degree and remaining endpoint work}}

Compute $d_e=|E_R(e)|$ and
$W_e=\sum_{(u,v)\in E_R(e)}R_{uv}+\alpha_r\cdot d_e$
for every endpoint $e$\;

\Comment{\textcolor{gray}{Endpoints that must make progress}}

$\delta\gets\max_e d_e$;\quad
$C\gets\{e\mid d_e=\delta\}$;\quad
$B\gets\max_e(a_e+W_e)$\;

\Comment{\textcolor{gray}{Estimate path-aware cost of each connection}}

\ForEach{$(u,v)\in E$}{
    $p_{uv}\gets\textsc{EarliestPath}(G,u,v,\mathcal{S})$\;
    $s_{uv}\gets\textsc{Start}(\mathcal{S},p_{uv})$\;
    $c_{uv}\gets s_{uv}-\alpha_r+\max(W_u,W_v)$\;
}

\Comment{\textcolor{gray}{Lowest-cost candidate set preserving progress}}

Find the smallest $H\geq B$ such that
$A=\{(u,v)\in E\mid c_{uv}\leq H\}$
admits a matching covering $C$\;

Order $A$ by decreasing $\max(W_u,W_v)$,
then decreasing $R_{uv}$,
then increasing $(u,v)$\;

$M\gets\emptyset$\;
\ForEach{$(u,v)\in A$ in order}{
    \If{$u,v$ are unused and
    $M\cup\{(u,v)\}$ can still be extended to cover $C$}{
        Append $(u,v)$ to $M$\;
    }
}

\Return{$M$}
\end{algorithm}

\subsection{Routing Selected Connections}
\label{subsec:tetris-routing}

The second step addresses \textit{\largeblackcircled{Q2}}: how should the selected matching be routed without unnecessarily delaying ongoing connections? A Bene\v{s} interconnect can realize every matching, but the same matching can often be realized through multiple internal configurations. These choices matter because different routes can require different internal $2\times2$ switches to change state. Figure~\ref{fig:tetris-without-rnb-routingawareness} illustrates the consequence: partial reconfiguration without \RNB-aware routing introduces additional delays even when compatible transmit and receive endpoints are available.

\myitem{Exploiting the recursive structure of Bene\v{s} routing:}
As discussed in \S\ref{subsec:architecture}, a Bene\v{s} interconnect recursively decomposes into two smaller Bene\v{s} subnetworks. At each recursive level, every selected connection must be assigned to either the upper or lower child.

Three constraints govern these assignments. The two ports connected to the same input-stage $2\times2$ switch must select opposite children; the two ports connected to the same output-stage switch must also select opposite children; and the source and destination of each selected connection must select the same child. These binary constraints form connected components. Each component admits two complementary legal orientations, corresponding to swapping its upper and lower assignments.

\myitem{Routing to protect endpoints with high remaining work:}
Algorithm~\ref{alg:tetris-routing} summarizes our recursive routing procedure.
For each legal orientation, \name evaluates how its boundary-switch decisions interact with the schedule already constructed. At the current recursive level, we combine the previously fixed ancestor traversals with the entry and exit switches implied by the orientation. This gives a partial path for each affected connection.

Suppose the resulting partial path for a connection involving endpoint~$e$ can become available at time~$t$. We estimate its endpoint cost as $t-\alpha_r+W_e$, using the same remaining-work estimate from \S\ref{subsec:tetris-selection}. For orientation~$q$, \name collects these values across all affected endpoints in $V_q$ and sorts them in decreasing order. We compare the two orientations lexicographically: first the largest endpoint cost, then the second largest, and so on. If both orientations produce the same endpoint-cost vector, \name prefers the orientation that changes fewer internal switch states.

Once an orientation is fixed, the matching separates into upper and lower child matchings, and the procedure recurses. Routing proceeds from the root of the Bene\v{s} interconnect toward its recursive children, making one local decision at each level rather than enumerating all complete routing configurations. The remaining-work estimates $W_e$ remain fixed while routing a selected matching, so every routing decision is evaluated against the same residual-demand state.

\subsection{Scheduling with Partial Reconfigurations}
\label{subsec:tetris-placement}

The third step addresses \textit{\largeblackcircled{Q3}}: how can \name preserve the progress guarantee of matching decomposition while allowing partial reconfiguration? We assume a centralized controller that maintains the communication demand, computes the schedule, and controls the internal $2\times2$ switches of the Bene\v{s} interconnect. Matchings are used only during schedule construction; once selected and routed, their connections execute independently as soon as their endpoints and routed paths become available.

Runtime execution is event-driven rather than tied to precise absolute times. When a connection completes, its transmit and receive endpoints, switches, and links are released. Any subsequently scheduled connection whose dependencies are now satisfied becomes eligible to proceed. The start times computed during schedule construction capture these dependencies for scheduling decisions, while runtime execution follows actual connection-completion and switch-reconfiguration events.

\myitem{Reconfiguring only switches whose states change:}
A routed connection traverses a sequence of internal $2\times2$ switches, but establishing a new circuit does not require reconfiguring the entire interconnect. If a switch is already in the required \emph{bar} or \emph{cross} state, the new circuit can reuse that state once its lanes become available; only switches whose states change incur reconfiguration delay $\alpha_r$.

To track these dependencies, $\mathcal{S}$ records, for each switch $s$, its last planned state $\sigma_s$, the completion time $F_s$ of its latest scheduled user, and the completion time $Q_s$ of the reconfiguration that established $\sigma_s$. For input lane $i$ and output lane $j$, $I_{s,i}$ and $O_{s,j}$ denote their planned release times. For a traversal through $(s,i,j)$, the switch-state readiness is $q_{s,i,j}=Q_s$ when $\sigma_s=\operatorname{state}(i,j)$; otherwise, $q_{s,i,j}=F_s+\alpha_r$, since the controller must wait for the previous user to complete before reconfiguring the switch.

A routed connection can proceed only after every traversal along its path is ready. For path $p$, \name estimates its earliest feasible start time as
$
\operatorname{Start}(\mathcal{S},p)
=
\max_{(s,i,j)\in p}
\{I_{s,i},O_{s,j},q_{s,i,j}\}.
$
This is the earliest time at which every required input lane, output lane, and switch state along the path is available. The value is a planning estimate rather than a requirement for precisely timed execution; at runtime, the same dependency order is enforced by completion and reconfiguration events.

\myitem{Matching boundaries do not become execution barriers:}
Once routed, each connection is scheduled independently of the others in its planning matching. A connection can begin while connections from the same or earlier matchings continue elsewhere in the interconnect. When an ongoing connection completes, the controller can immediately trigger any newly enabled switch reconfigurations and establish the next circuit once its complete path is ready. For a connection $(u,v)$ with transmission time $R_{uv}$, \name records its routed path and the dependencies that must complete before that path can be established. At runtime, the controller configures only switches whose states must change, starts transmission once the path is ready, and releases the corresponding endpoints and switch lanes when the connection completes.

Figure~\ref{fig:tetris-without-partial-reconfig} illustrates the cost of retaining matching-wide barriers: endpoints remain idle after their individual connections complete. Figure~\ref{fig:tetris-full} shows the complete \name design, where degree coverage guarantees progress, bottleneck-aware matching selection prioritizes constrained endpoints, and \RNB-aware routing enables partial reconfiguration while preserving ongoing circuits whenever possible.

\smallskip
\takeaway{\name selects matchings in a bottleneck-aware manner to decide which demands must make progress; routing and partial reconfiguration determine how that progress is realized without imposing matching-wide barriers.}

\section{Evaluation}
\label{sec:eval}

\begin{figure*}[t]
    \centering
    \begin{subfigure}{0.74\linewidth}
        \centering
        \includegraphics[width=\linewidth]{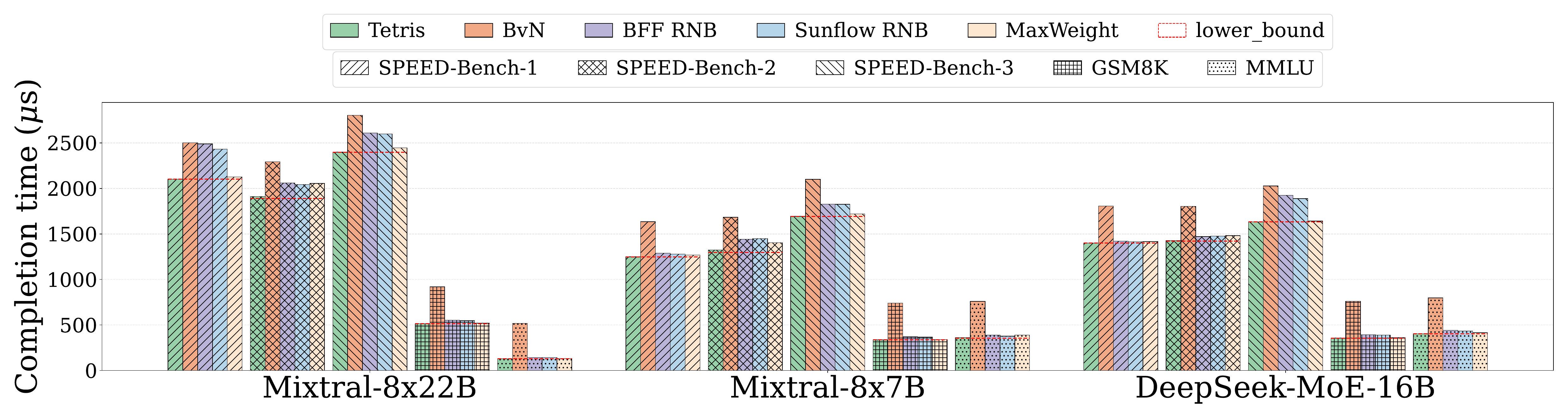}
        \caption{Completion time across workloads}
        \label{fig:workload-bar}
    \end{subfigure}%
    \hfill
    \begin{subfigure}{0.25\linewidth}
        \centering
        \includegraphics[width=\linewidth]{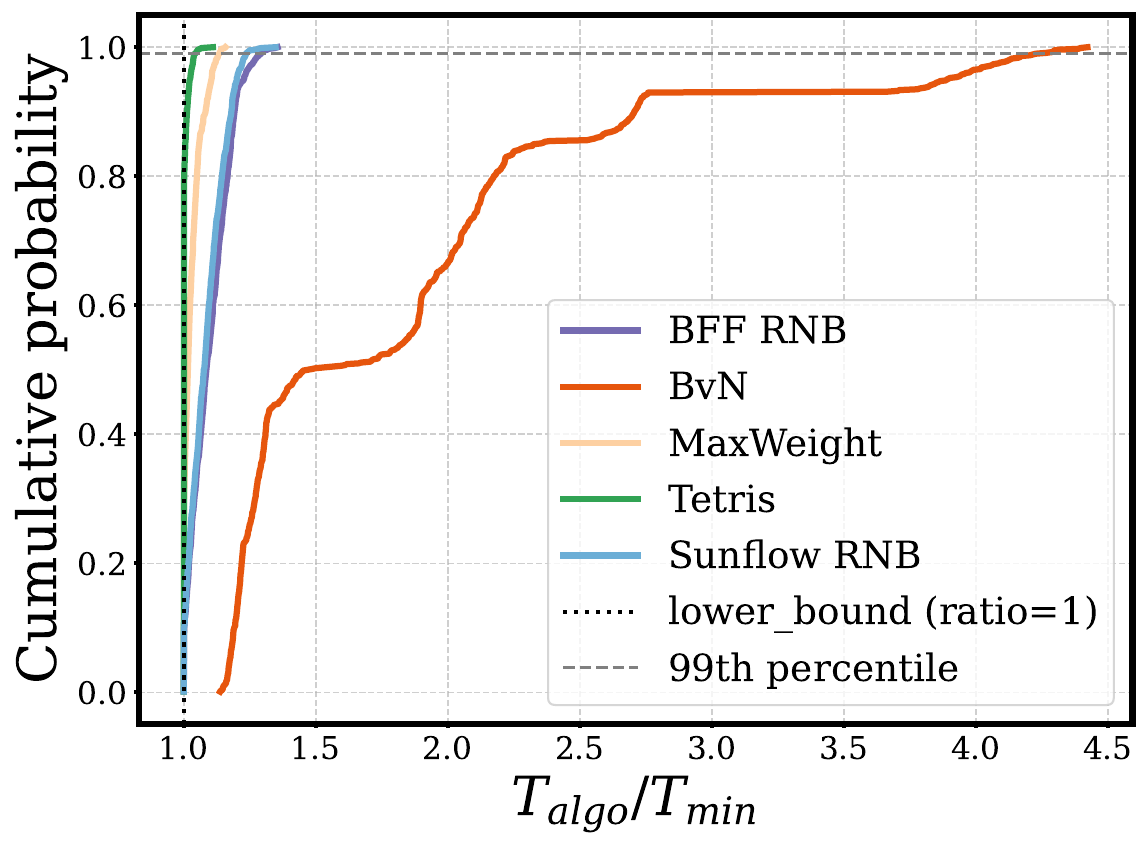}
        \caption{Normalized gap to optimal}
        \label{fig:cdf}
    \end{subfigure}
    \vspace{-2mm}
    \caption{\name's performance across different workloads (left), and the distribution of its normalized completion time relative to optimal across all evaluated workloads (right).}
    \label{fig:workload-bar-cdf}
    \vspace{-4mm}
\end{figure*}

We evaluate \name using simulation and hardware emulation, focusing on the completion time of All-to-All workloads.

\subsection{Setup}
\label{subsec:evaluation-setup}

\myitem{Metric:}
Our primary metric is demand completion time, measured from the beginning of communication until the entire demand matrix has been served. This metric captures both transmission time and the delays introduced by interconnect reconfiguration.

\myitem{Baselines:}
We compare \name against Birkhoff--von Neumann (BvN) decomposition~\cite{BIRKHOFF1946TresLineal,Liu2015SchedulingNetworks}, MaxWeight decomposition as used in systems such as ProjecToR~\cite{ProjecToR} and Helios~\cite{Helios}, Best First Fit (BFF)~\cite{BFF}, and Sunflow~\cite{Sunflow}. BvN and MaxWeight execute matchings with matching-wide reconfiguration barriers, while BFF and Sunflow exploit partial reconfiguration under a strictly non-blocking assumption. We also report the lower bound on demand completion time for each traffic matrix as an ideal reference point. For a fair comparison, we execute all schedules over the same Bene\v{s} interconnect and enforce the same \RNB routing constraints.

\myitem{Workloads:}
We evaluate using All-to-All communication traces collected from Mixture-of-Experts workloads running on an 8$\times$RTX~6000 GPU server. We consider DeepSeek-MoE-16B, Mixtral-$8\times7$B, and Mixtral-$8\times22$B, using inputs from SPEED-Bench~\cite{abramovich2026speedbenchunifieddiversebenchmark}, MMLU~\cite{hendryckstest2021,hendrycks2021ethics}, and GSM8K~\cite{cobbe2021gsm8k}. For each workload, we record the communication demand generated by the MoE layers during each dispatch and combine phases, capturing the input-dependent and skewed traffic patterns discussed in \S\ref{subsec:challenges-scheduling}. We additionally use synthetic demand matrices with controlled traffic skew to isolate how traffic asymmetry affects the different scheduling approaches.

\myitem{System configurations:}
We vary the number of GPUs from $8$ to $64$, the per-port link bandwidth from $100$ to $800$ Gbps, and the switch reconfiguration delay $\alpha_r$ from $10$ ns to $10$ ms, spanning different fast electro-optic and slower MEMS-based switching technologies. Unless stated otherwise, we use $800$ Gbps per-port bandwidth and a $10\ \mu s$ reconfiguration delay. These configurations let us evaluate \name across different system scales and identify the regimes in which fine-grained partial reconfiguration provides the largest benefit.

\myitem{Simulation:}
We built a custom simulator that executes schedules over an \RNB Bene\v{s} interconnect at the granularity of individual $2\times2$ switches. It tracks active paths, switch states, and transmit/receive endpoint availability. Blocking schedulers such as BvN and MaxWeight execute each matching as a synchronized round, while partial-reconfiguration schedulers allow completed connections to release their endpoints independently so that later connections can overlap with ongoing traffic. Every scheduled connection is explicitly routed through the Bene\v{s} interconnect, and the simulator applies reconfiguration delay $\alpha_r$ whenever required by the schedule and finally reports demand completion time.

\myitem{Hardware emulation:}
We view this work as an initial step toward a complete system.
Building a complete reconfigurable photonic interconnect requires optical switching hardware, transceivers, and control infrastructure, making a full end-to-end implementation a costly undertaking for an academic research effort. \name focuses on the scheduling problem, while important systems challenges such as controller design and transceiver locking under rapid reconfiguration remain open.

To capture realistic communication costs, we complement simulation with hardware emulation on our GPU server. For each demand in the MoE traces, we measure the corresponding one-hop GPU-to-GPU communication time on real hardware. We execute each schedule using these measured transmission times and inject the reconfiguration delays incurred by the scheduler. This preserves measured GPU communication behavior while emulating the timing impact of photonic reconfiguration and scheduling decisions.

\begin{figure*}[t]
   \begin{subfigure}{0.23\linewidth}
    \includegraphics[width=\linewidth]{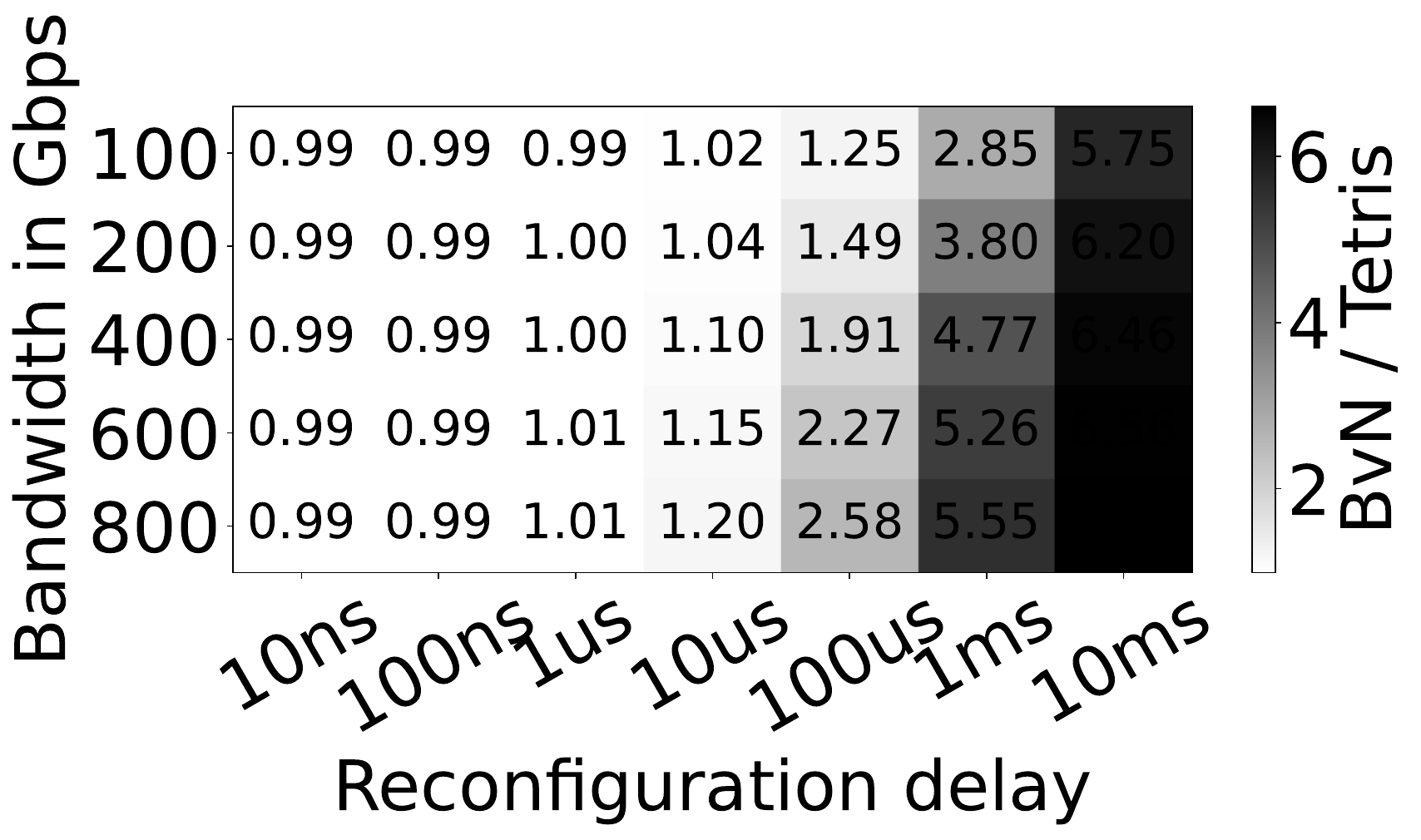}
    \caption{BvN against \name}
    \label{fig:heatmap-bvn}
  \end{subfigure}%
  \hfill
   \begin{subfigure}{0.23\linewidth}
    \includegraphics[width=\linewidth]{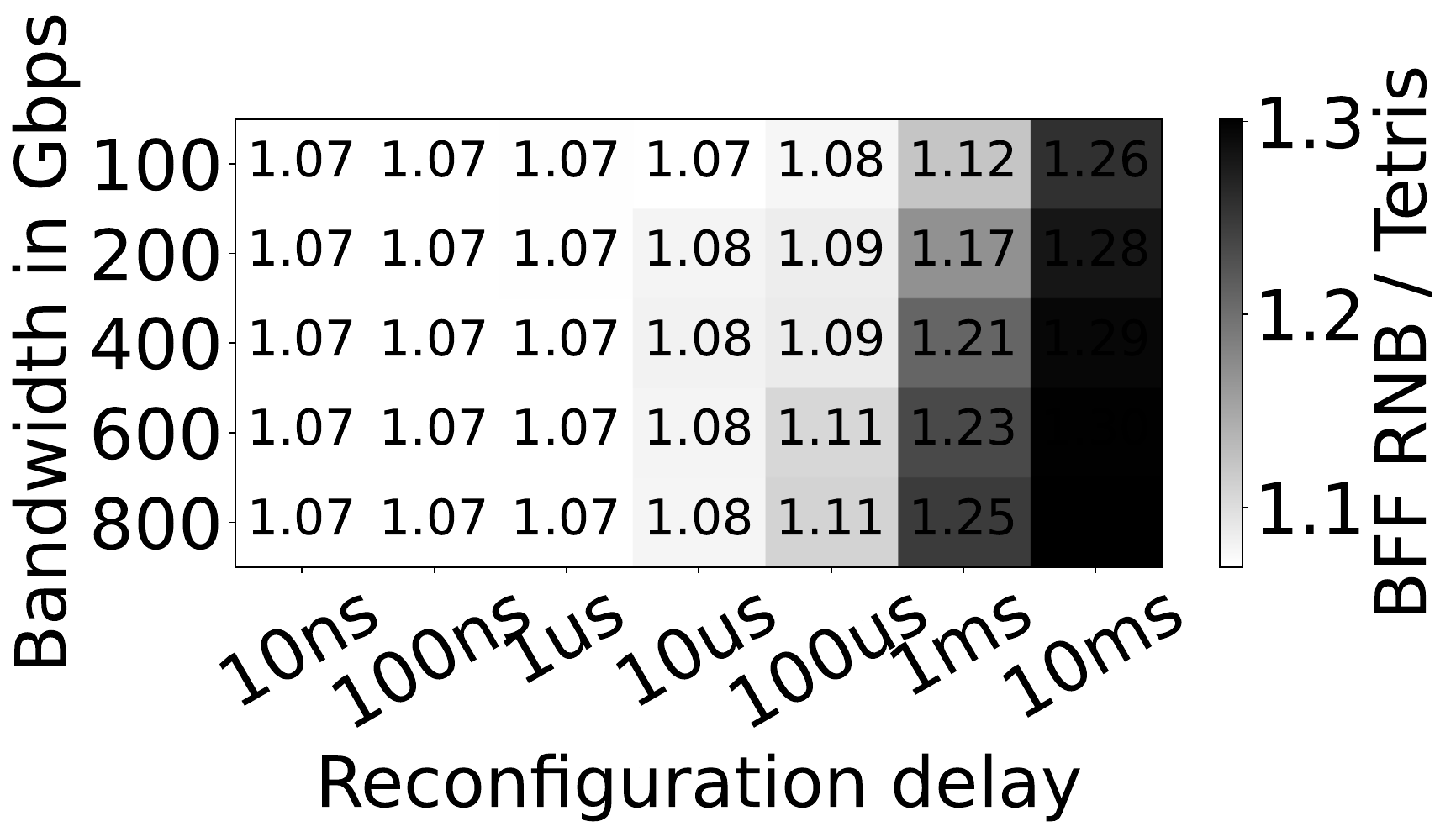}
    \caption{BFF against \name}
    \label{fig:heatmap-bff}
  \end{subfigure}
    \hfill
   \begin{subfigure}{0.23\linewidth}
    \includegraphics[width=\linewidth]{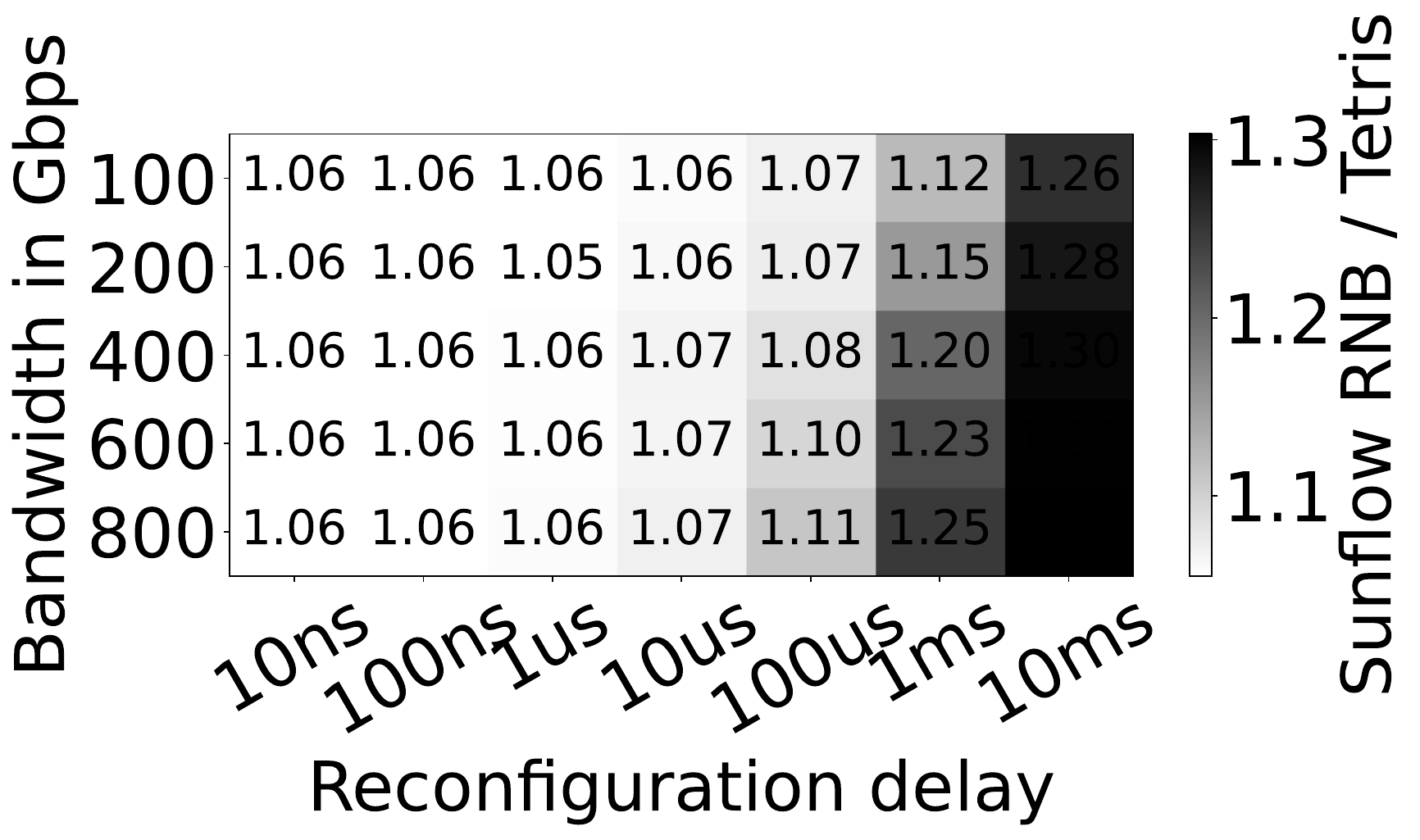}
    \caption{Sunflow against \name}
    \label{fig:heatmap-sunflow}
  \end{subfigure}
   \hfill
   \begin{subfigure}{0.23\linewidth}
    \includegraphics[width=\linewidth]{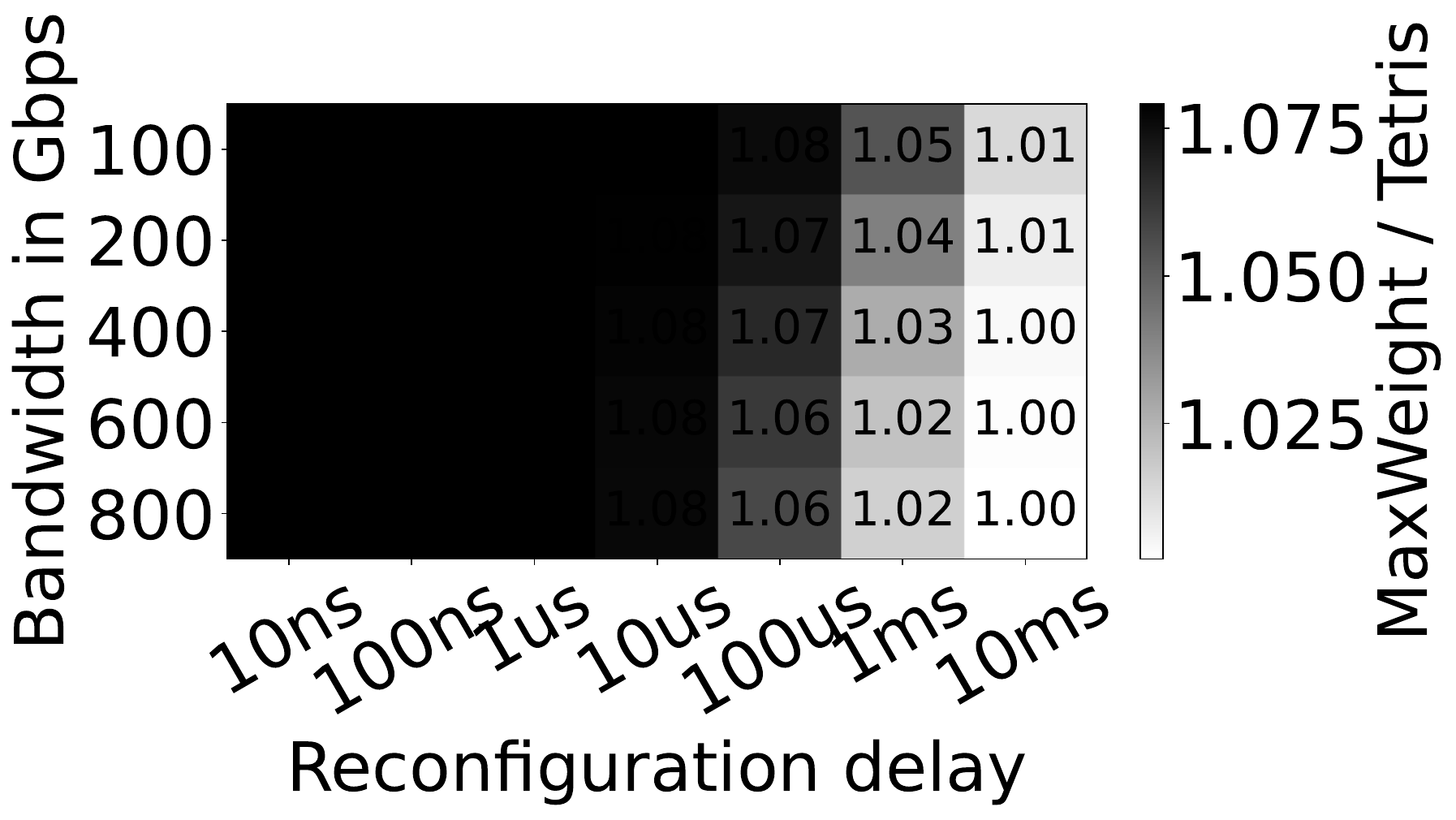}
    \caption{MaxWeight against \name}
    \label{fig:heatmap-maxweight}
  \end{subfigure}
    \caption{Sensitivity of \name's advantage over each baseline to link bandwidth and reconfiguration delay for one complete inference pass of Mixtral-8x22B.}
    \vspace{-2mm}
    \label{fig:heatmap}
    \vspace{-4mm}
\end{figure*}

\emph{Due to space constraints, additional results appear in \S\ref{app:additional-results}.}

\subsection{Results}

\myitem{\protect{\name} significantly outperforms existing approaches:}
We first compare All-to-All demand, completion time across different MoE models, prompt datasets, and layers. Figure~\ref{fig:makespan} shows results for Mixtral-$8\times22$B on the SPEED-Bench throughput $1$K and $2$K workloads, using $800$Gbps link bandwidth and a reconfiguration delay of $10\mu s$. Across layers with different communication demands and message volumes, \name consistently achieves lower completion time than all our baselines. We next evaluate whether these gains generalize beyond a single model and workload. Figure~\ref{fig:workload-bar} compares \name against the baselines across Mixtral-$8\times7$B, Mixtral-$8\times22$B, and DeepSeek-MoE-$16$B over different prompt datasets. Across these workloads, \name continues to outperform the baseline schedulers and remains close to the lower bound on completion time. Specifically, we see that \name outperforms BvN and MaxWeight by $20\%$ and $6\%$ respectively, while outperforming BFF and Sunflow by $8\%$ and $7\%$ respectively. 
We omit the lower-bound bars from Figure~\ref{fig:workload-bar} for visual clarity and quantify the gap separately. To measure this gap directly, we normalize the completion time of each scheduler by the lower bound for the corresponding demand matrix. Figure~\ref{fig:cdf} shows the resulting distribution across all evaluated matrices. \name remains consistently close to the lower bound, while the baseline approaches exhibit larger slowdowns.

Together, these results show that the gains of \name are not tied to a particular model, layer, prompt dataset, or message size. Across the evaluated MoE communication workloads, \name consistently reduces All-to-All completion time and remains close to optimal.

\myitem{\name adapts to bandwidth and reconfiguration delay:}
We next evaluate how the relative performance of \name changes with link bandwidth and switch reconfiguration delay. Figure~\ref{fig:heatmap} varies bandwidth from $100$ to $800$ Gbps and reconfiguration delay from $10$ ns to $10$ ms for one complete inference pass of Mixtral-$8\times22$B on SPEED-Bench throughput-$1$K. Each cell reports the ratio between the baseline completion time and that of \name; values greater than $1$ indicate that \name completes the demand faster.

Figure~\ref{fig:heatmap-bvn} highlights the sensitivity of BvN to reconfiguration delay. When reconfiguration is nearly free ($10$--$100$ns), BvN and \name are within $\approx 1\%$ of each other, with BvN occasionally completing slightly earlier. As $\alpha_r$ increases, however, BvN's large number of matching-wide reconfigurations quickly dominates completion time. At $800$ Gbps, \name's speedup against BvN grows from $0.99$ at $10$\,ns to $2.58$ at $100\mu s$, $5.55$ at $1$ ms, and $6.60$ at $10$ ms. The effect also becomes stronger at higher bandwidth because transmission time shrinks while reconfiguration delay remains fixed.

Figures~\ref{fig:heatmap-bff} and~\ref{fig:heatmap-sunflow} show a different trend for BFF and Sunflow. Both exploit partial reconfiguration, but do not account for the routing dependencies introduced by the \RNB interconnect. \name is already about $1.06$--$1.07\times$ faster when reconfiguration is inexpensive, and its advantage grows as reconfiguration becomes more costly, reaching up to $1.30\times$ for both approaches. This trend reflects the increasing cost of partial-reconfiguration decisions that disturb ongoing paths, and highlights key observation in \name i.e., bottleneck-awareness is not simply the largest demand, but one that contributed to overall progress.

MaxWeight exhibits the opposite sensitivity (Figure~\ref{fig:heatmap-maxweight}. Because it uses matching-wide barriers but avoids the large number of reconfigurations incurred by BvN, its performance becomes increasingly competitive as $\alpha_r$ grows. \name is up to $1.08\times$ faster at low and moderate reconfiguration delays by filling the idle gaps within MaxWeight matchings, while the two approaches converge as reconfiguration becomes very expensive.

Across the full spectrum, \name outperforms BFF, Sunflow, and MaxWeight for nearly all configurations and loses only marginally to BvN when reconfiguration is effectively free. More importantly, no single scheduler performs well across all operating regimes: BvN is effective only when reconfiguration is inexpensive, while blocking MaxWeight becomes competitive when reconfiguration dominates. By incorporating both reconfiguration cost and \RNB routing constraints into its scheduling decisions, \name remains competitive across the entire range of bandwidths and reconfiguration delays.

\begin{figure}[t]
\centering
\includegraphics[width=0.9\linewidth]{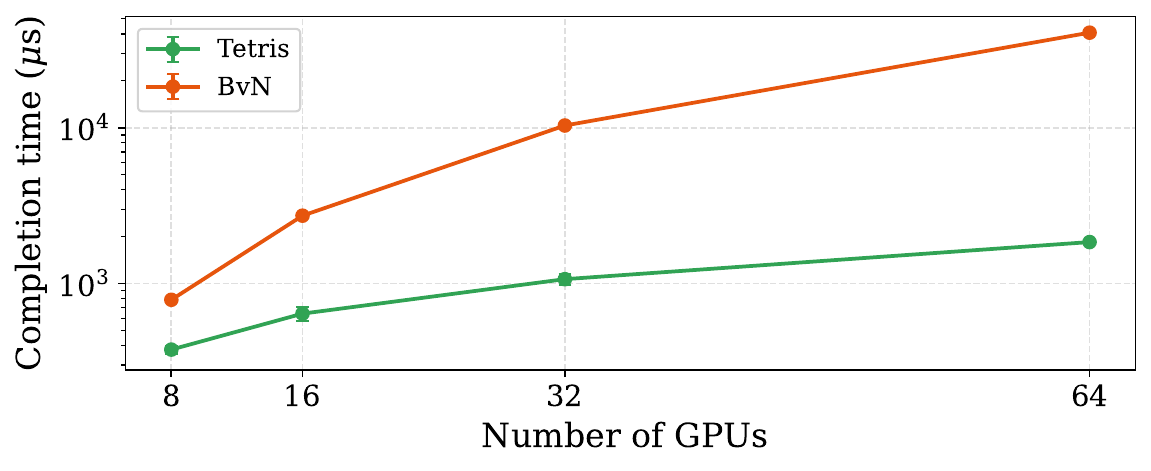}
\vspace{-2mm}
\caption{\name's completion time as the number of GPUs scales. Y-axis is log-scaled.}
\label{fig:scaling}
\vspace{-1mm}
\end{figure}

\begin{figure*}[t]
   \begin{subfigure}{0.23\linewidth}
    \includegraphics[width=\linewidth]{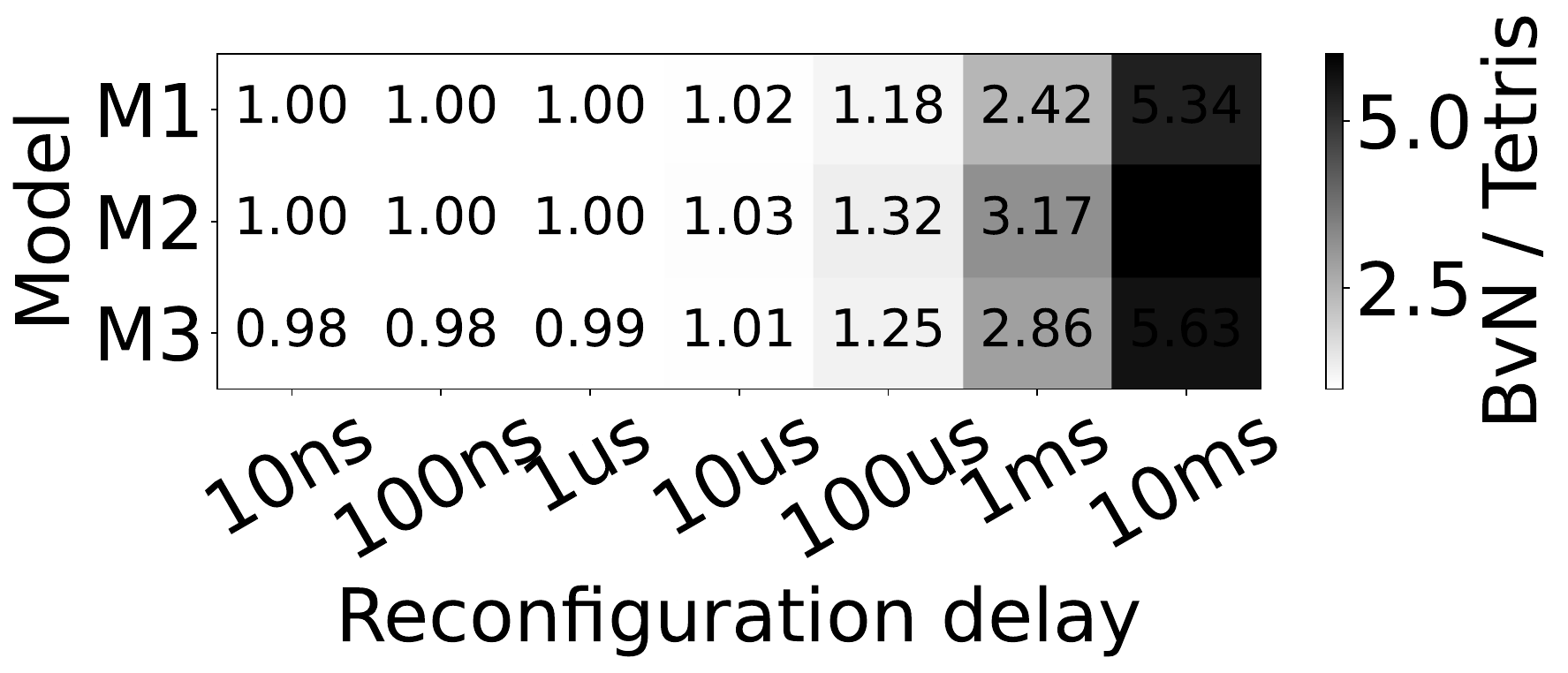}
    \caption{BvN against \name}
    \label{fig:heatmap-hardware-bvn}
  \end{subfigure}%
  \hfill
   \begin{subfigure}{0.23\linewidth}
    \includegraphics[width=\linewidth]{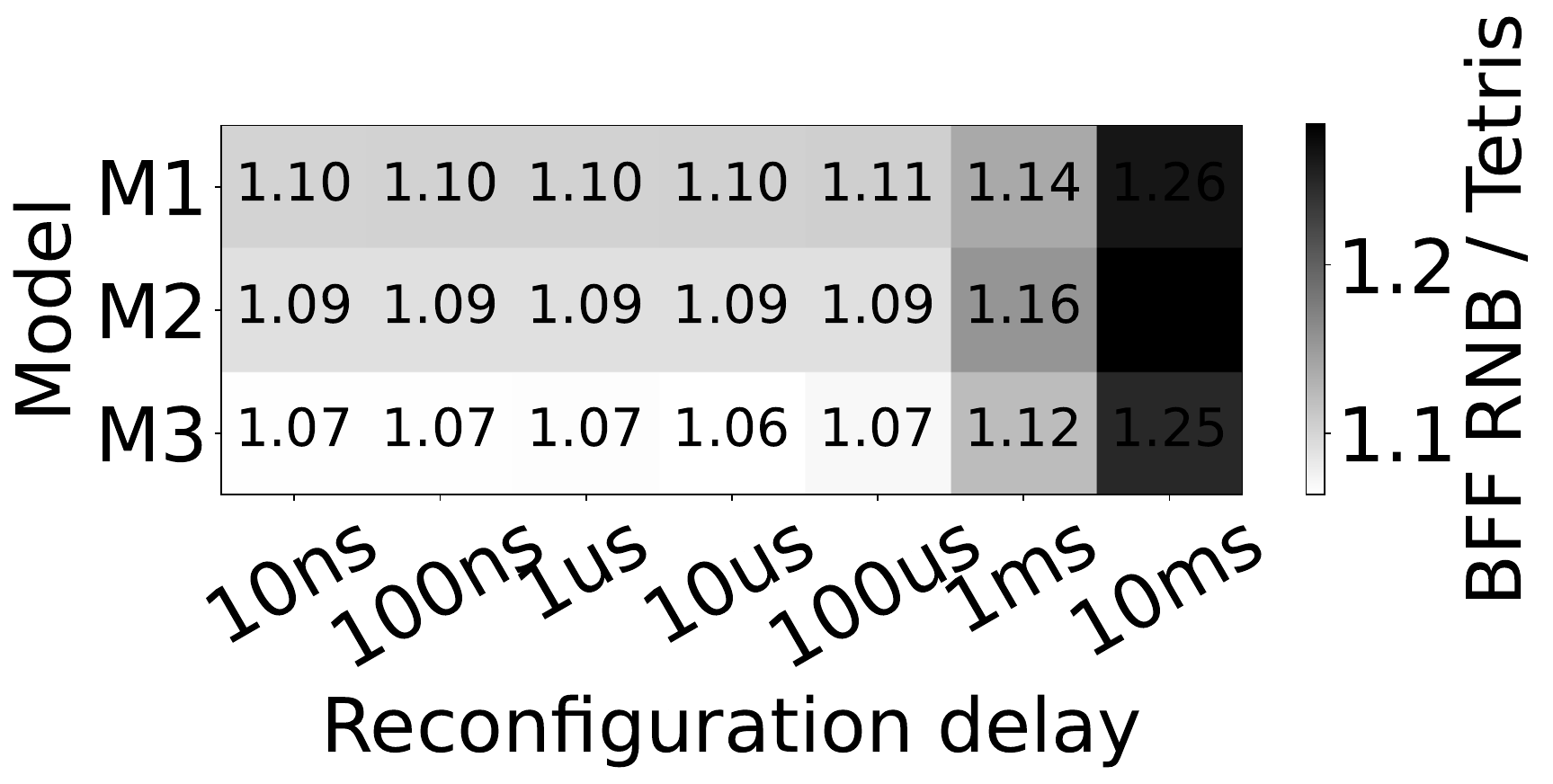}
    \caption{BFF against \name}
    \label{fig:heatmap-hardware-bff}
  \end{subfigure}
    \hfill
   \begin{subfigure}{0.23\linewidth}
    \includegraphics[width=\linewidth]{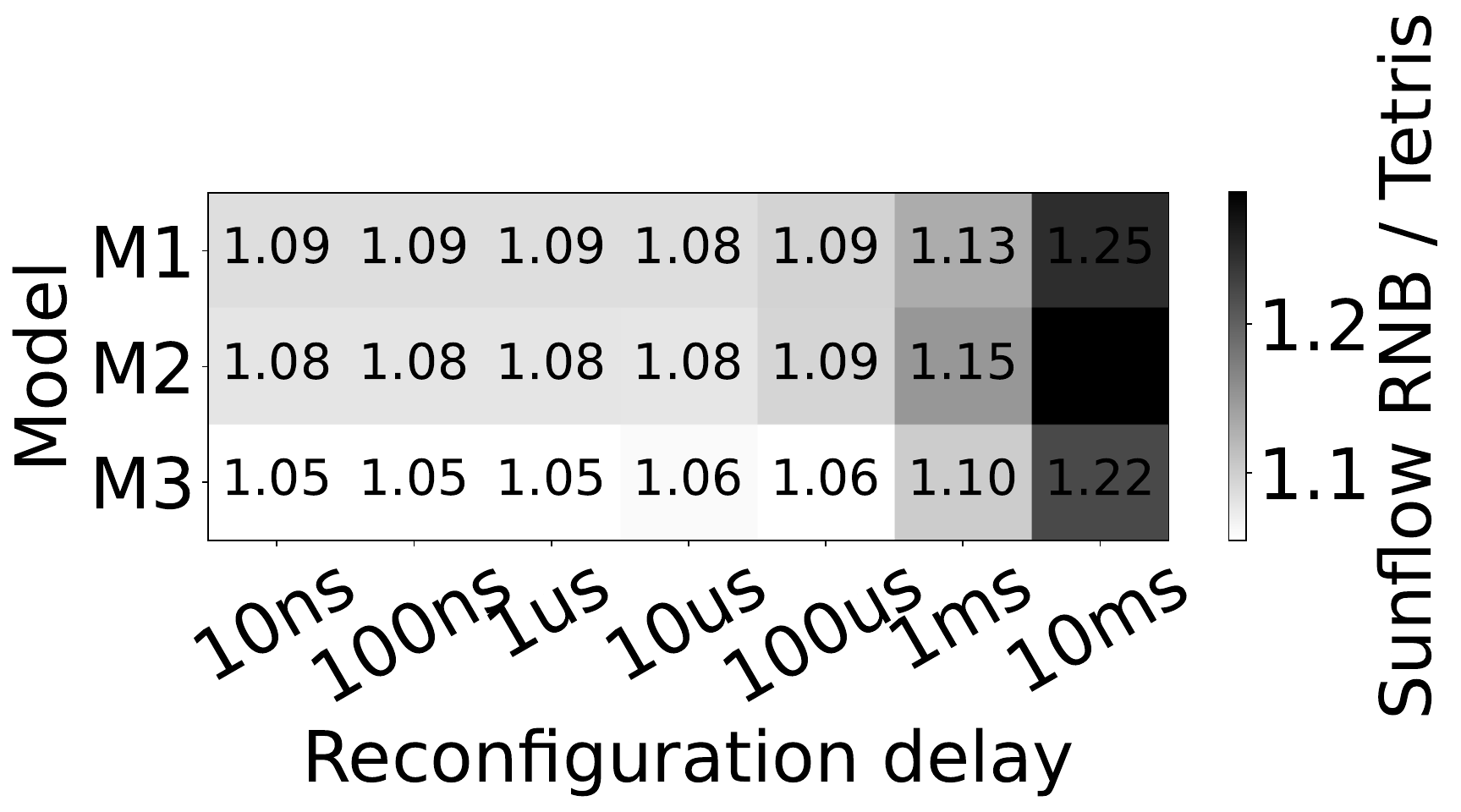}
    \caption{Sunflow against \name}
    \label{fig:heatmap-hardware-sunflow}
  \end{subfigure}
   \hfill
   \begin{subfigure}{0.23\linewidth}
    \includegraphics[width=\linewidth]{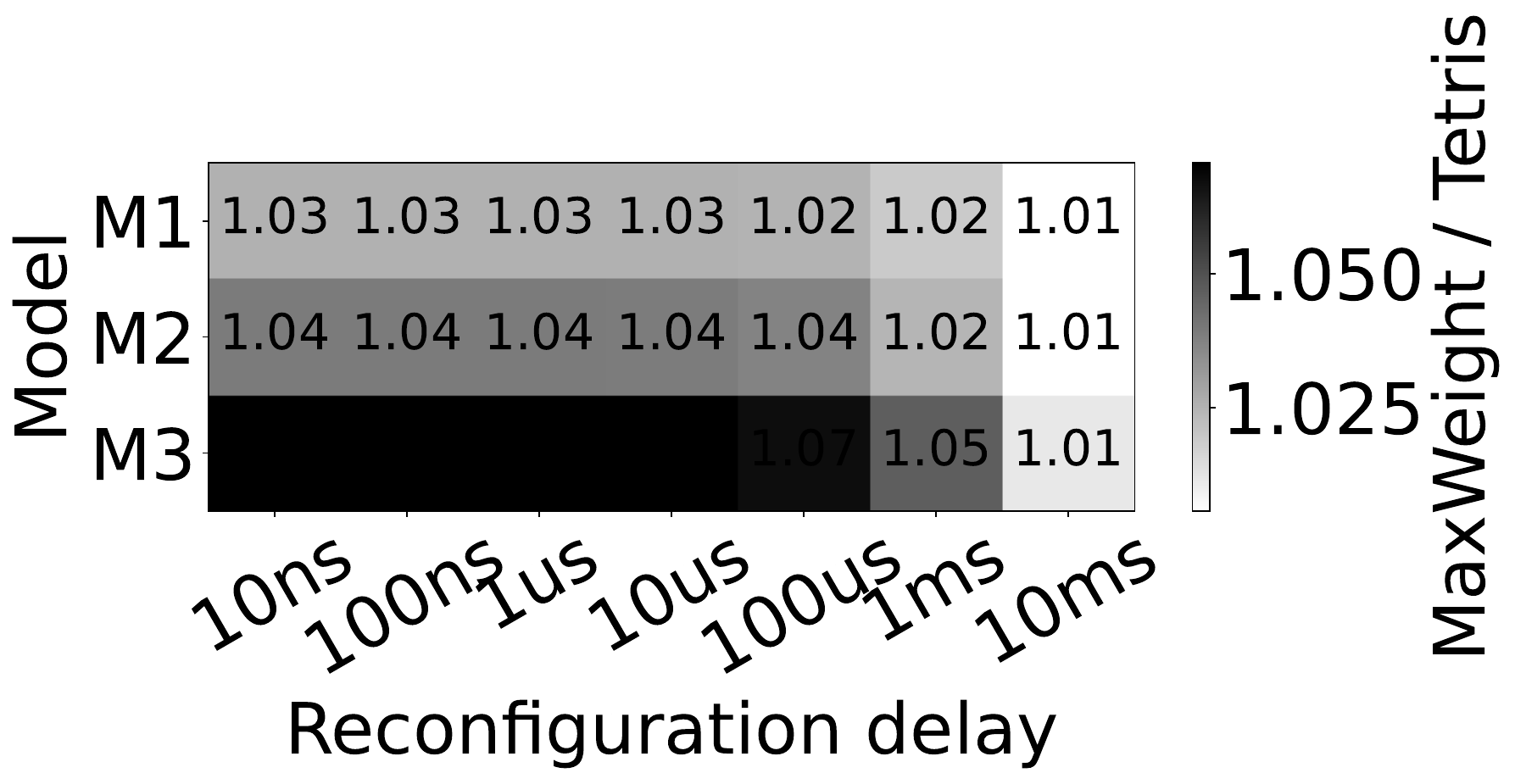}
    \caption{MaxWeight against \name}
    \label{fig:heatmap-hardware-maxweight}
  \end{subfigure}
  \vspace{-2mm}
    \caption{[Hardware emulation] Sensitivity of \name's performance advantage over each baseline to link bandwidth and switch reconfiguration delay for one complete inference pass of Mixtral-$8\times22$B (M1), Mixtral-$8\times7$B (M2) and DeepSeek-MoE-16B (M3).}
\label{fig:heatmap-hardware}
\vspace{-5mm}
\end{figure*}

\myitem{\name achieves consistent improvement across interconnect scale:}
We evaluate scaling using synthetic All-to-All demand matrices, since collecting real traces at larger scales would require increasingly costly GPU clusters. The generator preserves the skew and layer-to-layer variation observed in our MoE traces and uses the target model's hidden-state dimension to keep message sizes realistic. Figure~\ref{fig:scaling} compares \name against BvN from $8$ to $32$ GPUs. The gap widens quickly with scale: BvN takes roughly $2\times$, $5\times$, and nearly $10\times$ longer at $8$, $16$, and $32$ GPUs, respectively. The reason is that BvN fragments demand across more matchings as the interconnect grows, accumulating matching-wide reconfiguration delays. \name limits this overhead through degree coverage and partial reconfiguration, with completion time growing from only $\approx 400 \mu$s at $8$ GPUs to $1.1$ ms at $32$ GPUs.

\myitem{\name's performance gains persist under hardware emulation:}
We next evaluate whether the trends observed in simulation persist when communication times are measured on real GPUs. Figure~\ref{fig:heatmap-hardware} reports the completion-time ratio between each baseline and \name across reconfiguration delays from $10$ ns to $10$ ms for all three MoE models. The hardware results closely follow the simulation trends. BvN remains competitive when reconfiguration is nearly free, but degrades rapidly as $\alpha_r$ increases: at $10$ ms, BvN is $5.34\times$, $6.00\times$, and $5.63\times$ slower than \name across the three models (Figure~\ref{fig:heatmap-hardware-bvn}). BFF and Sunflow show smaller but consistent gaps. \name is already $1.05$--$1.10\times$ faster at low reconfiguration delays, and its advantage grows to as much as $1.29\times$ as reconfiguration becomes more expensive (Figures~\ref{fig:heatmap-hardware-bff} and~\ref{fig:heatmap-hardware-sunflow}). MaxWeight exhibits the expected complementary behavior: \name provides up to a $1.07\times$ improvement when reconfiguration is inexpensive, while the two approaches converge at larger reconfiguration delays as avoiding additional reconfigurations becomes increasingly important (Figure~\ref{fig:heatmap-hardware-maxweight}). Figure~\ref{fig:alpha-beta} shows that direct GPU-to-GPU transmission times on hardware closely follow the classic $\alpha$-$\beta$ cost model~\cite{rogerwhockney}, with an initial overhead for small messages, and linearly scaling as the message size grows. These results confirm that \name's gains \name preserves its advantage across models and reconfiguration regimes, while exhibiting the same trade-offs observed in simulation.

\section{Discussion and Future Research Directions}
\label{sec:discussion} 

\name is only a first step toward circuit scheduling that explicitly accounts for rearrangeably non-blocking interconnects. Exposing routing dependencies to the scheduler opens several broader questions spanning multi-tenancy, routing, scheduling objectives, and interconnect design. We discuss these directions below.

\myitem{Efficient algorithms for interconnect-aware scheduling:}
\name shows the benefit of treating demand decomposition and interconnect routing as a joint scheduling problem. A matching that appears attractive at the demand level may be costly to realize once current switch states and occupied paths are considered. By selecting matchings together with realizable routes, \name avoids decompositions that later require disruptive remapping.
The broader question is how to extend this idea beyond Bene\v{s} interconnects. Bene\v{s} routing admits a recursive treatment through local branch dependencies, while other \RNB topologies expose different routing constraints. One possible abstraction is a conflict graph whose vertices represent candidate connections or paths and whose edges capture incompatibility under the current interconnect state. \name is only a first step: designing efficient joint decomposition and routing algorithms for broader classes of \RNB interconnects, with low scheduling overhead, remains an open problem.

\myitem{Alternative scheduling objectives:}
Partial reconfiguration in \RNB interconnects also motivates objectives beyond demand completion time. One is to minimize how often internal $2\times2$ switches change state. Reusing \emph{bar}\ /\ \emph{cross} configurations for longer periods can reduce control overhead and, for technologies such as MEMS, limit mechanical wear; prior work has considered related objectives for extending switch lifetime~\cite{Bianco2016SchedulingFabrics}. Another objective is to preserve routing flexibility across successive configurations by preferring matchings and routes that leave more future connections realizable without disturbing ongoing circuits. These objectives may trade off with minimum completion time, but can be preferable when switch lifetime, control overhead, or future reconfiguration flexibility matter operationally. Understanding how to jointly optimize these goals with completion time remains an open direction.

\myitem{Partial reconfiguration for multi-tenant isolation:}
A particularly interesting extension of \name is to use the internal routing structure of an \RNB interconnect for multi-tenant isolation. The recursive structure of a Bene\v{s} interconnect naturally partitions it into smaller subnetworks, allowing different tenants to be constrained to disjoint sets of internal switches. If two tenants use non-overlapping switch regions, reconfiguration for one cannot alter the switch states used by the other's active circuits. This provides isolation directly through physical routing and can prevent changing demand from one tenant from disrupting another, which may be especially useful when latency-sensitive and throughput-oriented workloads share the same photonic interconnect. The main trade-off is reduced routing flexibility. Partitioning limits the paths available to each tenant and can leave capacity unused when demands are imbalanced. Open questions include how to allocate subnetworks across tenants, how to adapt these allocations as demand changes, and when to relax isolation to improve overall performance. More broadly, the routing dependencies of \RNB interconnects may serve not only as a scheduling constraint, but also as a mechanism for multi-tenant isolation.

\section{Related Work}
\label{sec:relatedwork}

Circuit scheduling for optical interconnects has been studied extensively, spanning both demand-aware~\cite{BojjaVenkatakrishnan2016CostlyTheorems,Bianco2016SchedulingFabrics,Sunflow,Helios,285119} and demand-oblivious~\cite{MARS,Rotornet,10.1145/3651890.3672273,10.1145/3519935.3520020,Breaking_VLB} designs, with growing interest in AI datacenters~\cite{10.1145/3579371.3589350,10.1145/3748273.3749203,10.1145/3789240.3829166,light_will,6008516,pccl,case_server,10.1145/3779212.3790238}. Much of this literature assumes blocking execution, where each matching is separated by a matching-wide reconfiguration barrier. Early demand-aware approaches build on Birkhoff--von Neumann decomposition~\cite{BIRKHOFF1946TresLineal,BojjaVenkatakrishnan2016CostlyTheorems,Liu2015SchedulingNetworks}, while greedy MaxWeight decomposition has also been explored~\cite{ProjecToR,Helios,10.1145/3789240.3822610}, including recently for electrical-optical hybrid networks targeting Mixture-of-Experts workloads~\cite{10.1145/3718958.3750465}. Google's Jupiter network uses optical switching with relatively infrequent reconfiguration based on traffic-engineering decisions~\cite{Jupiter}, while TPU supercomputing systems continue to use optical switching for large-scale connectivity, including support for multi-tenancy and reallocation under failures~\cite{10.1145/3579371.3589350,295551}. A separate line of work studies demand-oblivious interconnects, where a predetermined sequence of matchings is executed periodically without adapting to instantaneous demand~\cite{MARS,Shale,10.1145/3519935.3520020,Rotornet,246486,Sirius}. More recent work explores demand-aware periodic switching~\cite{vermilion}. For distributed training and inference workloads, several systems construct circuit schedules around collective communication patterns~\cite{10.1145/3789240.3829166,10.1145/3748273.3749210,10.1145/3748273.3749203,pccl,light_will,305352}, while others optimize communication jointly with the overall workload~\cite{285119,10.1145/3712285.3759842}.

Most of these approaches still retain a matching-by-matching execution model. Partial reconfiguration has received comparatively less attention. Sunflow~\cite{Sunflow}, Best Fit First (BFF)~\cite{BFF}, and NEIP~\cite{Interruptible} allow individual connections to be reconfigured as endpoints become available, but assume a strictly non-blocking interconnect in which establishing a new connection does not disturb ongoing circuits. \name targets a different regime: rearrangeably non-blocking (\RNB) photonic interconnects, where partial reconfiguration depends on the internal routing of both existing and newly established circuits. Recent photonic interconnect designs increasingly explore \RNB architectures because they can provide high-radix connectivity with lower switching complexity than strictly non-blocking alternatives~\cite{6008516,Dupuis:18,9658267,hu2025silicon,CHU:18}. \name complements this hardware direction by revisiting circuit scheduling under the routing dependencies introduced by \RNB interconnects.

\section{Conclusion}

We revisited circuit scheduling for rearrangeably non-blocking photonic interconnects, where partial reconfiguration depends on both communication demand and internal routing. We presented \name, which combines bottleneck-driven matching selection, interconnect-aware routing, and independent connection scheduling to exploit partial reconfiguration without matching-wide barriers. Our results show that accounting for \RNB constraints can recover much of the performance otherwise lost under idealized scheduling abstractions. More broadly, we believe practical \RNB interconnects open a rich set of questions spanning scheduling, routing, controller overhead, and multi-tenant isolation.

\label{bodyLastPage}

\bibliographystyle{plainurl}
\bibliography{reference}

@misc{missionapollo,
      title={Mission Apollo: Landing Optical Circuit Switching at Datacenter Scale}, 
      author={Ryohei Urata and Hong Liu and Kevin Yasumura and Erji Mao and Jill Berger and Xiang Zhou and Cedric Lam and Roy Bannon and Darren Hutchinson and Daniel Nelson and Leon Poutievski and Arjun Singh and Joon Ong and Amin Vahdat},
      year={2022},
      eprint={2208.10041},
      archivePrefix={arXiv},
      primaryClass={cs.NI},
      url={https://arxiv.org/abs/2208.10041}, 
}

@inproceedings{case_server,
author = {Kumar, Abhishek Vijaya and Devraj, Arjun and Bunandar, Darius and Singh, Rachee},
title = {A case for server-scale photonic connectivity},
year = {2024},
isbn = {9798400712722},
publisher = {Association for Computing Machinery},
address = {New York, NY, USA},
url = {https://doi.org/10.1145/3696348.3696856},
doi = {10.1145/3696348.3696856},
booktitle = {Proceedings of the 23rd ACM Workshop on Hot Topics in Networks},
pages = {290–299},
numpages = {10},
location = {Irvine, CA, USA},
series = {HotNets '24}
}

@ARTICLE{TeraPhy,
  author={Wade, Mark and Anderson, Erik and Ardalan, Shahab and Bhargava, Pavan and Buchbinder, Sidney and L. Davenport, Michael and Fini, John and Lu, Haiwei and Li, Chen and Meade, Roy and Ramamurthy, Chandru and Rust, Michael and Sedgwick, Forrest and Stojanovic, Vladimir and Van Orden, Derek and Zhang, Chong and Sun, Chen and Shumarayev, Sergey Y. and O'Keeffe, Conor and Hoang, Tim T. and Kehlet, David and Mahajan, Ravi V. and Guzy, Matthew T. and Chan, Allen and Tran, Tina},
  journal={IEEE Micro}, 
  title={TeraPHY: A Chiplet Technology for Low-Power, High-Bandwidth In-Package Optical I/O}, 
  year={2020},
  volume={40},
  number={2},
  pages={63-71},
  doi={10.1109/MM.2020.2976067}}

@ARTICLE{6008516,
  author={Chen, L. and Hall, E. and Theogarajan, L. and Bowers, J.},
  journal={IEEE Photonics Journal}, 
  title={Photonic Switching for Data Center Applications}, 
  year={2011},
  volume={3},
  number={5},
  pages={834-844},
  doi={10.1109/JPHOT.2011.2166994}}

@article{bogaerts2020programmable,
  title={Programmable photonic circuits},
  author={Bogaerts, Wim and P{\'e}rez, Daniel and Capmany, Jos{\'e} and Miller, David AB and Poon, Joyce and Englund, Dirk and Morichetti, Francesco and Melloni, Andrea},
  journal={Nature},
  volume={586},
  number={7828},
  pages={207--216},
  year={2020},
  publisher={Nature Publishing Group UK London}
}

@article{Dupuis:18,
author = {Nicolas Dupuis and Benjamin G. Lee},
journal = {J. Lightwave Technol.},
number = {3},
pages = {763--772},
publisher = {Optica Publishing Group},
title = {Impact of Topology on the Scalability of Mach--Zehnder-Based Multistage Silicon Photonic Switch Networks},
volume = {36},
month = {Feb},
year = {2018},
url = {https://opg.optica.org/jlt/abstract.cfm?URI=jlt-36-3-763},
}

@inproceedings{10.1145/3718958.3750468,
author = {Shou, Chenchen and Liu, Guyue and Nie, Hao and Meng, Huaiyu and Zhou, Yu and Jiang, Yimin and Lv, Wenqing and Xu, Yelong and Lu, Yuanwei and Chen, Zhang and Yu, Yanbo and Shen, Yichen and Zhu, Yibo and Jiang, Daxin},
title = {InfiniteHBD: Building Datacenter-Scale High-Bandwidth Domain for LLM with Optical Circuit Switching Transceivers},
year = {2025},
isbn = {9798400715242},
publisher = {Association for Computing Machinery},
address = {New York, NY, USA},
url = {https://doi.org/10.1145/3718958.3750468},
doi = {10.1145/3718958.3750468},
booktitle = {Proceedings of the ACM SIGCOMM 2025 Conference},
pages = {1–23},
numpages = {23},
location = {S{\~a}o Francisco Convent, Coimbra, Portugal},
series = {SIGCOMM '25}
}

@inproceedings{10.1145/3452296.3472900,
author = {Khani, Mehrdad and Ghobadi, Manya and Alizadeh, Mohammad and Zhu, Ziyi and Glick, Madeleine and Bergman, Keren and Vahdat, Amin and Klenk, Benjamin and Ebrahimi, Eiman},
title = {SiP-ML: high-bandwidth optical network interconnects for machine learning training},
year = {2021},
isbn = {9781450383837},
publisher = {Association for Computing Machinery},
address = {New York, NY, USA},
url = {https://doi.org/10.1145/3452296.3472900},
doi = {10.1145/3452296.3472900},
booktitle = {Proceedings of the 2021 ACM SIGCOMM 2021 Conference},
pages = {657–675},
numpages = {19},
location = {Virtual Event, USA},
series = {SIGCOMM '21}
}

@inproceedings{10.1145/3779212.3790238,
author = {Vijaya Kumar, Abhishek and Ding, Eric and Devraj, Arjun and Bunandar, Darius and Singh, Rachee},
title = {Reconfigurable Torus Fabrics for Multi-tenant ML},
year = {2026},
isbn = {9798400723599},
publisher = {Association for Computing Machinery},
address = {New York, NY, USA},
url = {https://doi.org/10.1145/3779212.3790238},
doi = {10.1145/3779212.3790238},
booktitle = {Proceedings of the 31st ACM International Conference on Architectural Support for Programming Languages and Operating Systems, Volume 2},
pages = {1547–1565},
numpages = {19},
location = {USA},
series = {ASPLOS '26}
}

@ARTICLE{9658267,
  author={Ben Yoo, S. J.},
  journal={Journal of Lightwave Technology}, 
  title={Prospects and Challenges of Photonic Switching in Data Centers and Computing Systems}, 
  year={2022},
  volume={40},
  number={8},
  pages={2214-2243},
  doi={10.1109/JLT.2021.3136570}}

@article{hu2025silicon,
  title={Silicon photonic MEMS switches based on split waveguide crossings},
  author={Hu, Yinpeng and Sun, Yi and Lu, Ye and Li, Huan and Liu, Liu and Shi, Yaocheng and Dai, Daoxin},
  journal={Nature Communications},
  volume={16},
  number={1},
  pages={331},
  year={2025},
  publisher={Nature Publishing Group UK London},
  doi = {https://doi.org/10.1038/s41467-024-55528-9}
}

@inproceedings{10.1145/3579371.3589350,
author = {Jouppi, Norm and Kurian, George and Li, Sheng and Ma, Peter and Nagarajan, Rahul and Nai, Lifeng and Patil, Nishant and Subramanian, Suvinay and Swing, Andy and Towles, Brian and Young, Clifford and Zhou, Xiang and Zhou, Zongwei and Patterson, David A},
title = {TPU v4: An Optically Reconfigurable Supercomputer for Machine Learning with Hardware Support for Embeddings},
year = {2023},
isbn = {9798400700958},
publisher = {Association for Computing Machinery},
address = {New York, NY, USA},
url = {https://doi.org/10.1145/3579371.3589350},
doi = {10.1145/3579371.3589350},
booktitle = {Proceedings of the 50th Annual International Symposium on Computer Architecture},
articleno = {82},
numpages = {14},
location = {Orlando, FL, USA},
series = {ISCA '23}
}

@inproceedings {295551,
author = {Yazhou Zu and Alireza Ghaffarkhah and Hoang-Vu Dang and Brian Towles and Steven Hand and Safeen Huda and Adekunle Bello and Alexander Kolbasov and Arash Rezaei and Dayou Du and Steve Lacy and Hang Wang and Aaron Wisner and Chris Lewis and Henri Bahini},
title = {Resiliency at Scale: Managing {Google{\textquoteright}s} {TPUv4} Machine Learning Supercomputer},
booktitle = {21st USENIX Symposium on Networked Systems Design and Implementation (NSDI 24)},
year = {2024},
isbn = {978-1-939133-39-7},
address = {Santa Clara, CA},
pages = {761--774},
url = {https://www.usenix.org/conference/nsdi24/presentation/zu},
publisher = {USENIX Association},
month = apr
}

@inproceedings{CHU:18,
author = {Tao CHU and Lei QIAO and Weijie TANG and Defeng GUO and Weike WU},
booktitle = {Optical Fiber Communication Conference},
journal = {Optical Fiber Communication Conference},
pages = {Th1J.4},
publisher = {Optica Publishing Group},
title = {Fast, High-radix Silicon Photonic Switches},
year = {2018},
url = {https://opg.optica.org/abstract.cfm?URI=OFC-2018-Th1J.4},
doi = {10.1364/OFC.2018.Th1J.4},
}

@inproceedings{light_will,
author = {Addanki, Vamsi},
title = {When Light Bends to the Collective Will: A Theory and Vision for Adaptive Photonic Scale-up Domains},
year = {2025},
isbn = {9798400722806},
publisher = {Association for Computing Machinery},
address = {New York, NY, USA},
url = {https://doi.org/10.1145/3772356.3772395},
doi = {10.1145/3772356.3772395},
booktitle = {Proceedings of the 24th ACM Workshop on Hot Topics in Networks},
pages = {326–334},
numpages = {9},
location = {UMD Campus, College Park, MD, USA},
series = {HotNets '25}
}

@inproceedings{10.1145/3789240.3822610,
author = {Amponsah, Eliezer and Addanki, Vamsi},
title = {Birkhoff Decompositions and Photonic Interconnects Wait! Don’t Forget the Compute!},
year = {2026},
isbn = {9798400724671},
publisher = {Association for Computing Machinery},
address = {New York, NY, USA},
url = {https://doi.org/10.1145/3789240.3822610},
doi = {10.1145/3789240.3822610},
booktitle = {Proceedings of the Workshop on Hot Topics in Optical Technologies and Applications in Networking},
pages = {2135–2141},
numpages = {7},
location = {Colorado Convention Center, Denver, CO, USA},
series = {HotOptics '26}
}

@inproceedings{Sirius,
author = {Ballani, Hitesh and Costa, Paolo and Behrendt, Raphael and Cletheroe, Daniel and Haller, Istvan and Jozwik, Krzysztof and Karinou, Fotini and Lange, Sophie and Shi, Kai and Thomsen, Benn and Williams, Hugh},
title = {Sirius: A Flat Datacenter Network with Nanosecond Optical Switching},
year = {2020},
isbn = {9781450379557},
publisher = {Association for Computing Machinery},
address = {New York, NY, USA},
url = {https://doi.org/10.1145/3387514.3406221},
doi = {10.1145/3387514.3406221},
booktitle = {Proceedings of the Annual Conference of the ACM Special Interest Group on Data Communication on the Applications, Technologies, Architectures, and Protocols for Computer Communication},
pages = {782–797},
numpages = {16},
location = {Virtual Event, USA},
series = {SIGCOMM '20}
}

@inproceedings{Rotornet,
author = {Mellette, William M. and McGuinness, Rob and Roy, Arjun and Forencich, Alex and Papen, George and Snoeren, Alex C. and Porter, George},
title = {RotorNet: A Scalable, Low-complexity, Optical Datacenter Network},
year = {2017},
isbn = {9781450346535},
publisher = {Association for Computing Machinery},
address = {New York, NY, USA},
url = {https://doi.org/10.1145/3098822.3098838},
doi = {10.1145/3098822.3098838},
booktitle = {Proceedings of the Conference of the ACM Special Interest Group on Data Communication},
pages = {267–280},
numpages = {14},
location = {Los Angeles, CA, USA},
series = {SIGCOMM '17}
}

@article{MARS,
author = {Addanki, Vamsi and Avin, Chen and Schmid, Stefan},
title = {Mars: Near-Optimal Throughput with Shallow Buffers in Reconfigurable Datacenter Networks},
year = {2023},
issue_date = {March 2023},
publisher = {Association for Computing Machinery},
address = {New York, NY, USA},
volume = {7},
number = {1},
url = {https://doi.org/10.1145/3579312},
doi = {10.1145/3579312},
journal = {Proc. ACM Meas. Anal. Comput. Syst.},
month = mar,
articleno = {2},
numpages = {43}
}

@inproceedings{Shale,
author = {Amir, Daniel and Saran, Nitika and Wilson, Tegan and Kleinberg, Robert and Shrivastav, Vishal and Weatherspoon, Hakim},
title = {Shale: A Practical, Scalable Oblivious Reconfigurable Network},
year = {2024},
isbn = {9798400706141},
publisher = {Association for Computing Machinery},
address = {New York, NY, USA},
url = {https://doi.org/10.1145/3651890.3672248},
doi = {10.1145/3651890.3672248},
booktitle = {Proceedings of the ACM SIGCOMM 2024 Conference},
pages = {449–464},
numpages = {16},
location = {Sydney, NSW, Australia},
series = {ACM SIGCOMM '24}
}

@inproceedings{Breaking_VLB,
author = {Wilson, Tegan and Amir, Daniel and Saran, Nitika and Kleinberg, Robert and Shrivastav, Vishal and Weatherspoon, Hakim},
title = {Breaking the VLB Barrier for Oblivious Reconfigurable Networks},
year = {2024},
isbn = {9798400703836},
publisher = {Association for Computing Machinery},
address = {New York, NY, USA},
url = {https://doi.org/10.1145/3618260.3649608},
doi = {10.1145/3618260.3649608},
booktitle = {Proceedings of the 56th Annual ACM Symposium on Theory of Computing},
pages = {1865–1876},
numpages = {12},
location = {Vancouver, BC, Canada},
series = {STOC 2024}
}

@inproceedings{ProjecToR,
author = {Ghobadi, Monia and Mahajan, Ratul and Phanishayee, Amar and Devanur, Nikhil and Kulkarni, Janardhan and Ranade, Gireeja and Blanche, Pierre-Alexandre and Rastegarfar, Houman and Glick, Madeleine and Kilper, Daniel},
title = {ProjecToR: Agile Reconfigurable Data Center Interconnect},
year = {2016},
isbn = {9781450341936},
publisher = {Association for Computing Machinery},
address = {New York, NY, USA},
url = {https://doi.org/10.1145/2934872.2934911},
doi = {10.1145/2934872.2934911},
booktitle = {Proceedings of the 2016 ACM SIGCOMM Conference},
pages = {216–229},
numpages = {14},
location = {Florianopolis, Brazil},
series = {SIGCOMM '16}
}

@article{Helios,
author = {Farrington, Nathan and Porter, George and Radhakrishnan, Sivasankar and Bazzaz, Hamid Hajabdolali and Subramanya, Vikram and Fainman, Yeshaiahu and Papen, George and Vahdat, Amin},
title = {Helios: a hybrid electrical/optical switch architecture for modular data centers},
year = {2010},
issue_date = {October 2010},
publisher = {Association for Computing Machinery},
address = {New York, NY, USA},
volume = {40},
number = {4},
issn = {0146-4833},
url = {https://doi.org/10.1145/1851275.1851223},
doi = {10.1145/1851275.1851223},
journal = {SIGCOMM Comput. Commun. Rev.},
month = aug,
pages = {339–350},
numpages = {12}
}

@inproceedings{10.1145/3651890.3672273,
author = {Mellette, William M. and Forencich, Alex and Athapathu, Rukshani and Snoeren, Alex C. and Papen, George and Porter, George},
title = {Realizing RotorNet: Toward Practical Microsecond Scale Optical Networking},
year = {2024},
isbn = {9798400706141},
publisher = {Association for Computing Machinery},
address = {New York, NY, USA},
url = {https://doi.org/10.1145/3651890.3672273},
doi = {10.1145/3651890.3672273},
booktitle = {Proceedings of the ACM SIGCOMM 2024 Conference},
pages = {392–414},
numpages = {23},
location = {Sydney, NSW, Australia},
series = {ACM SIGCOMM '24}
}

@inproceedings{vermilion,
  author = {Addanki, Vamsi and Rao, Deeksha P and Knabe, Goran Dario and Avin, Chen and Schmid, Stefan},
  title = {Vermilion: A Traffic-Aware Reconfigurable Optical Interconnect with Formal Throughput Guarantees},
  booktitle = {24th USENIX Symposium on Networked Systems Design and Implementation (NSDI 27)},
  year = {2027},
  address = {Providence, RI},
  publisher = {USENIX Association},
  url = {https://www.usenix.org/conference/nsdi27/presentation/addanki}
}

@INPROCEEDINGS{BFF,
  author={Liu, Liang and Gong, Long and Yang, Sen and Xu, Jun and Fortnow, Lance},
  booktitle={2018 IEEE 11th International Conference on Cloud Computing (CLOUD)}, 
  title={Best First Fit (BFF): An Approach to Partially Reconfigurable Hybrid Circuit and Packet Switching}, 
  year={2018},
  volume={},
  number={},
  pages={426-433},
  doi={10.1109/CLOUD.2018.00060}}

@inproceedings{Sunflow,
author = {Huang, Xin Sunny and Sun, Xiaoye Steven and Ng, T.S. Eugene},
title = {Sunflow: Efficient Optical Circuit Scheduling for Coflows},
year = {2016},
isbn = {9781450342926},
publisher = {Association for Computing Machinery},
address = {New York, NY, USA},
url = {https://doi.org/10.1145/2999572.2999592},
doi = {10.1145/2999572.2999592},
booktitle = {Proceedings of the 12th International on Conference on Emerging Networking EXperiments and Technologies},
pages = {297–311},
numpages = {15},
location = {Irvine, California, USA},
series = {CoNEXT '16}
}

@ARTICLE{Interruptible,
  author={Zhang, Shuai and Shao, Junyi and Chen, Baojun and Sun, Weiqiang and Hu, Weisheng},
  journal={Journal of Lightwave Technology}, 
  title={Interruptible Scheduling of Partially Re-Configurable Optical Switching in Data Center Networks}, 
  year={2024},
  volume={42},
  number={7},
  pages={2212-2224},
  doi={10.1109/JLT.2023.3341042}}

@inproceedings{Jupiter,
author = {Poutievski, Leon and Mashayekhi, Omid and Ong, Joon and Singh, Arjun and Tariq, Mukarram and Wang, Rui and Zhang, Jianan and Beauregard, Virginia and Conner, Patrick and Gribble, Steve and Kapoor, Rishi and Kratzer, Stephen and Li, Nanfang and Liu, Hong and Nagaraj, Karthik and Ornstein, Jason and Sawhney, Samir and Urata, Ryohei and Vicisano, Lorenzo and Yasumura, Kevin and Zhang, Shidong and Zhou, Junlan and Vahdat, Amin},
title = {Jupiter evolving: transforming google's datacenter network via optical circuit switches and software-defined networking},
year = {2022},
isbn = {9781450394208},
publisher = {Association for Computing Machinery},
address = {New York, NY, USA},
url = {https://doi.org/10.1145/3544216.3544265},
doi = {10.1145/3544216.3544265},
booktitle = {Proceedings of the ACM SIGCOMM 2022 Conference},
pages = {66–85},
numpages = {20},
location = {Amsterdam, Netherlands},
series = {SIGCOMM '22}
}

@inproceedings{10.1145/3603269.3604836,
author = {Liu, Hong and Urata, Ryohei and Yasumura, Kevin and Zhou, Xiang and Bannon, Roy and Berger, Jill and Dashti, Pedram and Jouppi, Norm and Lam, Cedric and Li, Sheng and Mao, Erji and Nelson, Daniel and Papen, George and Tariq, Mukarram and Vahdat, Amin},
title = {Lightwave Fabrics: At-Scale Optical Circuit Switching for Datacenter and Machine Learning Systems},
year = {2023},
isbn = {9798400702365},
publisher = {Association for Computing Machinery},
address = {New York, NY, USA},
url = {https://doi.org/10.1145/3603269.3604836},
doi = {10.1145/3603269.3604836},
booktitle = {Proceedings of the ACM SIGCOMM 2023 Conference},
pages = {499–515},
numpages = {17},
location = {New York, NY, USA},
series = {ACM SIGCOMM '23}
}

@article{Jonker1987AProblems,
    title = {{A shortest augmenting path algorithm for dense and sparse linear assignment problems}},
    year = {1987},
    journal = {Computing},
    author = {Jonker, R and Volgenant, A},
    number = {4},
    pages = {325--340},
    volume = {38},
    url = {https://doi.org/10.1007/BF02278710},
    doi = {10.1007/BF02278710},
    issn = {1436-5057}
}

@article{Crouse2016OnAlgorithms,
    title = {{On implementing 2D rectangular assignment algorithms}},
    year = {2016},
    journal = {IEEE Transactions on Aerospace and Electronic Systems},
    author = {Crouse, David F},
    number = {4},
    pages = {1679--1696},
    volume = {52},
    doi = {10.1109/TAES.2016.140952}
}

@article{Kuhn1955TheProblem,
    title = {{The Hungarian method for the assignment problem}},
    year = {1955},
    journal = {Naval research logistics quarterly},
    author = {Kuhn, Harold W},
    number = {1-2},
    pages = {83--97},
    volume = {2},
    publisher = {Wiley Online Library}
}

@inproceedings{BojjaVenkatakrishnan2016CostlyTheorems,
    title = {{Costly Circuits, Submodular Schedules and Approximate Carath{\'{e}}odory Theorems}},
    year = {2016},
    booktitle = {Proceedings of the 2016 ACM SIGMETRICS International Conference on Measurement and Modeling of Computer Science},
    author = {Bojja Venkatakrishnan, Shaileshh and Alizadeh, Mohammad and Viswanath, Pramod},
    pages = {75--88},
    series = {SIGMETRICS '16},
    publisher = {Association for Computing Machinery},
    url = {https://doi.org/10.1145/2896377.2901479},
    address = {New York, NY, USA},
    isbn = {9781450342667},
    doi = {10.1145/2896377.2901479}
}

@inproceedings{Liu2015SchedulingNetworks,
    title = {{Scheduling techniques for hybrid circuit/packet networks}},
    year = {2015},
    booktitle = {Proceedings of the 11th ACM Conference on Emerging Networking Experiments and Technologies},
    author = {Liu, He and Mukerjee, Matthew K and Li, Conglong and Feltman, Nicolas and Papen, George and Savage, Stefan and Seshan, Srinivasan and Voelker, Geoffrey M and Andersen, David G and Kaminsky, Michael and Porter, George and Snoeren, Alex C},
    series = {CoNEXT '15},
    publisher = {Association for Computing Machinery},
    url = {https://doi.org/10.1145/2716281.2836126},
    address = {New York, NY, USA},
    isbn = {9781450334129},
    doi = {10.1145/2716281.2836126}
}

@article{Bianco2016SchedulingFabrics,
    title = {{Scheduling traffic for maximum switch lifetime in optical data center fabrics}},
    year = {2016},
    journal = {Computer Networks},
    author = {Bianco, Andrea and Giaccone, Paolo and Ricca, Marco},
    pages = {75--88},
    volume = {105},
    url = {https://www.sciencedirect.com/science/article/pii/S1389128616301232},
    doi = {https://doi.org/10.1016/j.comnet.2016.05.002},
    issn = {1389-1286}
}

@ARTICLE{swithfabric,
  author={Lee, Benjamin G. and Dupuis, Nicolas},
  journal={Journal of Lightwave Technology}, 
  title={Silicon Photonic Switch Fabrics: Technology and Architecture}, 
  year={2019},
  volume={37},
  number={1},
  pages={6-20},
  doi={10.1109/JLT.2018.2876828}}

@article{BIRKHOFF1946TresLineal,
    title = {{Tres observaciones sobre el algebra lineal}},
    year = {1946},
    journal = {Univ. Nac. Tucuman, Ser. A},
    author = {BIRKHOFF, G},
    pages = {147--154},
    volume = {5},
    url = {https://cir.nii.ac.jp/crid/1570572699525842816}
}

@INPROCEEDINGS{piloss,
  author={Konoike, Ryotaro and Suzuki, Keijiro and Ikeda, Kazuhiro},
  booktitle={2022 Optical Fiber Communications Conference and Exhibition (OFC)}, 
  title={Path-Independent Insertion-Loss (PILOSS) 8 × 8 Silicon Photonics Switch with <8 nsec Switching Time}, 
  year={2022},
  volume={},
  number={},
  pages={1-3},
  doi={}}

@ARTICLE{4x4,
  author={Lu, Liangjun and Li, Xiaorui and Gao, Wei and Li, Xin and Zhou, Linjie and Chen, Jianping},
  journal={IEEE Photonics Journal}, 
  title={Silicon Non-Blocking 4 × 4 Optical Switch Chip Integrated With Both Thermal and Electro-Optic Tuners}, 
  year={2019},
  volume={11},
  number={6},
  pages={1-9},
  doi={10.1109/JPHOT.2019.2941960}}

@inproceedings{10.1145/3519935.3520020,
author = {Amir, Daniel and Wilson, Tegan and Shrivastav, Vishal and Weatherspoon, Hakim and Kleinberg, Robert and Agarwal, Rachit},
title = {Optimal oblivious reconfigurable networks},
year = {2022},
isbn = {9781450392648},
publisher = {Association for Computing Machinery},
address = {New York, NY, USA},
url = {https://doi.org/10.1145/3519935.3520020},
doi = {10.1145/3519935.3520020},
booktitle = {Proceedings of the 54th Annual ACM SIGACT Symposium on Theory of Computing},
pages = {1339–1352},
numpages = {14},
location = {Rome, Italy},
series = {STOC 2022}
}

@article{Clos1953ASO,
  title={A study of non-blocking switching networks},
  author={Charles Clos},
  journal={Bell System Technical Journal},
  year={1953},
  volume={32},
  pages={406-424},
  url={https://api.semanticscholar.org/CorpusID:111089194}
}

@article{10.1145/321439.321449,
author = {Waksman, Abraham},
title = {A Permutation Network},
year = {1968},
issue_date = {Jan. 1968},
publisher = {Association for Computing Machinery},
address = {New York, NY, USA},
volume = {15},
number = {1},
issn = {0004-5411},
url = {https://doi.org/10.1145/321439.321449},
doi = {10.1145/321439.321449},
journal = {J. ACM},
month = jan,
pages = {159–163},
numpages = {5}
}

@article{Sinkhorn1964AMatrices,
    title = {{A Relationship Between Arbitrary Positive Matrices and Doubly Stochastic Matrices}},
    year = {1964},
    journal = {The Annals of Mathematical Statistics},
    author = {Sinkhorn, Richard},
    number = {2},
    pages = {876--879},
    volume = {35},
    publisher = {Institute of Mathematical Statistics},
    url = {http://www.jstor.org/stable/2238545},
    issn = {00034851}
}

@inproceedings {316654,
author = {Yiran Lei and Dongjoo Lee and Liangyu Zhao and Daniar Kurniawan and Chanmyeong Kim and Heetaek Jeong and Changsu Kim and Hyeonseong Choi and Liangcheng Yu and Arvind Krishnamurthy and Justine Sherry and Eriko Nurvitadhi},
title = {{FAST}: An Efficient Scheduler for {All-to-All} {GPU} Communication},
booktitle = {23rd USENIX Symposium on Networked Systems Design and Implementation (NSDI 26)},
year = {2026},
isbn = {978-1-939133-54-0},
address = {Renton, WA},
pages = {2515--2531},
url = {https://www.usenix.org/conference/nsdi26/presentation/lei-yiran},
publisher = {USENIX Association},
month = may
}

@inproceedings{10.1145/3718958.3750465,
author = {Liao, Xudong and Sun, Yijun and Tian, Han and Wan, Xinchen and Jin, Yilun and Wang, Zilong and Ren, Zhenghang and Huang, Xinyang and Li, Wenxue and Tse, Kin Fai and Zhong, Zhizhen and Liu, Guyue and Zhang, Ying and Ye, Xiaofeng and Zhang, Yiming and Chen, Kai},
title = {MixNet: A Runtime Reconfigurable Optical-Electrical Fabric for Distributed Mixture-of-Experts Training},
year = {2025},
isbn = {9798400715242},
publisher = {Association for Computing Machinery},
address = {New York, NY, USA},
url = {https://doi.org/10.1145/3718958.3750465},
doi = {10.1145/3718958.3750465},
booktitle = {Proceedings of the ACM SIGCOMM 2025 Conference},
pages = {554–574},
numpages = {21},
location = {S{\~a}o Francisco Convent, Coimbra, Portugal},
series = {SIGCOMM '25}
}

@inbook{10.1145/3789240.3829166,
author = {Rahman, Mahir and Joseph, Samuel and Kodkani, Nihar and Arzani, Behnaz and Addanki, Vamsi},
title = {Harvest: Adaptive Photonic Switching Schedules for Collective Communication in Scale-up Domains},
year = {2026},
isbn = {9798400724671},
publisher = {Association for Computing Machinery},
address = {New York, NY, USA},
url = {https://doi.org/10.1145/3789240.3829166},
booktitle = {Proceedings of the ACM SIGCOMM 2026 Conference},
pages = {1576–1600},
numpages = {25}
}

@inproceedings{10.1145/3748273.3749210,
author = {Renganathan, Sundararajan and McKeown, Nick},
title = {Chronos: Prescheduled circuit switching for LLM training},
year = {2025},
isbn = {9798400720826},
publisher = {Association for Computing Machinery},
address = {New York, NY, USA},
url = {https://doi.org/10.1145/3748273.3749210},
doi = {10.1145/3748273.3749210},
booktitle = {Proceedings of the 2nd Workshop on Networks for AI Computing},
pages = {89–97},
numpages = {9},
location = {Coimbra, Portugal},
series = {NAIC '25}
}

@inproceedings{10.1145/3748273.3749203,
author = {Athapathu, Rukshani and Porter, George},
title = {Reconfigurability within Collective Communication Algorithms},
year = {2025},
isbn = {9798400720826},
publisher = {Association for Computing Machinery},
address = {New York, NY, USA},
url = {https://doi.org/10.1145/3748273.3749203},
doi = {10.1145/3748273.3749203},
booktitle = {Proceedings of the 2nd Workshop on Networks for AI Computing},
pages = {43–49},
numpages = {7},
location = {Coimbra, Portugal},
series = {NAIC '25}
}

@misc{pccl,
      title={PCCL: Photonic circuit-switched collective communication for distributed ML}, 
      author={Abhishek Vijaya Kumar and Arjun Devraj and Rachee Singh},
      year={2025},
      eprint={2509.15450},
      archivePrefix={arXiv},
      primaryClass={cs.DC},
      url={https://arxiv.org/abs/2509.15450}, 
}

@inproceedings{10.1145/3712285.3759842,
author = {Wu, Zhenguo and Klenk, Benjamin and Dennison, Larry and Bergman, Keren},
title = {ACTINA: Adapting Circuit-Switching Techniques for AI Networking Architectures},
year = {2025},
isbn = {9798400714665},
publisher = {Association for Computing Machinery},
address = {New York, NY, USA},
url = {https://doi.org/10.1145/3712285.3759842},
doi = {10.1145/3712285.3759842},
booktitle = {Proceedings of the International Conference for High Performance Computing, Networking, Storage and Analysis},
pages = {1211–1222},
numpages = {12},
location = {
},
series = {SC '25}
}

@inproceedings {305352,
author = {Liangyu Zhao and Siddharth Pal and Tapan Chugh and Weiyang Wang and Jason Fantl and Prithwish Basu and Joud Khoury and Arvind Krishnamurthy},
title = {Efficient {Direct-Connect} Topologies for Collective Communications},
booktitle = {22nd USENIX Symposium on Networked Systems Design and Implementation (NSDI 25)},
year = {2025},
isbn = {978-1-939133-46-5},
address = {Philadelphia, PA},
pages = {705--737},
url = {https://www.usenix.org/conference/nsdi25/presentation/zhao-liangyu},
publisher = {USENIX Association},
month = apr
}

@inproceedings {316650,
author = {Yiming Lei and Federico De Marchi and Jialong Li and Raj Joshi and Shu-Ting Wang and Xiaoqi Chen and Balakrishnan Chandrasekaran and Yiting Xia},
title = {{OpenOptics}: Enabling Open Research and Implementation of Optical Data Center Networks},
booktitle = {23rd USENIX Symposium on Networked Systems Design and Implementation (NSDI 26)},
year = {2026},
isbn = {978-1-939133-54-0},
address = {Renton, WA},
pages = {987--1004},
url = {https://www.usenix.org/conference/nsdi26/presentation/lei-optical},
publisher = {USENIX Association},
month = may
}

@inproceedings {316652,
author = {Yiming Lei and Jialong Li and Zhengqing Liu and Raj Joshi and Yiting Xia},
title = {{SyncWise}: {Error-Aware} Time Synchronization for Reconfigurable Data Center Networks},
booktitle = {23rd USENIX Symposium on Networked Systems Design and Implementation (NSDI 26)},
year = {2026},
isbn = {978-1-939133-54-0},
address = {Renton, WA},
pages = {951--967},
url = {https://www.usenix.org/conference/nsdi26/presentation/lei-syncwise},
publisher = {USENIX Association},
month = may
}

@article{rogerwhockney,
title = {The communication challenge for MPP: Intel Paragon and Meiko CS-2},
journal = {Parallel Computing},
volume = {20},
number = {3},
pages = {389-398},
year = {1994},
issn = {0167-8191},
doi = {10.1016/S0167-8191(06)80021-9},
author = {Roger W. Hockney}
}

@inproceedings {285119,
author = {Weiyang Wang and Moein Khazraee and Zhizhen Zhong and Manya Ghobadi and Zhihao Jia and Dheevatsa Mudigere and Ying Zhang and Anthony Kewitsch},
title = {{TopoOpt}: Co-optimizing Network Topology and Parallelization Strategy for Distributed Training Jobs},
booktitle = {20th USENIX Symposium on Networked Systems Design and Implementation (NSDI 23)},
year = {2023},
isbn = {978-1-939133-33-5},
address = {Boston, MA},
pages = {739--767},
url = {https://www.usenix.org/conference/nsdi23/presentation/wang-weiyang},
publisher = {USENIX Association},
month = apr
}

@inproceedings {246486,
author = {William M. Mellette and Rajdeep Das and Yibo Guo and Rob McGuinness and Alex C. Snoeren and George Porter},
title = {Expanding across time to deliver bandwidth efficiency and low latency},
booktitle = {17th USENIX Symposium on Networked Systems Design and Implementation (NSDI 20)},
year = {2020},
isbn = {978-1-939133-13-7},
address = {Santa Clara, CA},
pages = {1--18},
url = {https://www.usenix.org/conference/nsdi20/presentation/mellette},
publisher = {USENIX Association},
month = feb
}

@misc{abramovich2026speedbenchunifieddiversebenchmark,
      title={SPEED-Bench: A Unified and Diverse Benchmark for Speculative Decoding}, 
      author={Talor Abramovich and Maor Ashkenazi and Izzy Putterman and Benjamin Chislett and Tiyasa Mitra and Bita Darvish Rouhani and Ran Zilberstein and Yonatan Geifman},
      year={2026},
      eprint={2604.09557},
      archivePrefix={arXiv},
      primaryClass={cs.DC},
      url={https://arxiv.org/abs/2604.09557}, 
}

@article{hendryckstest2021,
  title={Measuring Massive Multitask Language Understanding},
  author={Dan Hendrycks and Collin Burns and Steven Basart and Andy Zou and Mantas Mazeika and Dawn Song and Jacob Steinhardt},
  journal={Proceedings of the International Conference on Learning Representations (ICLR)},
  year={2021}
}

@article{hendrycks2021ethics,
  title={Aligning AI With Shared Human Values},
  author={Dan Hendrycks and Collin Burns and Steven Basart and Andrew Critch and Jerry Li and Dawn Song and Jacob Steinhardt},
  journal={Proceedings of the International Conference on Learning Representations (ICLR)},
  year={2021}
}

@article{cobbe2021gsm8k,
  title={Training Verifiers to Solve Math Word Problems},
  author={Cobbe, Karl and Kosaraju, Vineet and Bavarian, Mohammad and Chen, Mark and Jun, Heewoo and Kaiser, Lukasz and Plappert, Matthias and Tworek, Jerry and Hilton, Jacob and Nakano, Reiichiro and Hesse, Christopher and Schulman, John},
  journal={arXiv preprint arXiv:2110.14168},
  year={2021}
}

\appendix

\clearpage
\begin{table}[t]
\centering
\small
\resizebox{\linewidth}{!}{%
\begin{tabular}{|p{2.7cm}|c|c|c|}
\hline
\textbf{Topology} &
\textbf{Blocking} &
\textbf{$2\times2$ Switches} &
\textbf{Stages} \\
\hline\hline

Butterfly / Banyan / Omega &
Blocking &
$\frac{n}{2}\cdot\log_2 n$ &
$\log_2 n$ \\
\hline

PILOSS &
SNB &
$n^2$ &
$n$ \\
\hline

Double-Layer Network (DLN) &
SNB &
$\frac{5}{4}\cdot n^2-2\cdot n$ &
$2\cdot\log_2 n-1$ \\
\hline

Bene\v{s} &
RNB &
$n\cdot\log_2 n-\frac{n}{2}$ &
$2\cdot\log_2 n-1$ \\
\hline

Spanke-Bene\v{s} &
RNB &
$\frac{n\cdot(n-1)}{2}$ &
$n$ \\
\hline

\end{tabular}
}
\caption{Comparison of representative $n\times n$ interconnects constructed from $2\times2$ switching elements. The Bene\v{s} interconnect provides rearrangeably non-blocking connectivity using $O(n\log n)$ switching elements and $O(\log n)$ stages.}
\label{tab:topo}
\end{table}

\begin{table}[t]
\centering
\small

\renewcommand{\arraystretch}{1.10}
\setlength{\tabcolsep}{5pt}
\arrayrulecolor{gray!45}

\begin{tabular}{@{}p{0.18\columnwidth}|p{0.75\columnwidth}@{}}
\hline
\rowcolor{black!12}
\textbf{Notation} & \textbf{Description} \\
\hline

\rowcolor{gray!10}
\multicolumn{2}{@{}l@{}}{\textcolor{red!55!black}{\textbf{Interconnect types}}} \\
\hline
\SNB & Strictly non-blocking \\
\hline
\RNB & Rearrangably non-blocking \\
\hline

\rowcolor{gray!10}
\multicolumn{2}{@{}l@{}}{\textcolor{red!55!black}{\textbf{Demand and schedule}}} \\
\hline
$\Demand_{uv}$ & Communication demand, in bytes, from transmitter $u$ to receiver $v$ \\
\hline
$b$ & Bandwidth of each transmit and receive link \\
\hline
$\Trans_{uv}$ & Transmission time of demand $(u,v)$, $\Trans_{uv}=\Demand_{uv}/b$ \\
\hline
$R_{uv}$ & Residual transmission time of demand $(u,v)$ \\
\hline
$\alpha_r$ & Reconfiguration delay of an internal $2\times2$ switch \\
\hline
$G$ & Bene\v{s} interconnect \\
\hline
$\mathcal{S}$ & Circuit schedule constructed so far \\
\hline
$M$ & Planning matching selected in one iteration \\
\hline

\rowcolor{gray!10}
\multicolumn{2}{@{}l@{}}{\textcolor{red!55!black}{\textbf{Matching selection}}} \\
\hline
$E_R(e)$ & Remaining demands incident on endpoint $e$ \\
\hline
$d_e$ & Residual degree of endpoint $e$, $d_e=|E_R(e)|$ \\
\hline
$W_e$ & Estimated remaining communication and reconfiguration work of endpoint $e$ \\
\hline
$a_e$ & Completion time of the latest scheduled connection involving endpoint $e$ \\
\hline
$\delta$ & Maximum residual endpoint degree, $\delta=\max_e d_e$ \\
\hline
$C$ & Endpoints with maximum residual degree, $C=\{e\mid d_e=\delta\}$ \\
\hline

\rowcolor{gray!10}
\multicolumn{2}{@{}l@{}}{\textcolor{red!55!black}{\textbf{Path and connection cost}}} \\
\hline
$p_{uv}$ & Candidate path used to estimate the cost of connection $(u,v)$ \\
\hline
$s_{uv}$ & Estimated earliest start time of candidate connection $(u,v)$ \\
\hline
$c_{uv}$ & Cost estimate used when selecting connection $(u,v)$ \\
\hline
$P_{uv}$ & Routed path assigned to selected connection $(u,v)$ \\
\hline
\end{tabular}

\arrayrulecolor{black}
\caption{Notations used in \S\ref{sec:tetris}.}
\label{tab:notation}
\end{table}

\setlength{\textfloatsep}{0.2cm}
\setlength{\floatsep}{0.2cm}
\begin{algorithm}[t]
\small
\DontPrintSemicolon
\SetAlgoLined
\caption{\textsc{Route}$(g,M,R,\mathcal{S},\Pi)$}
\label{alg:tetris-routing}

\KwIn{Bene\v{s} subnetwork $g$, matching $M$, residual matrix $R$, schedule $\mathcal{S}$, ancestor traversals $\Pi$}
\KwOut{Path $P_{uv}$ for every $(u,v)\in M$}

\If{$M=\emptyset$}{
\Return{$\emptyset$}
}

\If{$g$ is a $2\times2$ switch}{
\Return{required traversals joined with $\Pi$}
}

Build input-pair, output-pair, and connection branch constraints\;
\Comment{\textcolor{gray}{Each component has two legal orientations}}

\ForEach{constraint component $K$}{
\ForEach{legal orientation $q\in\{0,1\}$}{
$V_q\gets\emptyset$;\quad
$X_q\gets\emptyset$\;

\Comment{\textcolor{gray}{Evaluate one orientation}}

\ForEach{$(u,v)\in M$ belonging to $K$}{
    $p\gets$ traversals in $\Pi$ and current boundaries under $q$\;
    $t\gets\textsc{Start}(\mathcal{S},p)$\;

    Append $t-\alpha_r+W_u$ and $t-\alpha_r+W_v$ to $V_q$\;
    Add switches in $p$ requiring a state change to $X_q$\;
}

$L_q\gets\operatorname{sort}_{\downarrow}(V_q)$\;
\Comment{\textcolor{gray}{Endpoint costs; switch changes break ties}}

$J_q\gets(L_q,|X_q|)$\;
}

Choose $q^\star\gets\arg\min_q^{\mathrm{lex}}J_q$\;
Fix orientation $q^\star$ for component $K$\;
}

Split $M$ into upper and lower child matchings $M_U,M_L$\;
Extend $\Pi$ with selected boundary traversals to obtain $\Pi_U,\Pi_L$\;

\Comment{\textcolor{gray}{Recurse after fixing the current level}}

$P_U\gets\textsc{Route}(g_U,M_U,R,\mathcal{S},\Pi_U)$\;
$P_L\gets\textsc{Route}(g_L,M_L,R,\mathcal{S},\Pi_L)$\;

\Return{$P_U\cup P_L$}
\end{algorithm}

\section{Additional Results}
\label{app:additional-results}

\begin{figure*}[t]
    \centering
    \begin{subfigure}{0.49\textwidth}
        \centering
        \includegraphics[width=0.9\linewidth]{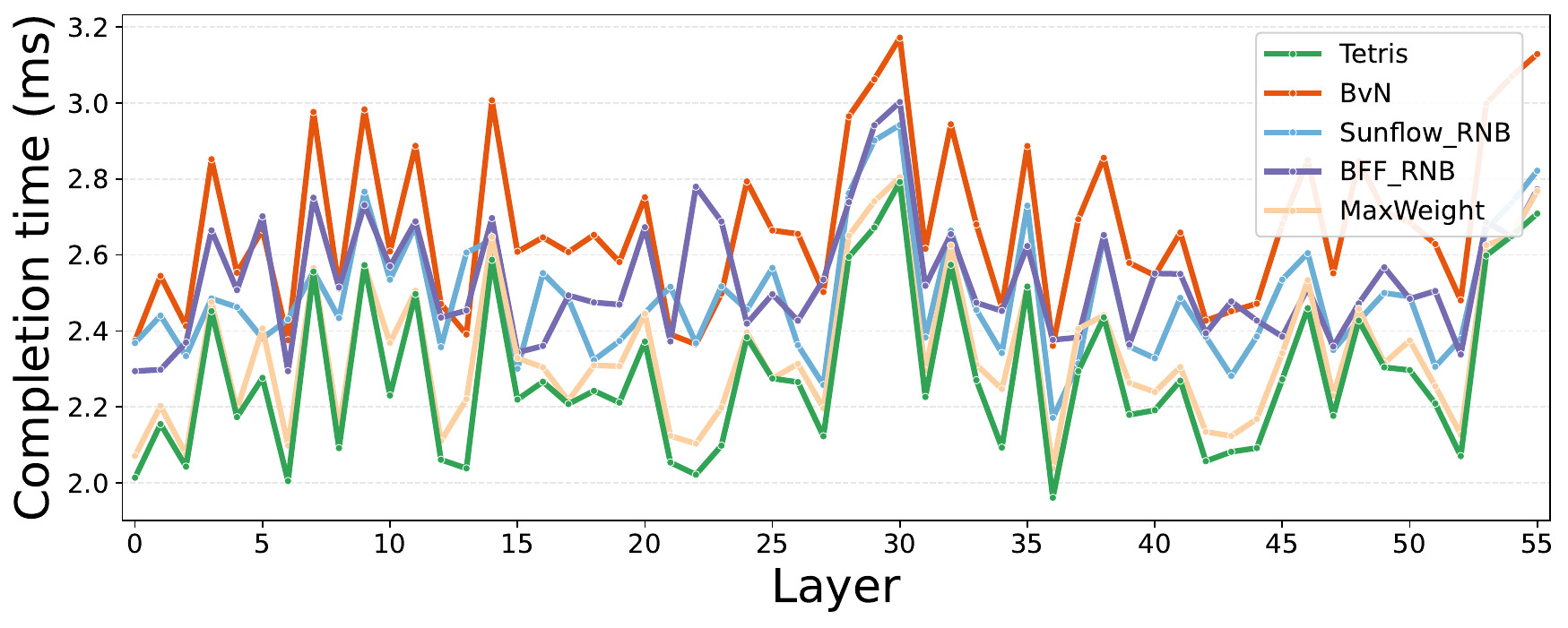}
        \caption{SPEED-Bench throughput-$1$K dataset} %
        \label{fig:makespan-1k}
    \end{subfigure}
    \hfill
    \begin{subfigure}{0.49\textwidth}
        \centering
        \includegraphics[width=0.9\linewidth]{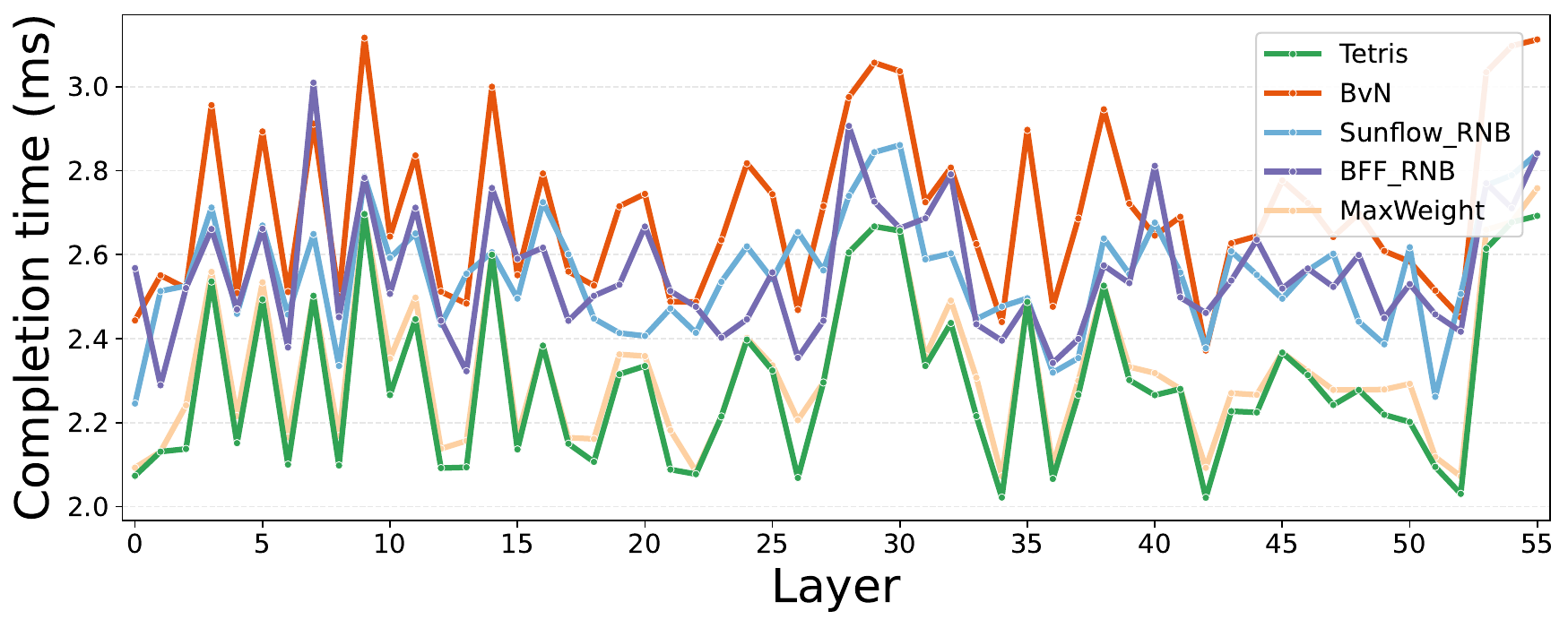}
        \caption{SPEED-Bench throughput-$2$K dataset}
        \label{fig:makespan-2k}
    \end{subfigure}
    \caption{Per-layer communication time for one inference pass of Mixtral-$8\times22$B, comparing \name with BvN, MaxWeight, Sunflow, and BFF, at $b=800$~Gbps and $\alpha_r=10\,\mu$s.}
    \label{fig:makespan}
\end{figure*}

\begin{figure*}
\centering
   \begin{subfigure}{0.23\linewidth}
    \includegraphics[width=\linewidth]{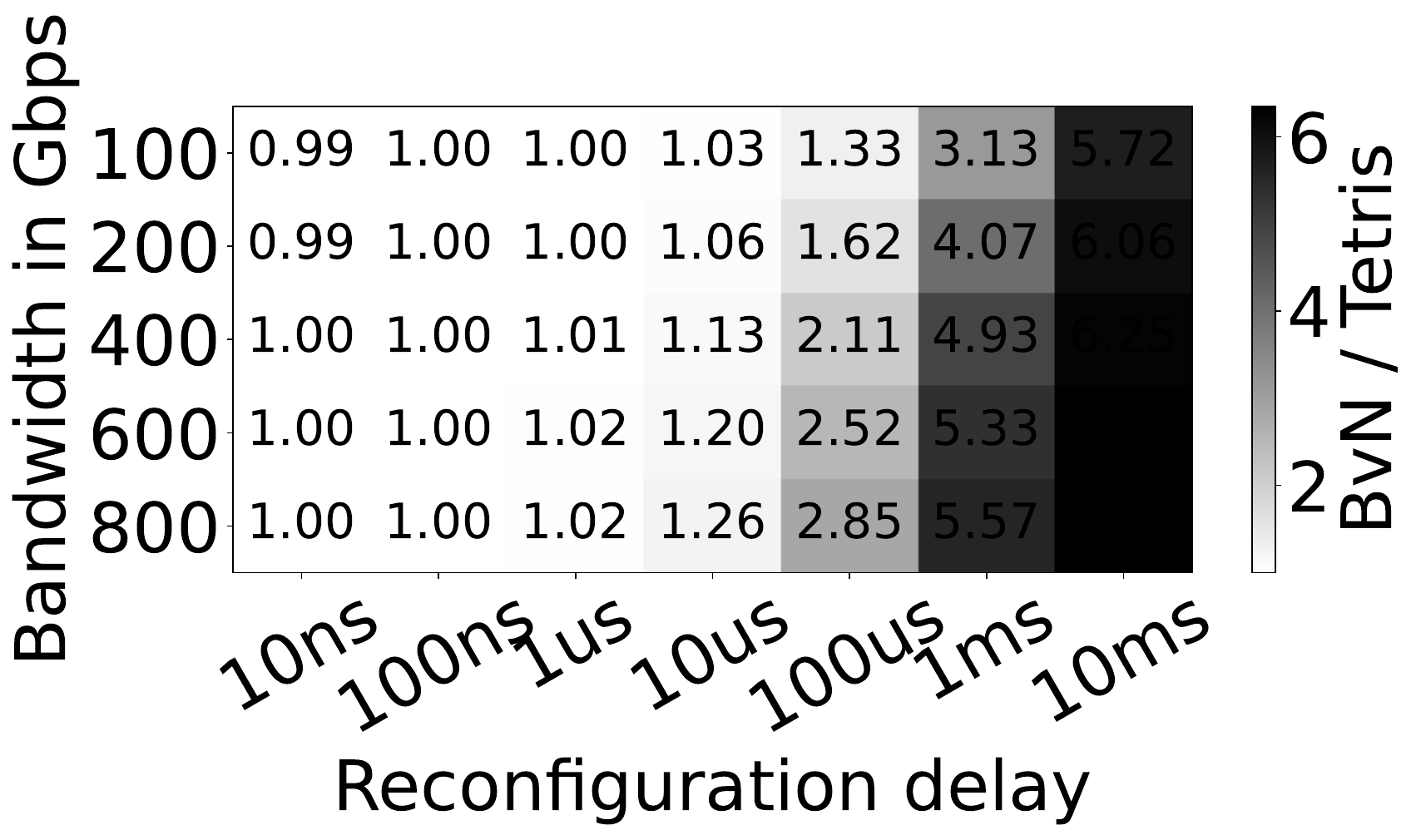}
    \caption{BvN against \name}
    \label{fig:heatmap-deepseek-bvn}
  \end{subfigure}%
  \hfill
   \begin{subfigure}{0.23\linewidth}
    \includegraphics[width=\linewidth]{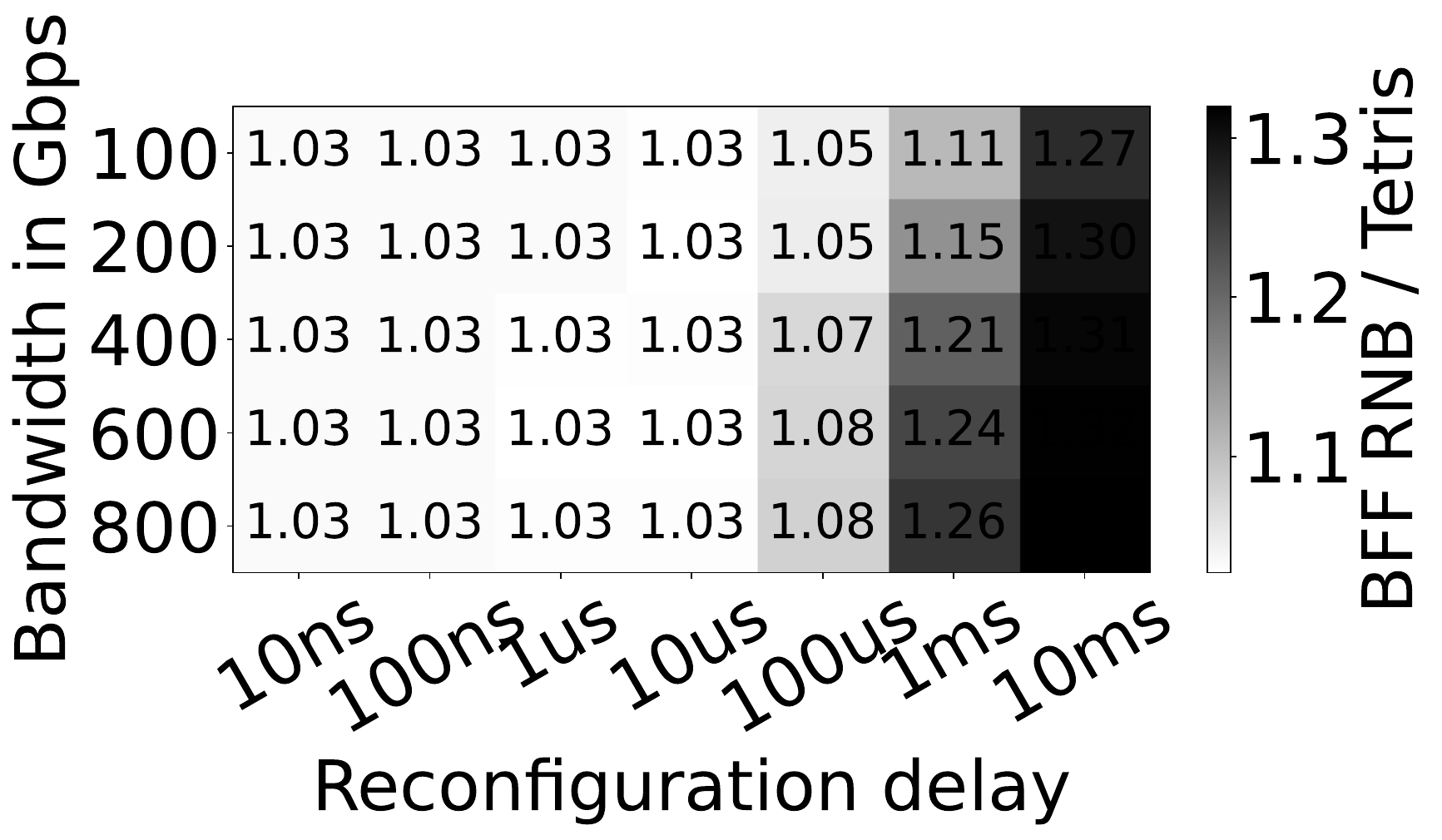}
    \caption{BFF against \name}
    \label{fig:heatmap-deepseek-bff}
  \end{subfigure}
    \hfill
   \begin{subfigure}{0.23\linewidth}
    \includegraphics[width=\linewidth]{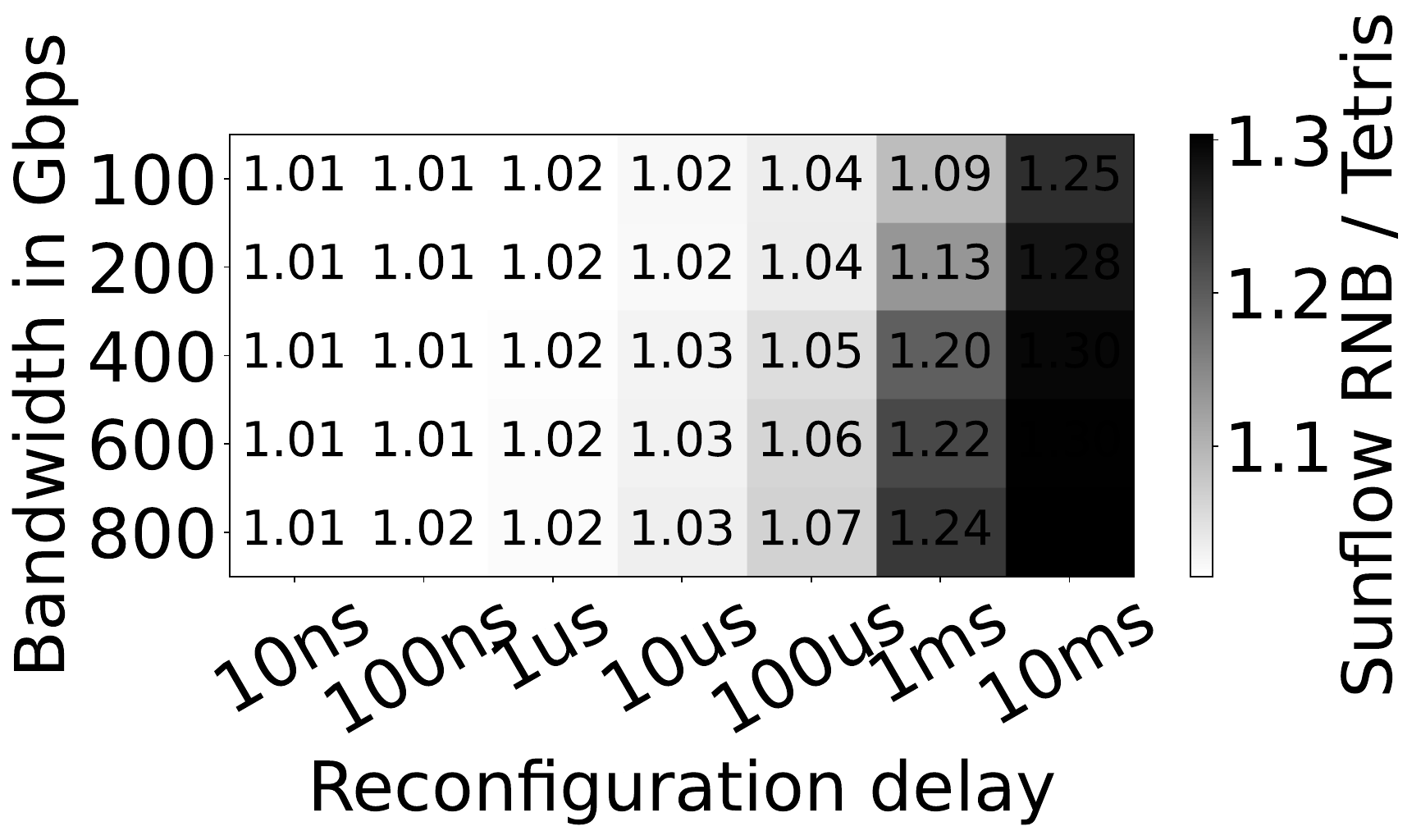}
    \caption{Sunflow against \name}
    \label{fig:heatmap-deepseek-sunflow}
  \end{subfigure}
   \hfill
   \begin{subfigure}{0.23\linewidth}
    \includegraphics[width=\linewidth]{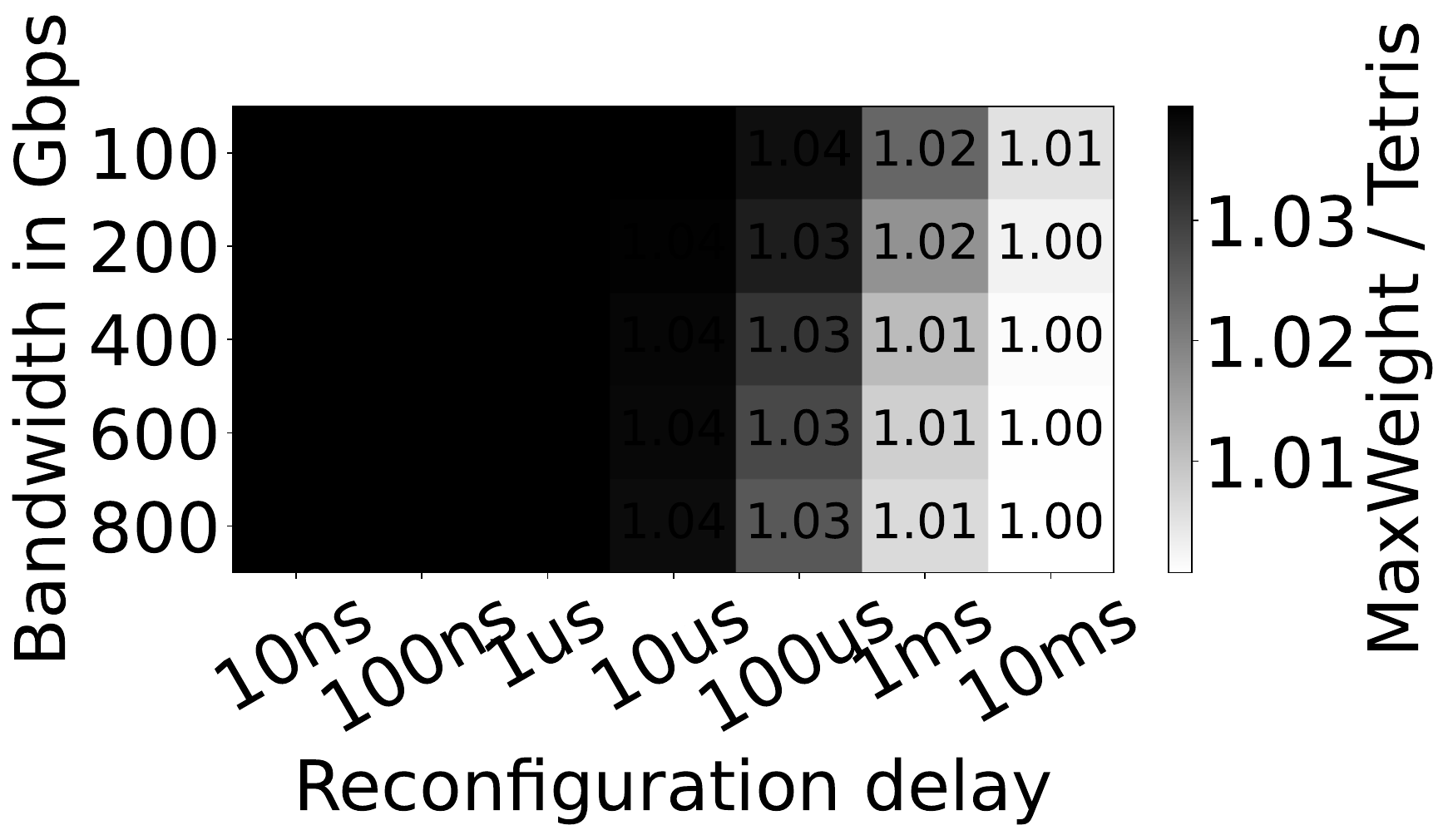}
    \caption{MaxWeight against \name}
    \label{fig:heatmap-deepseek-maxweight}
  \end{subfigure}

\vspace{1mm}

   \begin{subfigure}{0.23\linewidth}
    \includegraphics[width=\linewidth]{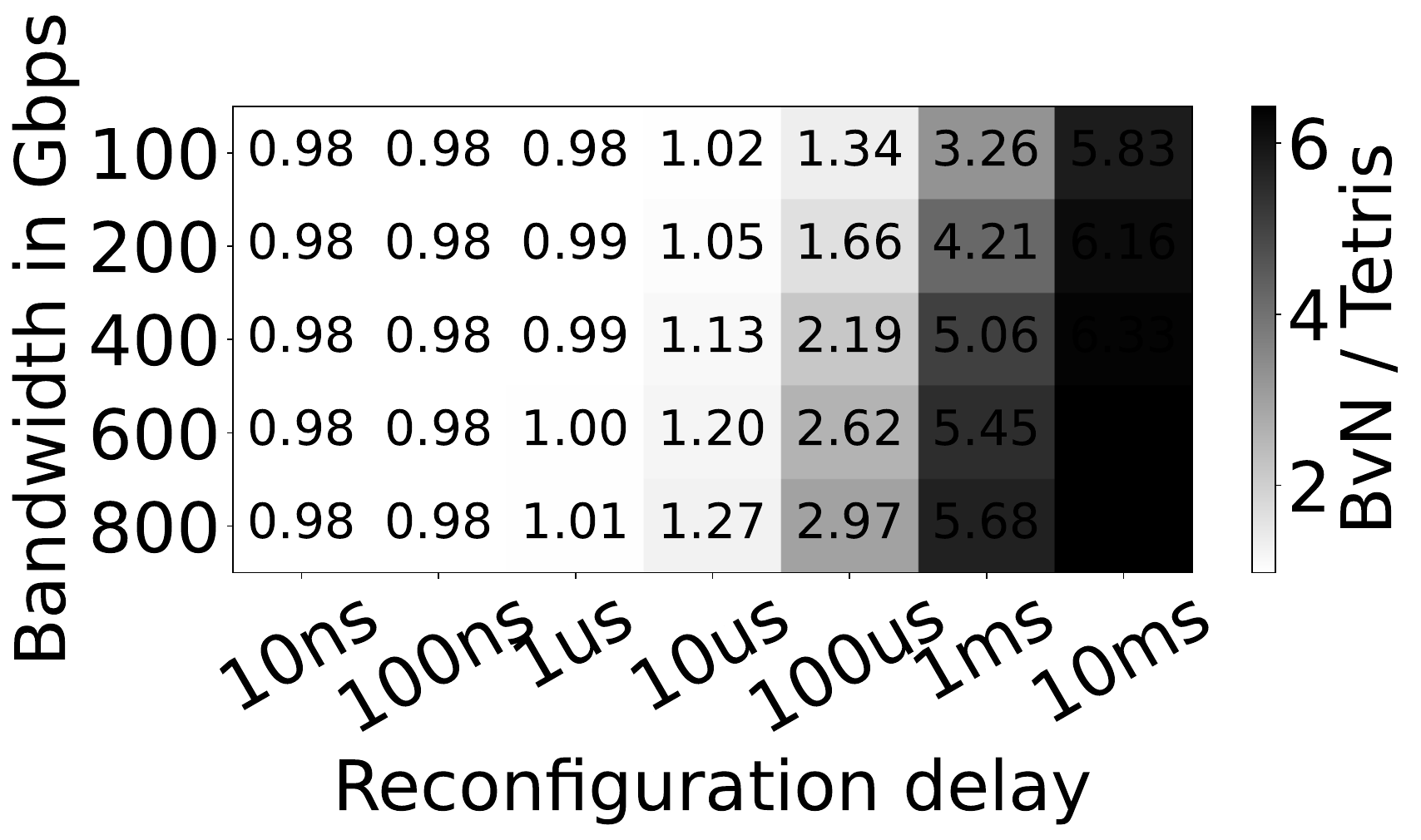}
    \caption{BvN against \name}
    \label{fig:heatmap-mixtral7b-bvn}
  \end{subfigure}%
  \hfill
   \begin{subfigure}{0.23\linewidth}
    \includegraphics[width=\linewidth]{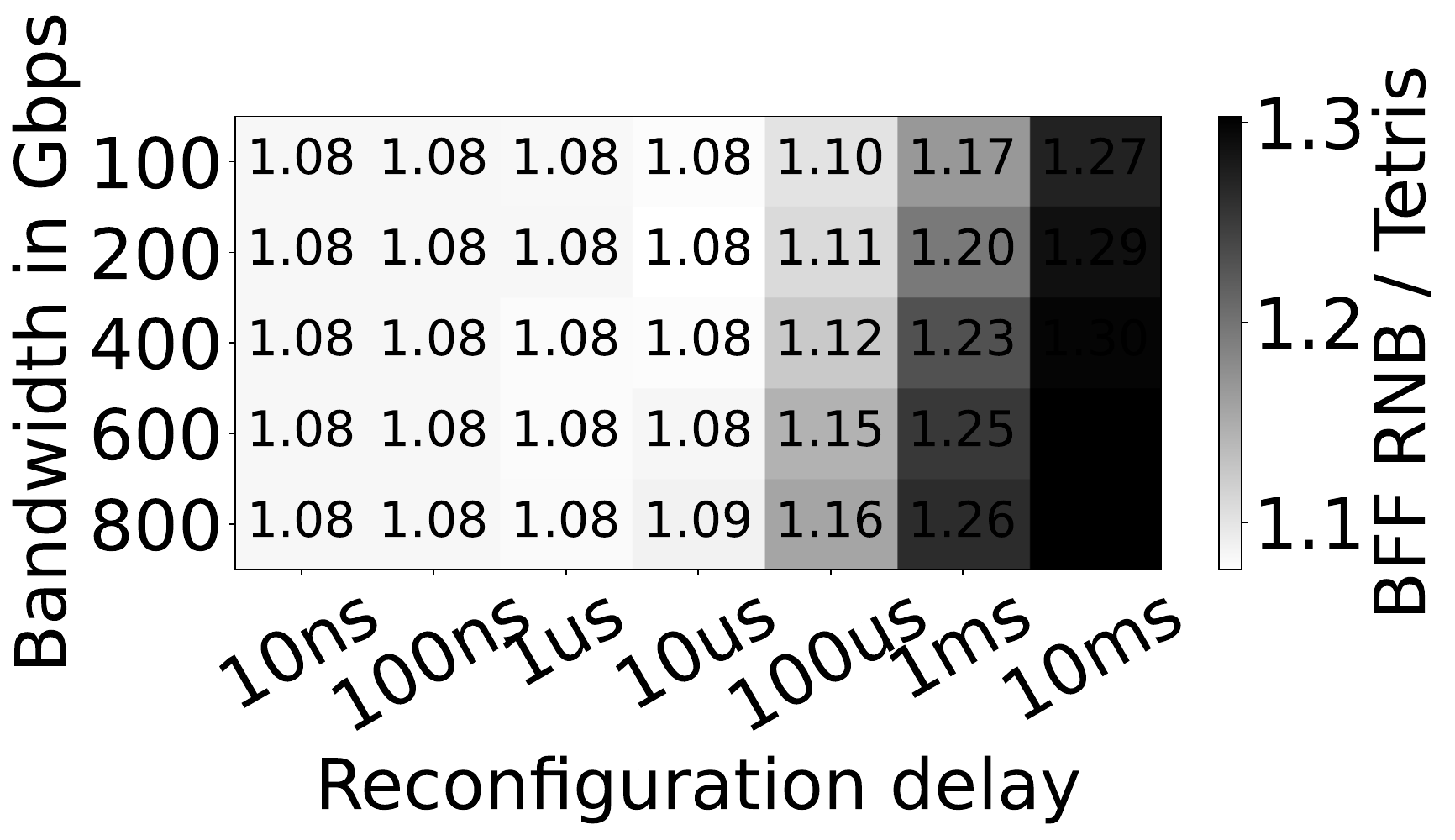}
    \caption{BFF against \name}
    \label{fig:heatmap-mixtral7b-bff}
  \end{subfigure}
    \hfill
   \begin{subfigure}{0.23\linewidth}
    \includegraphics[width=\linewidth]{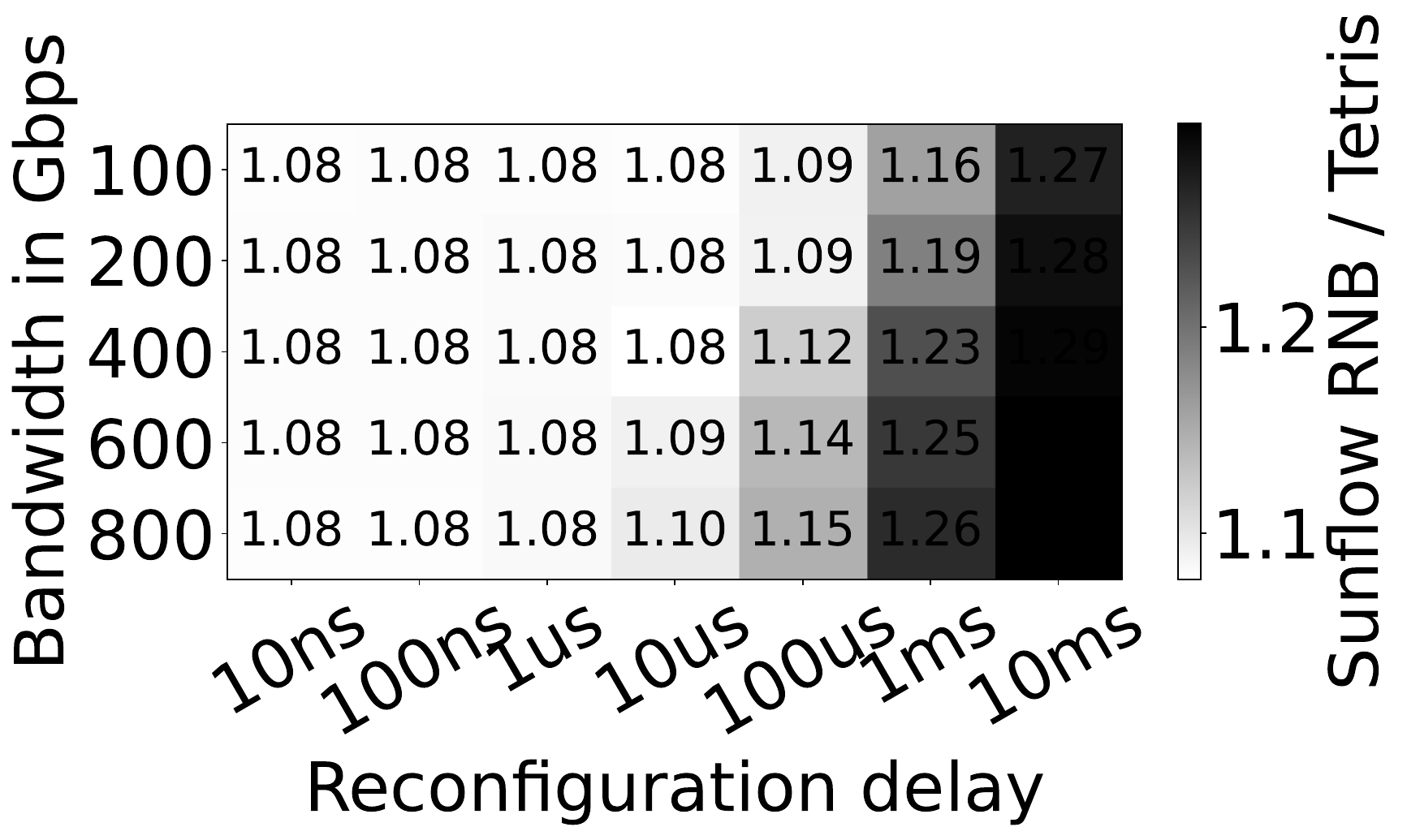}
    \caption{Sunflow against \name}
    \label{fig:heatmap-mixtral7b-sunflow}
  \end{subfigure}
   \hfill
   \begin{subfigure}{0.23\linewidth}
    \includegraphics[width=\linewidth]{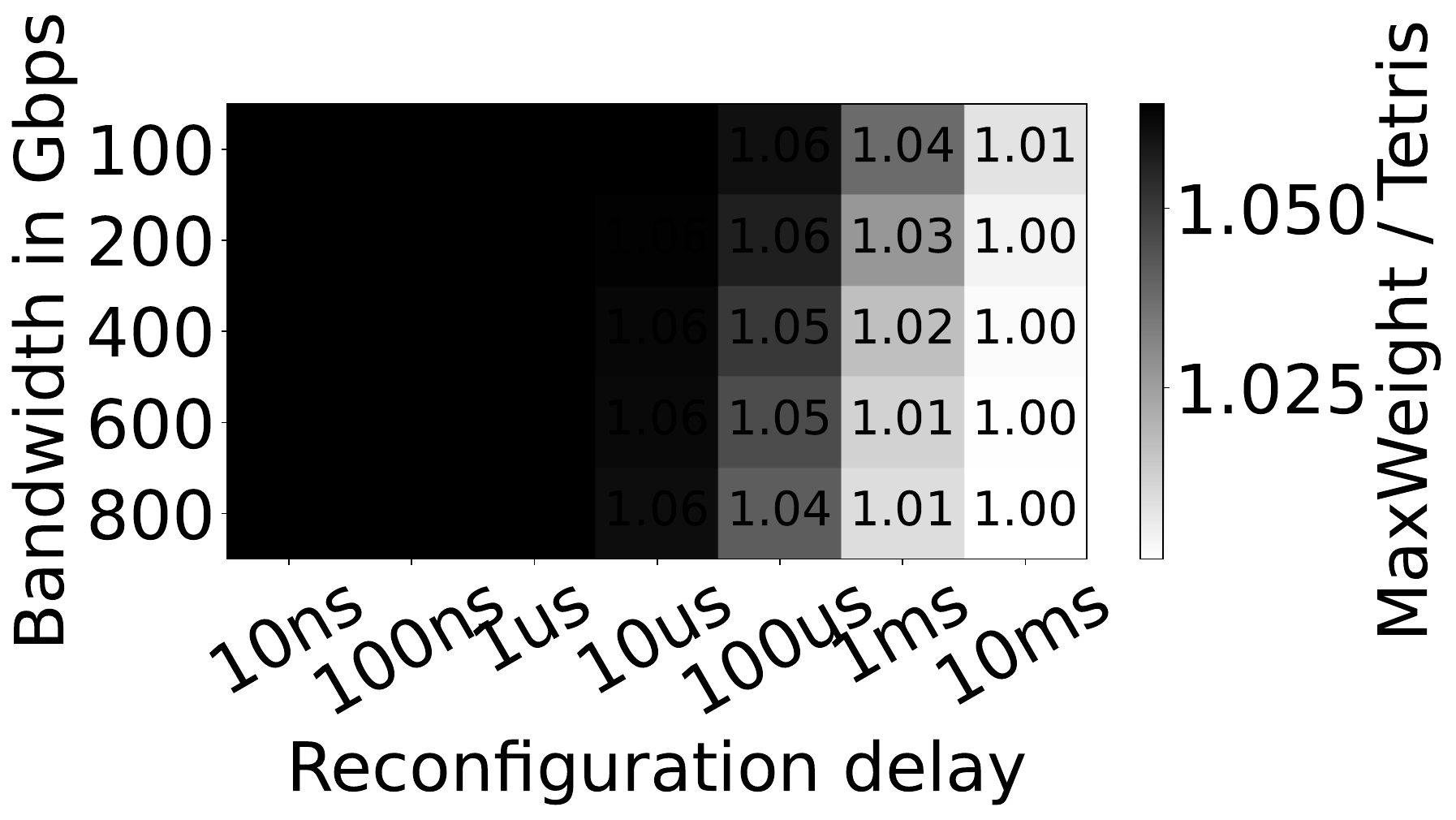}
    \caption{MaxWeight against \name}
    \label{fig:heatmap-mixtral7b-maxweight}
  \end{subfigure}

    \caption{Sensitivity of \name's advantage over each baseline to link
    bandwidth and reconfiguration delay, for one complete inference pass of
    DeepSeek-MoE-$16$B (top row) and Mixtral-$8\times7$B-Instruct (bottom
    row) on the SPEED-Bench throughput-$1$K dataset. The corresponding results
    for Mixtral-$8\times22$B appear in Figure~\ref{fig:heatmap}.}
    \label{fig:app:heatmap-models}
\end{figure*}

We report results omitted from \S\ref{sec:eval} for space, alongside the
interconnect comparison discussed in \S\ref{subsec:architecture}
(Table~\ref{tab:topo}), the notation used throughout \S\ref{sec:tetris}
(Table~\ref{tab:notation}), and \name's internal routing procedure (Algorithm~\ref{alg:tetris-routing}). Figure~\ref{fig:makespan} covers an entire forward pass of Mixtral8x22B and plots per layer completion time. Throughout the run, \name consistently achieves lower completion time than existing approaches, and remain close to optimal. Figure~\ref{fig:app:heatmap-models} shows \name's performance gains over BvN, MaxWeight, BFF, and Sunflow for DeepSeek-MoE-16B and Mixtral-$8\times7$B. The results are consistent with our observations in \S\ref{sec:eval}. Specifically, BvN remains competitive when reconfiguration is nearly free, but degrades rapidly as reconfiguration delay and bandwidth increase, becoming up to $6.35\times$ and $6.43\times$ slower than \name for DeepSeek-MoE-16B and Mixtral-$8\times7$B, respectively. BFF and Sunflow exhibit smaller but consistent gaps that grow with reconfiguration delay, reaching up to $1.32\times$ and $1.30\times$. MaxWeight shows the complementary trend: \name provides a modest advantage when reconfiguration is inexpensive, while the two converge as reconfiguration becomes dominant. These results indicate that the trends observed for Mixtral-$8\times22$B in \S\ref{sec:eval} generalize across different MoE models. 

\begin{figure*}
\centering
\begin{subfigure}{0.33\linewidth}
\centering
\includegraphics[width=1\linewidth]{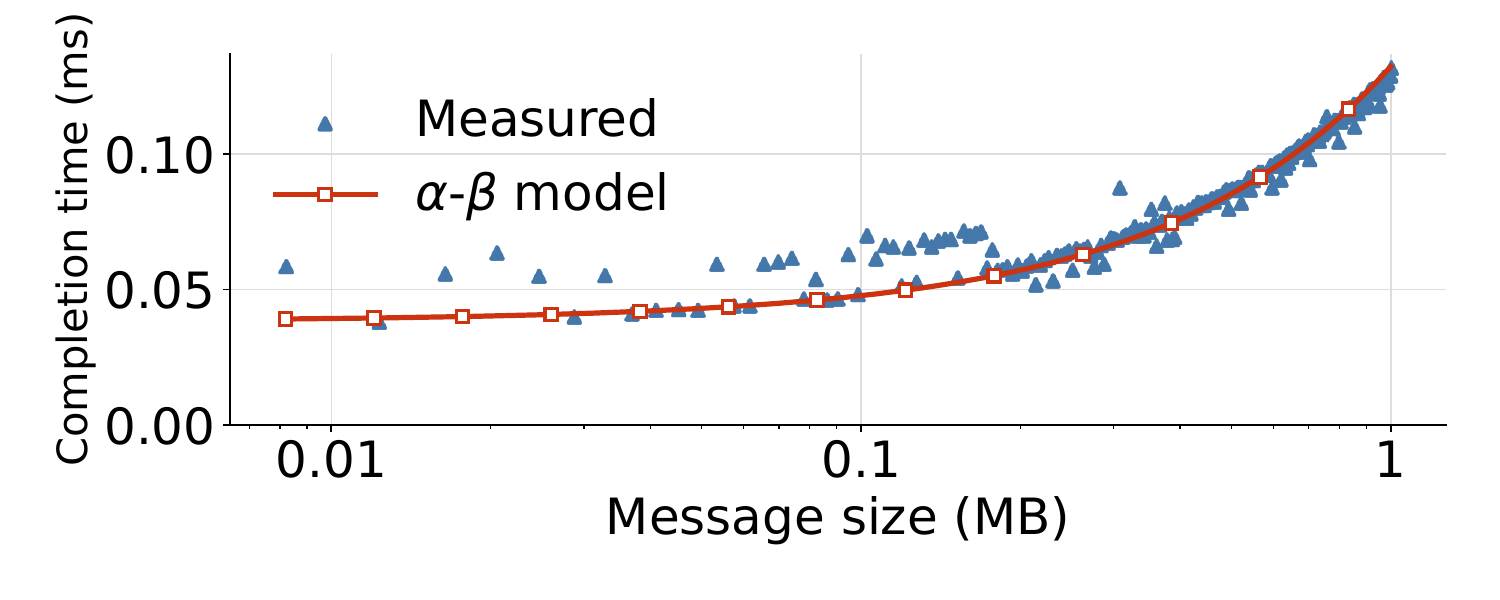}
\caption{Up to $1$ MB message sizes}
\label{fig:alpha-beta-small}
\end{subfigure}
\begin{subfigure}{0.33\linewidth}
\centering
\includegraphics[width=1\linewidth]{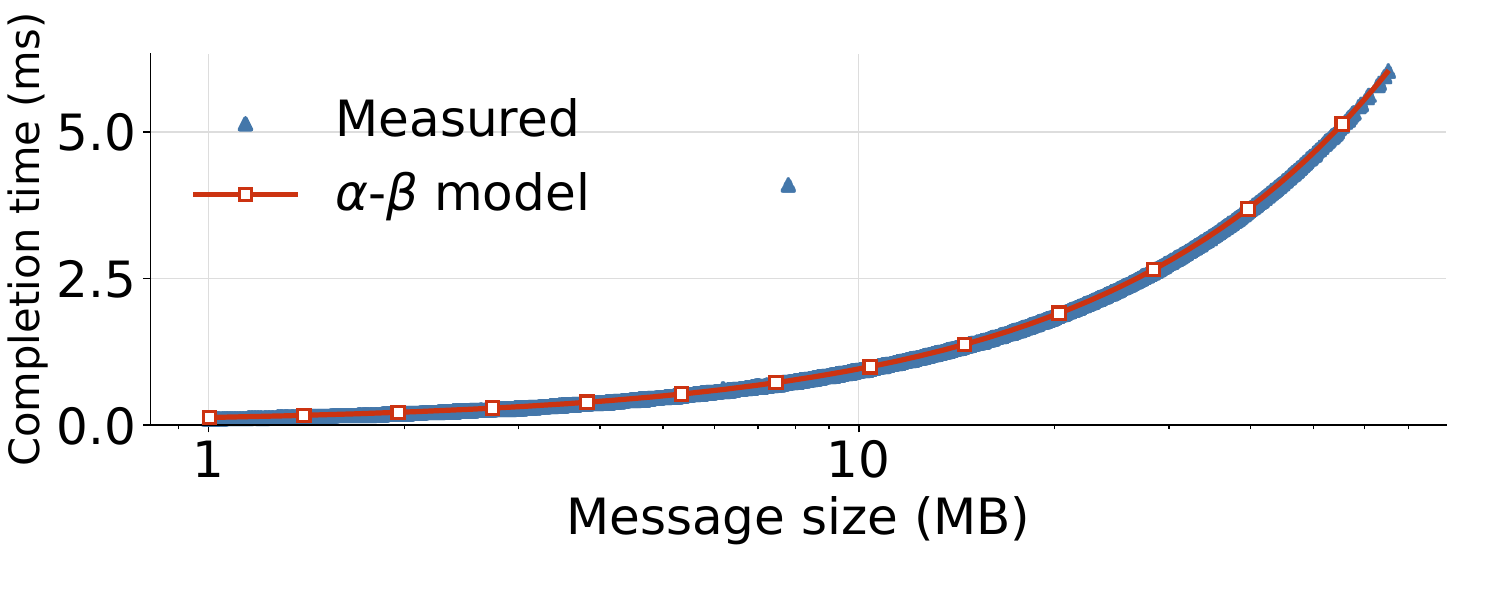}
\caption{$1$ MB -- $100$ MB message sizes}
\label{fig:alpha-beta-large}
\end{subfigure}
\begin{subfigure}{0.33\linewidth}
\centering
\includegraphics[width=1\linewidth]{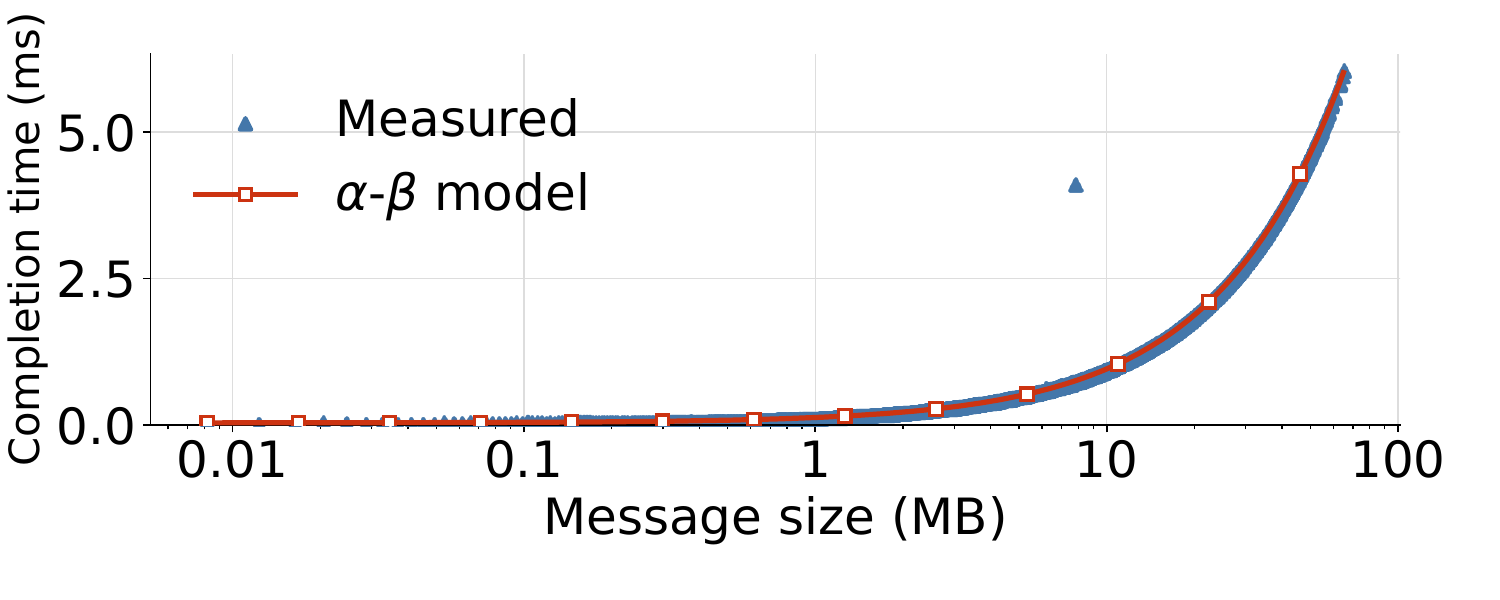}
\caption{Across all message sizes}
\label{fig:alpha-beta-all}
\end{subfigure}
\caption{GPU-to-GPU direct communication closely follows the $\alpha$-$\beta$
cost model~\cite{rogerwhockney}: message sizes beyond $1$ MB scale
near-linearly with the size of the transmission. We measure $\alpha \approx
38\mu s$ and $\frac{1}{\beta} \approx 86$ Gbps on a $100$ Gbps transceiver,
making our cost estimates a good proxy for GPU-to-GPU transmission times in
\name.}
\label{fig:alpha-beta}
\end{figure*}

\label{LastPage}

\end{document}